\documentclass[titlepage,12pt]{utarticle}   

\usepackage{graphics}
\usepackage{latexsym,amsmath,amssymb,amscd,cite}
\usepackage{epsf}

\DeclareFontFamily{U}{rsf}{}
\DeclareFontShape{U}{rsf}{m}{n}{<5> <6> rsfs5 <7> <8> <9> rsfs7 <10-> rsfs10}{}
\DeclareMathAlphabet\Scr{U}{rsf}{m}{n}

\newcommand {\PSL}{\rm{PSL}}
\newcommand{\nn}{ \nonumber }
\newcommand{\Z}{\mathbb{Z}}
\def\P{{\Bbb P}}
\def\R{{\Bbb R}}
\def\C{{\Bbb C}}
\def\deg{{\rm deg}}
\def\A{{\Bbb A}}

\newcommand {\e} {\epsilon}

\newcommand {\cF} {{\cal F}}

\newcommand {\cO} {{\cal O}}

\newcommand {\bbH} {\mathbb{H}}
\newcommand {\bbR} {\mathbb{R}}
\newcommand {\bbS} {\mathbb{S}}

\newcommand {\BS} {\Scr{B}}
\newcommand {\6} {\partial}
\newcommand {\td} {\tilde}
\newcommand {\bra}{\bigl\langle}
\newcommand {\ket}{\bigr\rangle}

\newcommand {\bulk}{{S^2}}
\newcommand {\ws}{{D^2}}
\newcommand {\bws}{{\partial\ws}}
\newcommand {\grd}{{\td\omega}}
\newcommand {\unit}{~1\!\!\!\!1~}
\newcommand {\rhs} {{(rhs)}}
\newcommand {\lhs} {{(lhs)}}

\def\cW{{\mathcal W}}

\def\cBB{\cF}
\def\cF{{\mathcal F}}
\def\cA{{\mathcal A}}
\def\BB#1{\cBB_{#1}}

\def\ssc#1{{^{(#1)}}}
\def\vp{\varphi}
\def\bl{\kappa}
\def\cR{{\mathcal R}}
\def\dl{\6}
\def\al{\alpha}
\def\bib#1#2{\newcommand{#1}{#2}}
\long\def\be #1 \ee{\begin{equation}#1\end{equation}}
\long\def\beq #1 \eeq{\begin{equation}#1\end{equation}}
\long\def\bea #1 \eea{\begin{eqnarray}#1\end{eqnarray}}
\long\def\del#1\enddel{}
\long\def\new#1\endnew{{\bf #1}}

\begin{document}        
\preprint{
  CERN--PH--TH/2004-31\\
  MAD-TH-04-2\\
  {\tt hep-th/0402110}\\
}
\title{Superpotentials, $A_\infty$ Relations and WDVV Equations
for Open Topological Strings}
\author{
Manfred Herbst${}^a$, Calin-Iuliu Lazaroiu${}^b$, Wolfgang Lerche${}^a$
} 
    \oneaddress{
     ${}^a$ Department of Physics, CERN\\
     Theory Division\\
     CH-1211 Geneva 23\\
     Switzerland\\
     \email{firstname.lastname@cern.ch}\\{~}\\
     ${}^b$ 5243 Chamberlin Hall\\
     University of Wisconsin\\
     1150 University Ave\\
     Madison, Wisconsin 53706, USA\\
     \email{calin@physics.wisc.edu}\\
    }

\Abstract{
We give a systematic derivation of the consistency conditions which constrain
open-closed disk amplitudes of topological strings. They include the
$A_\infty$ relations (which generalize associativity of the boundary product
of topological field theory), as well as certain homotopy versions of
bulk-boundary crossing symmetry and Cardy constraint.  We discuss integrability
of amplitudes with respect to bulk and boundary deformations, and write down
the analogs of WDVV equations for the space-time superpotential. 
We also study the structure of these equations from a string field theory
point of view.  As an application, we determine the
effective superpotential for certain families of D-branes in B-twisted
topological minimal models, as a function of both closed and open string
moduli.  This provides an exact description of tachyon condensation in such
models, which allows one to determine the truncation of the open string spectrum
in a simple manner.}
\date{}

\maketitle
\tableofcontents
\pagebreak


\section{Introduction and Summary}


Closed topological string theories have played a very important role in the
past \cite{WittenNLSM,WittenTG,Wittenmirror,Labastida,BCOV}.  They capture the
exactly solvable sector of $N=2$ supersymmetric string compactifications and govern
holomorphic quantities such as the prepotential of the effective action.
The tree-level amplitudes of topological strings can often be computed by
geometric methods, see for example: \cite{Candelas,KlemmA,
Greene,mirbookA,mirbookB,mirbookC}.  These amplitudes
satisfy a hierarchy of consistency conditions, such as crossing symmetry,
factorization constraints, the WDVV equations\cite{WDVV} and $t-t^*$
equations\cite{BCOV}, which are often strong enough to determine them.

At the level of (conformal) closed topological field theory (TFT) in two dimensions, one
considers correlators of unintegrated and BRST closed zero-form
observables. Such correlators must obey the sewing constraints of
\cite{Sonoda}, namely crossing symmetry on the sphere and modular invariance
on the torus. Restricting to tree-level, one is left with crossing symmetry
only, which encodes associativity of the operator product. The WDVV equations
are a consequence of this condition. They constrain tree level amplitudes,
which are correlators on the sphere containing three zero-form observables and
an arbitrary number of integrated two-form descendants. These relations can be
derived within (twisted $N=2$) TFT \cite{WDVV}, and are part of a broader
hierarchy of conditions known as the $t-t^*$ equations \cite{BCOV}, which
constrain topological string amplitudes at all genera.

The WDVV equations can be derived from the crossing symmetry relation as follows.
One starts with the constraint:
\begin{equation}
\label{intC}
\6_i C_{jkl}(t)\ =\ \6_j C_{ikl}(t)\ , 
\end{equation}
where
\begin{equation}
\nn
C_{ijk}(t)\ =\ \langle\phi_i\phi_j\phi_k
e^{\sum_{l} t_\ell\int\phi^{(2)}_\ell}\rangle~
\end{equation}
are the triple correlators on the sphere deformed by integrated descendants: 
\begin{equation}
\nn
\int\phi_\ell^{(2)}=
\int{dz\wedge d{\bar z} [G_{-1}, [{\bar G}_{-1},\phi_\ell]] }~~.
\end{equation} 
Here $t_i$ are the flat deformation parameters and $\6_i\equiv{\6\over\6 t_i}$.
\del
We shall loosely refer to the deformation parameters $t_i$ as moduli
even though they need not correspond to exactly marginal deformations
in the sense of conformal field theory.
\enddel
Equation (\ref{intC}) expresses the fact  that the four-point function
$C_{ijkl}\equiv \6_i C_{jkl}$ is completely symmetric in
all indices. This implies integrability of the deformed
triple correlator:
\begin{equation}
\nn
C_{ijk}(t)\ =\ \6_i\6_j\6_k \,\cF(t)\ ~~,
\end{equation}
where ${\cal F}$ is the generating function of genus zero amplitudes, 
known as the WDVV potential. In appropriate situations, this quantity can be
interpreted as the effective prepotential of a
Calabi-Yau compactification of the associated untwisted superstring theory.

The crossing symmetry of four-point correlators now gives a system of equations
for $\cF$:
\begin{equation}
\nn
\6_i\6_k\6_m\cF\,\eta^{mn}\,\6_n\6_j\6_l\cF
\ =\ 
\6_i\6_j\6_m\cF\,\eta^{mn}\,\6_n\6_k\6_l\cF\ ~~,
\end{equation}
where $\eta^{mn}$ denotes the inverse of the topological metric
$\eta_{mn}=C_{0mn}$~, which one can show to be independent of the deformation
parameters $t_i$. These are the famous associativity, or WDVV equations \cite{WDVV}. They
constrain the generating function ${\cal F}(t)$, and encode a Frobenius 
structure on the associated moduli space \cite{Dubrovin}. 

In principle, the WDVV equations allow one to determine ${\cal F}$ without 
directly computing all genus zero amplitudes\footnote{Some applications 
along these lines can be found in \cite{WDVV,LSW,VerlindeWarner,Dubrovin,KlemmTheisen}.}. 
Such amplitudes are affected by contact terms arising from colliding operators, and
therefore are hard to compute directly. The $t-t^*$ extension of the
WDVV system constrains higher genus contributions to the generating function,
and can be used to determine the gravitational $F$-terms of the corresponding
effective action \cite{BCOV}.

For open topological strings, which correspond to world-sheets
with boundaries, the situation is much more complicated.
Although geometric methods have been successfully applied
to computing non-trivial effective $N=1$ superpotentials from $D$-branes (see e.g., \cite{KachruA,KachruB,Vafa,Klemm,MayrFF,MayrLoc,LMWa,LMWb}) ,
they were tailored to specific D-brane geometries (notably 
non-compact toric Calabi-Yau manifolds and their mirrors) and it is not obvious how
these methods can be generalized to other classes of D-brane
backgrounds. 

Experience with closed string TFT suggests that a good strategy should be to first study the
consistency conditions constraining open-closed string amplitudes, and then
translate the results into a geometric structure associated with the moduli
space of D-branes.
The work of \cite{HofmanMaA} (see also \cite{HofmanMaB,HofmanB}) discussed 
particular cases of the relevant algebraic constraints for open-closed topological
strings (the basic conditions on pure
boundary amplitudes on the disk were given in \cite{Gaberdiel} in the
context of open string field theory). However, the full set of constraints
on open-closed amplitudes on the disk has not yet been worked out
completely, and in particular was  not clearly understood for deformation families
of amplitudes with integrated insertions.

A new aspect of open topological strings is the generic lack of integrability
of disk amplitudes with respect to boundary deformation parameters, and
relatedly, the lack of flat coordinates. This 
arises because such amplitudes are only cyclically symmetric 
with respect to boundary insertions. 
Another complication is the fact that open topological field theories in 
two dimensions have more sewing constraints than their closed counterparts. 
As shown in \cite{Lewellen,MooreSegal,Moore,CILtop}, one finds four 
sewing conditions involving boundary amplitudes, namely crossing
symmetry on the disk, 
two bulk-boundary crossing relations and a topological version of the Cardy
constraint (the other two conditions involve only bulk correlators).
As a consequence, one has more families of algebraic
and differential conditions on deformed tree-level open-closed amplitudes. 

The first constraint mentioned above encodes associativity of the boundary
operator product. Because the product fails to be commutative, 
the correct stringy generalization of this condition turns out to be
more complicated than for closed strings. Namely, one
finds a series of equations reflecting an $A_\infty$ structure, which can be
derived by using the Ward identities of the BRST operator.  
This means that the associated 'string products' (the stringy generalizations
of the boundary operator product) are associative 'up to homotopies'. This
$A_\infty$ structure is cyclic, due to cyclicity of disk amplitudes in the
boundary insertions; it is also unital and minimal. 
The fact that tree-level open string products obey $A_\infty$
constraints was originally pointed out in \cite{Gaberdiel} in the context of
open string field theory (see also \cite{Zwiebachoc}). It was further
discussed in \cite{CILc,CILd,CILe} as the underlying structure controlling 
D-brane superpotentials, following ideas originally put forward in
\cite{WittenCS}. Further discussion of such relations in the context of
bosonic string field theory can be found in \cite{KajiuraA,KajiuraB}.
Such constraints also played a role in \cite{HofmanMaA,HofmanB,Tomasiello}.
$A_\infty$ constraints are central to the homological mirror 
symmetry program 
\cite{KontsevichHMS,Fukayainfty,Fukayactg,Fukayabook,Fukayarev,FukayamirrorA,
FukayamirrorB,Polishchuk}, where they arise via open string field 
theory \cite{CILf} (see also \cite{CILa,CILb,Diaconescu,CILg,CILh}). 

Upon perturbing boundary disk amplitudes via bulk insertions,
the $A_\infty$ algebra deforms in a manner compatible with cyclicity.  Since
linearized deformations of cyclic $A_\infty$ algebras are controlled by their cyclic
complex, the first order approximation leads to a map which associates a cocycle of this complex
to each BRST-closed bulk insertion. 
In similar manner, the appropriate generalization of the remaining sewing constraints 
(namely bulk-boundary crossing symmetry
and the Cardy condition) is given by two countable sets of algebraic
conditions on bulk-boundary amplitudes on the disk.  

Our main purpose is to derive this series of constraints from
the Ward identities of a general, twisted $N=2$ topological 
theory on the disk.  We shall express these conditions as a countable
family of nonlinear algebraic and differential equations which constrain
the moduli-dependent disk amplitudes; these relations
constitute the open string analogs of the WDVV equations.

The paper is organized as follows. 
In Section \ref{sec:preliminaries}, we briefly recall some basic
features of open-closed topological field theories. 
In Section \ref{sec:corrfunc}, we define the genus zero, deformed open string amplitudes:
$$
 B_{a_0\ldots a_m;i_1\ldots i_n}\ =\
    (-1)^{\td a_1+\ldots +\td a_{m-1}}\bra~ \psi_{a_0} \psi_{a_1} ~
  P \int\psi^{(1)}_{a_2}\ldots \int\psi^{(1)}_{a_{m-1}} ~
  \psi_{a_m} \int \phi^{(2)}_{i_1}\ldots \int \phi^{(2)}_{i_n}~
  \ket_{\rm disk}\ ,
$$
and discuss their basic properties such as independence of the worldsheet
metric, cyclicity with respect to boundary
insertions and constancy of the boundary two-point function. 
A shift in the grading of the boundary fields, denoted by $\td a_i\equiv a_i+1$ mod $2$,
will play a crucial role. This shift reflects the change in the degree
of operators induced by super-integration over the moduli of boundary insertions, and 
gives a physical realization  of the ``suspension'' operation used
in the mathematics literature.

In Section \ref{sec:superpot}, we discuss the issue of integrability
with respect to the bulk and boundary perturbation parameters,
denoted by $t_i$ and $s_a$, respectively. We show that correlators
deformed by bulk operators integrate to disk amplitudes 
$\cF_{a_0\dots a_m}(t)$, so that:
\begin{equation}
B_{a_0\ldots a_m;i_1\ldots i_n}=\partial_{i_1}\dots\partial_{i_n}
{\cal F}_{a_0\dots a_m}(t)|_{t=0}~~.
\end{equation}
Because disk amplitudes are only cyclically
(rather than completely) symmetric in the boundary
fields, they are {\it a priori} not integrable with respect to the
boundary deformation parameters.  However, by promoting these 
to formal non-commutative variables ${\hat s}_a$, one can define a formal generating function
${\hat\cW}(\hat s,t)$ through the expansion:
\begin{equation}
{\hat {\cal W}}({\hat s},t)=\sum_{m\geq 1}{\frac{1}{m}{\hat s}_{a_m}\dots
  {\hat s}_{a_1}{\cal F}_{a_1\dots a_m}(t) }~~.
\end{equation}
This generating function encodes all information contained in the disk
correlators, and can be used to define a sort of formal noncommutative
Frobenius (super)manifold.  To recover a physically meaningful
quantity, one can impose (super)commutativity of ${\hat s}_a$ by
working modulo the ring generated by their commutators. Denoting
the resulting equivalence classes by $s_a$,  this gives a quantity
${\cal W}(s,t)$ which in the appropriate framework can be identified
with the effective $N=1$ superpotential of a four-dimensional
superstring compactification with D-branes, where $s_a$ and $t_i$
correspond to vacuum expectation values.

Notice the difference to the bulk effective
prepotential $\cBB(t)$: its derivatives directly yield all bulk topological
amplitudes, while the $s_a$-derivatives
of $\cW(s,t)$ give only sums of correlators, namely the (super)symmetrized
version of the boundary amplitudes:
\begin{equation}
\nn
\cA_{a_0...a_m}(t)\equiv m!\,\cBB_{(a_0...a_m)}(t)
\end{equation}
The effective superpotential ${\cal W}(s,t)$ thus contains {\it less} information
than that provided by the full collection of disk amplitudes. This is exemplified by the
consistency relations when applied to the super-symmetrized quantities
$\cA_{a_0...a_m}(t)$: sometimes these equations are nearly
empty, in contrast to the full constraints on the original amplitudes 
$\cF_{a_0...a_m}(t)$, which are only cyclically symmetric.
The conditions on $\cW(s,t)$ induced by the consistency conditions
can be viewed as a weak form of the generalized WDVV equations.

In Section \ref{sec:Ainfty}, we begin
our analysis of the Ward identities, 
by first deriving a series of conditions on undeformed boundary amplitudes on the
disk, which generalize associativity  of the boundary product. These take the form:
\begin{equation}
\nn
  \sum 
  (-1)^{\td a_1 + \ldots + \td a_{l-2}}
  B^b{}_{a_1\ldots a_{l-2}ca_{k+1}\ldots a_m} B^c{}_{a_{l-1}\ldots a_k}
  = 0~~,
\end{equation}
and encode a so-called minimal $A_\infty$ structure. Due to cyclicity of 
amplitudes with respect to boundary insertions, this is in fact a cyclic
$A_\infty$ algebra; we also show that it admits a unit.  

Section \ref{sec:def_constraints} considers general bulk-boundary amplitudes on the disk.
We discuss deformations induced by an arbitrary number of 
bulk insertions, and show that the deformed boundary amplitudes $\cF_{a_0\dots a_m}(t)$
preserve a weak, unital and cyclic $A_\infty$ structure. 
In Subsection \ref{sec:bb_crossing}, we extract certain identities
generalizing the bulk-boundary crossing symmetry constraint of two-dimensional TFT.
These identities take the form:
\begin{eqnarray}
  &&\6_i \6_j \6_k \cF(t) ~\eta^{kl}~ \6_l \cF_{a_0 a_1 \ldots a_m}(t) =
 \nonumber \\[5pt]
  &=&\sum
  (-1)^{\tilde a_{m_1+1}+\ldots +\tilde a_{m_3}}
  \cF_{a_0\ldots a_{m_1}ba_{m_2+1}\ldots a_{m_3}ca_{m_4+1}\ldots a_m}(t)
  ~\6_i \cF^b{}_{a_{m_1+1}\ldots a_{m_2}}(t)
  ~\6_j \cF^c{}_{a_{m_3+1}\ldots a_{m_4}}(t)~~.
   \nonumber~
\end{eqnarray}
Subsection \ref{sec:Cardy} discusses the generalization of 
the topological Cardy constraint, which is given by the following series of equations:
\begin{eqnarray}
  &&\6_i \cF_{a_0 \ldots a_n}(t) \eta^{ij}
  \6_j \cF_{b_0 \ldots b_m}(t)  = \nonumber\\[5pt]
  &=&
  \sum  (-1)^{s+\td c_1 +\td c_2}~
  \omega^{c_1d_1}~ \omega^{c_2d_2}~
  \cF_{a_0 \ldots a_{n_1} d_1 b_{m_1\!\!+\!1} \ldots b_{m_2} c_2 a_{n_2\!+\!1}
  \ldots a_{n}}(t) ~
  \cF_{b_0 \ldots b_{m_1} c_1 a_{n_1\!\!+\!1} \ldots a_{n_2} d_2 b_{m_2\!+\!1}
  \ldots b_{m}}(t)~~.\nonumber~
\end{eqnarray}

In Section \ref{sec:TLG}, we demonstrate the power of the open
string consistency conditions by applying them to topological minimal
models with a boundary. We find that they give a highly overdetermined
system of equations which uniquely determines all disk amplitudes
(as functions of both open and closed string deformation parameters)
thereby fixing the effective superpotential $\cW(s,t)$. As a consistency
check, we show that the critical loci of $\cW(s,t)$ correspond to
a factorized Landau-Ginzburg superpotential, which is known
\cite{KontsevichLG,KapA,BHLS,KapB,KapC,CILLG} to be the criterion
for unbroken supersymmetry. Starting from the unperturbed theory
and moving along these loci describes an exactly solvable, topological
version of tachyon condensation, which truncates the open string
spectrum in a computable manner.


\section{Preliminaries}

\label{sec:preliminaries}

\subsection{Bulk topological conformal field theory}
\label{WDVV_bulk}

Let us start with a
brief review of bulk topological conformal field theory
\cite{WDVV,BCOV}. This will prove useful later on, when we discuss
the boundary extension. We shall consider a closed conformal field theory with 
topological conformal symmetry on the worldsheet. 
Denoting by $T(z)$ and ${\bar T}({\bar z})$ the left and right moving
components of the stress-energy tensor, this implies the existence of odd scalar charges
$Q_0, {\bar Q}_0$ and spin two fermionic currents 
$G(z)$ and ${\bar G}(\bar z)$ such that $Q_0^2={\bar Q}_0^2=0$ and:
\begin{eqnarray}
\label{eq:EMtensor}
T(z)=[Q_0,G(z)]~~,
\end{eqnarray}
with a similar relations for the right movers.
We also have $U(1)$ charges $J$ and ${\bar J}$ 
with the property:
\begin{equation}
\nn
[J,T(z)]=0~~,~~[J,Q_0]=Q_0~~,~~[J,G(z)]=-G(z)~~,
\end{equation}
and a similar relation for the left movers. 
Here and below we use $[\cdot,\cdot]$ to denote the {\em super}commutator. 
This data defines what is generally called a {\em string background}
\cite{KSV,Voronov}.

Using the mode expansions:
\begin{eqnarray}
\nn T(z)&=&\sum_{n\in \Z}{L_n z^{-n-2}}\\
\nn G(z)&=&\sum_{n\in \Z}{G_n z^{-n-2}}~~,
\end{eqnarray}
one finds the algebra:
\begin{eqnarray}
\label{algebra}
\nn\left[L_m,L_n\right]=(m-n)L_{m+n}~~&,&~~\left[L_m,G_n\right]=(m-n)G_{m+n}\\
\left[L_m,Q_0\right]=0~~&,&~~\left[L_m,J_0\right]=0\\
\nn\left[G_m, Q_0\right]=L_{m}~~&,&~~\left[J_0,J_0\right]=0\\
\nn\left[J_0,G_n\right]=G_{n}~~&,&~~\left[J_0,Q_0\right]=-Q_0~~,
\end{eqnarray}
with similar relations constraining the right movers.

We let $\phi_i(z,{\bar z})$ ($i=0..h_c-1$) be a collection of zero-form operators which satisfy:
\begin{equation}
\nn
[Q_0,\phi_i]=[{\bar Q}_0,\phi_i]=0~~.
\end{equation}
We shall assume that this system is complete in the sense that it descends to
a basis of the space of on-shell observables\footnote{Through the
operator-state correspondence, $H_c$ can be identified with the space of
on-shell oscillation states of the closed topological string. Then $\phi_i$ can be viewed
as linear operators on $H_c$.}  $H_c$, which is
the double BRST cohomology computed with $Q_0$ and ${\bar Q}_0$.
We choose $\phi_i$ such that $\phi_0$ coincides with the bulk identity operator $1_c$. For
simplicity, we shall also assume that  each $\phi_i$ is Grassmann even. 
This simplifies certain sign prefactors in later sections and suffices for our
main application, which concerns topological Landau-Ginzburg models. With this
assumption, $H_c$ is a complex vector space of dimension $h_c$ (as opposed to
a super-vector space, which is the general case).

Given such operators, one can construct their descendants by using 
relation (\ref{eq:EMtensor}) and its counterpart for the right movers, which imply:
\begin{equation}
\nn
[Q_0,G_{-1}] = L_{-1}~~,~~[{\bar Q}_0, {\bar G}_{-1}]={\bar L}_{-1}~~.
\end{equation} 
Since the commutator with $L_{-1}$ and ${\bar L}_{-1}$ acts as $\frac{\partial}{\partial z}$ and 
$\frac{\partial}{\partial {\bar z}}$, we find that the operators:
\begin{eqnarray}
\nn\phi_i^{(1,0)}&=&[G_{-1},\phi_i]dz\\
\nn\phi_i^{(0,1)}&=&[{\bar G}_{-1},\phi_i]d{\bar z}\\
\nn\phi_i^{(1,1)}&=&[G_{-1},[{\bar G}_{-1},\phi_i]]dz\wedge d{\bar z}= [{\bar
G}_{-1},[G_{-1},\phi_i]]d{\bar z}\wedge dz~
\end{eqnarray}
satisfy the descent equations:
\begin{eqnarray}
  \label{eq:descendants}
\nn [Q_0, \phi_i^{(1,0)}] = \partial \phi_i~~
    &,&~~[{\bar Q}_0, \phi_i^{(1,0)}]=0\\[5pt]
    [Q_0,  \phi_i^{(0,1)}] =0~~
    &,&~~ [{\bar Q}_0, \phi_i^{(0,1)}]={\bar \partial}\phi_i \\[5pt]
\nn[Q_0, \phi_i^{(1,1)}] =\partial \phi_i^{(0,1)}=d \phi_i^{(0,1)}~~
    &,&~~ [{\bar Q}_0, \phi_i^{(1,1)}]={\bar\partial}
                        \phi_i^{(1,0)}=d\phi_i^{(1,0)}~~.
\end{eqnarray}
Notice that $\phi_i^{(1,0)}$ and $\phi_i^{(0,1)}$ are operator-valued sections
  of the canonical and anticanonical line bundles over $\P^1$, while $\phi_i^{(1,1)}$
is an operator-valued two-form. The integrated operators:
\begin{equation}
  \label{eq:intbulk}
  \int_{S^2} \phi_i^{(1,1)}
\end{equation}
are both $Q_0$- and ${\bar Q}_0$-closed. 
One can also write down BRST-closed loop integrals of the one-form descendants
$\phi_i^{(0,1)}+\phi_i^{(1,0)}$, but those observables do not play a
role for what follows.

The fundamental correlators of the topological field theory at tree level are 
the two- and tree-point functions $\eta_{ij}:=\langle \phi_i\phi_j\rangle$ and 
$C_{ijk}=\langle \phi_i\phi_j\phi_k\rangle$ of zero-form observable. 
Notice that $\eta_{ij}=C_{0ij}$. It is known that this quantity defines a
non-degenerate symmetric pairing on the space of zero-form
observables.  Any tree-level correlator of the form $\langle \phi_{i_1}\dots
\phi_{i_n}\rangle$ depends only on the BRST cohomology classes of
$\phi_{i_j}$, is completely symmetric under permutations of these
operators, and factorizes as a sum of products of three-point
functions, with insertions of the inverse $\eta^{ij}$ of the
topological metric. At tree level, 
compatibility of various factorizations of a given $n$-point function 
is assured by a single sewing constraint, namely crossing symmetry of the four
point function on the sphere\cite{Sonoda}. This condition says that the three distinct
factorizations of this function must agree:
\begin{equation}
  \label{eq:fact4}
\langle \phi_i\phi_j\phi_k\phi_l\rangle=
C_{ijm} \eta^{mn} C_{nkl}=C_{ikm}
\eta^{mn} C_{njl}=C_{ilm}\eta^{mn}C_{njk}~~.
\end{equation}
Due to symmetry of $C_{ijk}$ under permutations of its indices, 
only one of the two equalities to the right is an independent condition.

We next consider tree-level amplitudes\footnote{Throughout this paper, we
  shall use the term {\em amplitudes} for correlators containing integrated
  insertions of descendants. Correspondingly, 
correlators without integrated descendants will be referred to as {\em TFT
  correlators}. With this distinction, amplitudes can be represented as
  integrals of correlators over the moduli space of the underlying Riemann
  surface with punctures. Notice that TFT correlators can be described
  entirely in the simple algebraic framework of topological field theory, which
  reduces them to algebraic building blocks (see
  \cite{MooreSegal,CILtop,Moore} for a complete analysis in the open-closed
  case). On the other hand, amplitudes are related to 
  scattering in the topological string
  theory built by considering such a TFT on the worldsheet. It is such
  amplitudes which form the main focus of the present paper. }, which have the form:
\begin{equation}
  \label{eq:bulkfund}
  C_{i_1 \ldots i_n} := 
\bra 
\phi_{i_1} \ldots \phi_{i_3}   \int \phi^{(1,1)}_{i_4} \ldots \int \phi^{(1,1)}_{i_n}\ket_{\bulk}~~.
\end{equation}
These differ from topological field theory correlators since they contain
  integrated insertions of higher form operators. Such insertions implement
  integration of the underlying zero-form correlator 
$\langle \phi_{i_1} \dots \phi_{i_n}\rangle_{\bulk}$
over an appropriate compactification of the configuration space of $n$ points
  on the sphere. 

To derive the WDVV equations, one uses conformal invariance and the Ward
identities for the supercurrents $G(z)$ and ${\bar G}(\bar z)$. Though in this paper we
assume full twisted $N=2$ topological symmetry, it is worth noting that the 
derivation of \cite{WDVV} only depends on the following properties:

(A) $Q_0$ and ${\bar Q}_0$ are symmetries of the theory

(B) $G_{-1},G_0,G_1$ and their right-moving counterparts are symmetries.

Through equations (\ref{eq:EMtensor}), this implies that the $\PSL(2,\C)$
group generated by $L_{-1},L_0$ and $L_{1}$ and their right-moving
counterparts is also a symmetry. Using these assumptions, one can show that \cite{WDVV}:

(I) The tree-level string amplitudes $C_{0i_2\dots i_n}$ vanish for $n\geq 3$.

(II) The amplitudes $C_{i_1\dots i_n}$ are symmetric
    under permutations of $i_1\dots i_n$.

Let us define perturbed string amplitudes by the expression:
\begin{eqnarray}
\nn
C_{i_1\dots i_n}(t)=
\langle \phi_{i_1}\phi_{i_2}\phi_{i_3}
\int_{S^2}{\phi_{i_4}^{(1,1)}}\dots \int_{S^2}{\phi_{i_n}^{(1,1)}}
 e^{\sum_{p=0}^{h_c-1}{t_p\int_{S^2}{\phi_p^{(1,1)}}}}\rangle~~,
\end{eqnarray}
which is understood as the formal power series:
\begin{equation}
\nn
C_{i_1\dots i_n}(t)=\sum_{N_0\dots N_{h_c-1}=0}^\infty
{\prod_{p=0}^{h_c-1}{\frac{t_p^{N_p}}{N_p!}}
\langle 
\phi_{i_1}\phi_{i_2}\phi_{i_3}
\int_{S^2}{\phi_{i_4}^{(1,1)}}\dots \int_{S^2}{\phi_{i_n}^{(1,1)}}  
\prod_{p=0}^{h_c-1}{\left[\int_{S^2} \phi^{(1,1)}_{p}\right]^{N_p}} \rangle}~~.
\end{equation}
Here $t=(t_0\dots t_{h_c-1})$ is a collection of complex-valued parameters.
Using property (II), we can express all deformed amplitudes on the sphere 
with at least four insertions as partial derivatives of
the deformed three-point function: 
\begin{equation}
C_{i_1\dots i_n}(t)=\partial_{i_4}\dots
\partial_{i_n}C_{i_1i_2i_3}(t)|_{t=0}~~{\rm ~for~}n\geq 3~~.\nn
\end{equation}  
Here and below we use the notation $\partial_i:=\frac{\partial}{\partial
  t_i}$.

Then property (I) shows that the perturbed topological metric
$\eta_{ij}(t):=C_{0ij}(t)$ is independent of the parameters $t$. 
On the other hand, property (II) implies that $C_{i_1\dots i_n}(t)$
are symmetric under all permutations of indices. In particular, 
we find that $\partial_i C_{jkl}(t)$ is completely symmetric in $i,j,k$
and $l$. This is an integrability property allowing us to write the
deformed three-point correlator as a
triple derivative of a function ${\cal F}(t)$:
\begin{equation}
\label{pot}
C_{ijk}(t)=\partial_i\partial_j\partial_k {\cal F}(t)~~.
\end{equation}
The generating function ${\cal F}$ is known as the WDVV potential. In the
appropriate geometric set-up, it can be interpreted as the prepotential of the
effective space-time theory associated with an $N=2$ Calabi-Yau
compactification of a superstring model (to which the topological worldsheet
theory is related by twisting). 

The WDVV equations \cite{WDVV} are the conditions that the quantities $\eta_{ij}$
and $C_{ijk}(t)$ can be viewed as the two and three point functions
of a deformed topological field theory.
Since we work at tree-level, this amounts to the requirement that the deformed 
three-point functions satisfy the sewing constraint (\ref{eq:fact4}) {\em for finite t}~:
\begin{equation}
\label{crossing_t}
C_{ijm}(t) \eta^{mn} C_{nkl}(t)=C_{ikm}(t)\eta^{mn} C_{njl}(t)~~.
\end{equation} 
Using relation (\ref{pot}), this gives a system of second order, quadratic 
partial differential equations for the prepotential:
\begin{equation}
  \label{eq:WDVV}
  \6_i \6_j \6_m \cF ~\eta^{mn} ~\6_n \6_k \6_l \cF=
\6_i \6_k \6_m \cF ~\eta^{mn} ~\6_n \6_j \6_l \cF~~. 
\end{equation}
These are the well-known associativity, or WDVV relations \cite{WDVV}. 

\paragraph{\bf Remark} The sewing
  constraints (\ref{crossing_t}) can be viewed as integrability conditions for
  the existence of a deformed topological field theory at finite $t$. 
  They must be satisfied if deforming the
  worldsheet action by the infinitesimal term:
\begin{equation}
\label{delta_S}
\delta S=\sum_{i=0}^{h_c-1}{t_i \int{\phi_i^{(1,1)}}}~~
\end{equation} 
is to lead to a quantum worldsheet theory which satisfies the (tree-level) topological sewing
  constraint, when such deformations are extended to finite $t$. One case in which this is guaranteed is for
  those perturbations which are {\em exactly} marginal and preserve
  the symmetries $Q, G_{-1}, G_0$ and $G_{1}$ as well as their right-moving
  counterparts, or a deformation thereof which satisfy the same algebra. 
  In this case, relation (\ref{eq:EMtensor}) and its
  right-moving counterpart continue to hold in the deformed theory, which
  therefore is topological. Since the variation (\ref{delta_S}) is BRST closed
  but not exact, the topological character of the theory cannot generally
  be preserved without modifying the BRST operator. Similarly, the generators 
  $Q, G_{-1}, G_0$ and $G_{1}$ must change in a $t$-dependent manner. As a
  consequence, the descendants $\phi_i^{(1,1)}$ will also depend on $t$
  (though their cohomology classes need not), and
  extending (\ref{delta_S}) to finite $t$ becomes nontrivial. 
  This generally makes it difficult to construct appropriate deformations of the
  worldsheet theory. The WDVV equations (\ref{eq:WDVV}) are the minimal 
  (tree-level) requirement for such deformations to exist. In applications, it
  is often nontrivial to build appropriate deformations of the worldsheet
  model, even in those situations where the WDVV potential can be determined 
  independently from equations (\ref{eq:WDVV}). A well-known example is the 
  topological B-model of \cite{Wittenmirror}, for which the full WDVV
  potential (including the part depending on odd parameters $t$) was 
  constructed in \cite{BK} and the underlying deformation of the worldsheet
  theory was given only relatively recently in \cite{JS}.


\subsection{Adding a boundary}


Extending the analysis of TCFT to worldsheets with boundaries induces 
several profound changes in the structure of the theory. An important new 
aspect concerns the integrated operators (\ref{eq:intbulk}), 
which fail to be $Q$-closed due to the presence of the boundary. 
We shall return to this point later, after studying the boundary conditions. 
As above, we concentrate on tree-level amplitudes, which in this context are
the (integrated) correlation functions on the disk, or equivalently the upper half-plane.
We let $z=\tau+i\sigma$ with $\tau,\sigma$ the real coordinates of the complex
plane. 

We shall require that the boundary conditions preserve the transformations
generated by $Q:=Q_0+{\bar Q}_0$ and $J=J_0+{\bar J}_0$ and that
the following condition holds at the boundary: 
\begin{eqnarray}
\label{bc_G}\nn
G(z)&=&\bar G(\bar z) ~~{\rm for}~z={\bar z} ~~.
\end{eqnarray}
Due to equation~(\ref{eq:EMtensor}), these constraints ensure that there is no flow
of energy across the boundary, i.e. we have $T(z)=\bar T(\bar z)$ for $z={\bar
z}$.  Let us define:
\begin{eqnarray}
\nn
Q:=Q_0+{\bar Q}_0~~,\quad G:=G_{-1}+{\bar G}_{-1}~~.
\end{eqnarray}
Then $Q^2=0$ and the following identity holds:
\begin{equation}
\label{eq:EMtensor_bdry}
[Q,G]=L_{-1}+{\bar L}_{-1}~~.
\end{equation}
Also notice that the transformations generated by 
$Q$ and $G, G_{0}+{\bar G}_{0}, G_1+{\bar G}_1$ as well 
as $L_{-1}+{\bar L}_{-1}, L_{0}+{\bar L}_{0}, L_1+{\bar L}_1$ are symmetries of
the theory defined on the disk. This follows from our boundary conditions and
from properties $(A)$ and $(B)$ discussed in the previous subsection. The last three
operators generate the group $\PSL(2,\R)$ of global conformal symmetries of the disk.

\subsubsection{Zero-form observables and sewing constraints}
\label{sewing_bdry}

When constructing observables, we must consider both bulk zero-form operators
$\phi_i$ and zero-form operators $\psi_a$ ($a=0\dots h_o-1$) supported on the boundary.
The bulk zero-forms are as before. For the boundary operators,
we assume $[Q,\psi_a]=0$ and choose a collection which induces a basis of
the $Q$-cohomology $H_o$ of boundary operator-valued zero-forms. 
We shall allow the space $H_o$ to carry a $\Z_2$-grading denoted by $|\cdot|$\footnote{In
  general, this is the sum of the Grassmann degree and another $\Z_2$ grading
  arising from a `brane-antibrane' structure in the boundary sector. The later
  occurs by describing boundary data through a topological version of tachyon
  condensation between 'elementary D-branes'. Examples are discussed in
  \cite{CILb,CILg,Diaconescu,CILh}
  for topological sigma models and in \cite{BHLS,KapC,CILLG} for topological
  Landau-Ginzburg models. We stress that such a grading on the space of
  boundary observables seems to be essential in almost any realistic model, at
  least if one wishes to consider reasonably general D-branes. As a
  consequence, the boundary sector of the worldsheet topological field theory
  must be described in the $\Z_2$-graded framework discussed in
  \cite{CILtop}, which generalizes the analysis performed in
  \cite{MooreSegal,Moore} for the ungraded case. }. 
This complex super-vector space of dimension $h_o$ becomes an associative superalgebra 
over the complex numbers when endowed with the composition given by the boundary product:
\begin{equation}
\label{defDabc}
(\psi_a, \psi_b)\rightarrow \psi_a\psi_b=D_{ab}^c\psi_c~~,
\end{equation}
where $D_{ab}^c\in \C$ are the structure constants of the boundary OPE. 
Associativity follows from one of the sewing constraints, namely boundary
crossing symmetry:
\begin{equation}
\label{assoc}
D^d_{ab}D^{e}_{dc}=D^e_{ad}D^d_{bc}~~.
\end{equation}
In particular, we notice compatibility between the grading and boundary product:
\begin{equation}
\nn
|\psi_a\psi_b|=|\psi_a|+|\psi_b|~~.
\end{equation}
For ease of notation, we shall also denote the degree of $\psi_a$ by
$|a|$:
\begin{equation}
\label{Grass_degree}
|a|:=|\psi_a|\in \Z_2~~.
\end{equation}
The boundary two-point function on the disk:
\begin{equation}
\label{2point}
\omega(\psi_a,\psi_b):=\langle \psi_a \psi_b\rangle:=\omega_{ab}~~
\end{equation}
defines a non-degenerate bilinear form on $H_o$, which satisfies the graded
symmetry property:
\begin{equation}
\label{sym}
\omega_{ab}=(-1)^{|a||b|}\omega_{ba}
\end{equation}
and the selection rule:
\begin{equation}
\label{selrule}
\omega_{ab}=0~~{\rm~unless~}|a|+|b|=|\omega|~{\rm mod}~2~~,
\end{equation}
where $|\omega|\in \Z_2$ is a model-dependent degree. This bilinear form is
known as the {\em boundary topological metric} \cite{Moore,CILtop}.
We shall denote its inverse by $\omega^{ab}$.  

One has the compatibility property:
\begin{equation}
\nn
\omega(\psi_a,\psi_b\psi_c)=\omega(\psi_a\psi_b,\psi_c)~~,
\end{equation}
which amounts to cyclicity of $\omega$ when combined with (\ref{sym}):
\begin{equation}
\omega(\psi_a,\psi_b\psi_c)=(-1)^{|c|(|a|+|b|)}\omega(\psi_c,\psi_a\psi_b)~~.\nn
\end{equation}
Defining $D_{abc}:=\omega_{ae}D^e_{bc}$, the last relation becomes:
\begin{equation}
D_{abc}=(-1)^{|c|(|a|+|b|)}D_{cab}~~.\nn
\end{equation}
As explained in \cite{CILtop}, the boundary algebra admits a unit $1_o\in H_o$
(which automatically has even degree). We shall chose $\psi_a$ such that
$\psi_0:=1_o$. With this choice, we have:
\begin{equation}
D^a_{0b}=D^a_{b0}=\delta^a_b~~.\nn
\end{equation}

Other important data are the bulk-boundary two-point function on the disk,
which can be related to the boundary and bulk topological metrics with the
help of the so-called bulk-boundary and boundary-bulk maps\footnote{The
  bulk-boundary map $e$ is a map from $H_c$ to $H_o$, while the boundary-bulk
  map $f$ is a map from $H_o$ to $H_c$. Both maps are complex-linear.} of \cite{CILtop}:
\begin{equation}
\langle \phi_i \psi_a\rangle_{\rm disk}
=\omega(e(\phi_i), \psi_a)=\eta(\phi_i, f(\psi_a))~~.\nn
\end{equation}
Writing $e(\psi_i)=e_i^a\psi_a$ and $f(\psi_a)=f_a^i\phi_i$, we find the
adjunction relation:
\begin{equation}
e_{ia}=f_{ai}~~,\nn
\end{equation}
where:
\begin{equation}
e_{ia}:=e_i^b\omega_{ba}~~,~~f_{ai}:=\eta_{ij}f_a^j~~.\nn
\end{equation}
One has $e(1_c)=1_o$, i.e:
\begin{equation}
e_0^a=\delta_0^a~~.\nn
\end{equation}

As discussed in \cite{CILtop}, the data $(D,e)$ is subject to the
constraints:
\begin{equation}
\label{bb1}
D_{ab}^ce_i^b=e_i^bD_{ba}^c~~
\end{equation}
\begin{equation}
\label{bb2}
D_{ab}^ce_i^ae_j^b=C_{ij}^k e_k^c~~,
\end{equation}
which encode the two basic bulk-boundary sewing conditions. These relations 
mean that $H_o$ becomes a (unital) associative 
superalgebra over the bulk ring $H_c$ if the external multiplication is defined through:
\begin{equation}
\phi_i\psi_a:=e(\phi_i)\psi_c=e_i^bC_{ba}^c\psi_c~~.\nn
\end{equation}
The remaining sewing constraint involving boundary data is the Cardy condition, which can be
expressed as follows:
\begin{equation}
\label{cardy}
\eta_{ij}e_{ia}e_{jb}=(-1)^{s(c,d)}D_{ad}^cD_{cb}^d~~,
\end{equation}
where $s(c,d)$ is a model-dependent sign and summation over $d$ is understood.

\subsubsection{Descendants}

In the presence of the boundary, the descent equations for bulk operators
must be taken with respect to the BRST charge $Q=Q_0+{\bar Q}_0$. In particular, we 
shall consider a single one-form descendant for each
$\phi_i$. Adapting the construction of the previous section, we
define: 
\begin{eqnarray}
\nn\phi_i^{(1)}&=&\phi_i^{(1,0)}+\phi_i^{(0,1)}\\
\nn\phi_i^{(2)}&=&\phi_i^{(1,1)}~~.
\end{eqnarray}
Using equations (\ref{eq:descendants}), we find the descent relations:
\begin{eqnarray}
\label{eq:bulk_descendants}
\nn [Q,\phi_i^{(1)}]&=&d\phi_i~~\\~
 [Q,\phi_i^{(2)}]&=&d\phi_i^{(1)}~~.
\end{eqnarray}
The last of these equations implies:
\begin{equation}
  \label{eq:Qvardesc}
  [Q,\int\limits_\ws \phi_i^{(2)}] = \int\limits_\bws \phi_i^{(1)}~~.
\end{equation}
Notice the presence of a boundary term on the right-hand side. Defining:
\begin{equation}
\nn
\psi_a^{(1)}:=[G,\psi_a]d\tau~~
\end{equation}
and using relation (\ref{eq:EMtensor_bdry}), we also find the boundary descent equation:
\begin{equation}
  \label{eq:bounddesc}
  [Q, \psi_a^{(1)}] = (\frac{d}{d\tau} \psi_a) d\tau~~.
\end{equation}
Since operators will be inserted on
the boundary in cyclic order, the typical
integral of  a descendant $\psi_a^{(1)}$ runs from the 
insertion to its left to the insertion to its right:
\begin{equation}
  \label{eq:intbound}
  \int\limits_{\tau_l}^{\tau_r} \psi_a^{(1)}~~.
\end{equation}
Here `left' and `right' should be understood in the sense of the cyclic order
on the boundary of the disk, which is determined by the orientation on the
boundary induced from the orientation of the interior. 
As a consequence, we find that the BRST variation of
(\ref{eq:intbound}) need not vanish:
\begin{equation}
  \label{eq:Qvardescbound}
  [Q, \int\limits_{\tau_l}^{\tau_r} \psi_a^{(1)}] ~=~ 
  \psi_a ~\biggr|_{\tau_l}^{\tau_r}~~.
\end{equation}

Notice that the Grassmann degree of $\psi_a^{(1)}$ 
is opposite to that of $\psi_a$. It is convenient to
take this into account by introducing a new 
grading on the boundary algebra $H_o$:
\begin{equation}
\nn
\deg~\psi=|\psi|+1~~({\rm mod}~2)~~.
\end{equation}
For ease of notation, this shifted, or ``suspended'' grade of $\psi_a$ will be denoted by a tilde: 
\begin{equation}
\label{suspended}
{\tilde a}:=\deg~\psi_a = |a| + 1~~({\rm mod}~2)~~. 
\end{equation}
In terms of the suspended grade, the selection rule (\ref{selrule}) becomes:
\begin{equation}
\label{selrule_shifted}
\omega_{ab}=0~~{\rm unless}~~\td a+\td b ={\tilde \omega}+1 ~({\rm mod}~2)~~,
\end{equation}
where ${\tilde \omega}=|\omega|+1 \in \Z_2$. 
Moreover, the graded symmetry property (\ref{sym}) of the boundary 2-point function
takes the form:
\begin{equation}
  \label{eq:symmsympl}
  \omega_{ab} = (-1)^\grd(-1)^{\td a \td b}\omega_{ba}~~.
\end{equation}

\paragraph{\bf Remark} In string theory we have in addition to open
strings attached to a single D-brane (or a stack of D-branes),
also strings which are stretched between two different
D-branes. It is well-known \cite{Cardyir} that the former correspond
to boundary preserving operators $\psi_a^{AA}$, whereas the latter
correspond to boundary condition changing operators $\psi_a^{AB}$
(since they mediate between two different boundary conditions (D-branes) 
labeled by $A$ and $B$). Of course, all operators of the topological conformal algebra 
(\ref{algebra}) are boundary preserving, since they are related by
a single condition on the boundary. The action of the charges on
boundary condition changing operators can be written in a form which is very
similar to that relevant for the boundary preserving sector. For example:
\begin{equation}
  \label{eq:boundchanging}
  [G, \psi_a]^{AB} = G^{AA}~\psi_a^{AB} -(-1)^{|a|} \psi_a^{AB}~G^{BB} :=
  \oint (G(z) ~\psi_a)^{AB}~~,
\end{equation}
where --- using the doubling trick --- the left- and right-moving currents
are joined according to the boundary conditions $A$ and $B$ on the respective
side of $\psi_a^{AB}$.  This allows us to treat boundary and boundary
  condition changing operators in identical manner (the difference
  being akin to that between the adjoint and bi-fundamental representation of
  the same Lie or vertex operator algebra).  In particular, all 
relations derived in this paper are also true if one includes boundary
  changing sectors, provided that one adds labels for the various
  boundary sectors in the appropriate places. Note that for each boundary component,
the boundary labels must be "cyclically closed" in correlation functions,
for example correlators such as $\bra \psi_{a_1} \ldots \psi_{a_n}\ket$ should
  be expanded to $\bra \psi_{a_1}^{A_1A_2} \ldots \psi_{a_n}^{A_nA_1}\ket$
  when restoring boundary labels. In the presence of boundary condition
  changing sectors, the various algebraic structures extracted in this paper 
  are promoted to their category-theoretic counterparts. This follows in 
  standard manner by  viewing D-branes (a.k.a boundary sectors) as objects of
  a category and identifying  boundary and boundary condition changing
  operators with endomorphisms and morphisms between distinct objects.


\section{Immediate properties of tree-level amplitudes}

\label{sec:corrfunc}


In this section, we discuss the most basic
 properties of open-closed amplitudes on
the disk. After explaining the regularization used in later sections, we show
that two basic forms for such amplitudes are equal up to sign and independent
of the positions of boundary insertions and, more generally,  of the worldsheet
metric. Moreover, we check that such amplitudes are cyclic with respect to
boundary insertions and completely symmetric with respect to insertions of
bulk operators. All properties established in this section are elementary,
though the precise proofs in conformal field theory are not always obvious. 
The main point of interest for later sections is the regularization of
open-closed tree-level amplitudes, which will play a central role in our
discussion of the algebraic constraints.

\subsection{The regularized amplitudes}

\label{regularization}

Since disk amplitudes with integrated boundary 
descendants are affected by contact divergences, the conformal field theory 
arguments of later sections will require a regulator. 
In this paper, we  shall use a version of 
point-splitting for integrated bulk operators
approaching the boundary of the disk and for integrated
boundary operators approaching each other. This regularization 
is essential only for the arguments of Sections 5 and 6.1.

Given bulk descendants $\phi_{i_k}^{(2)}$ with $k = 1\ldots n$, we will
choose their integration domain as follows: 
\begin{equation}
  \label{eq:bulkdomain}
  \bbH_n = \left\{ (z_1, \ldots, z_n)\in \C^{n}~|~ 
  \Im(z_k) \in ( k\epsilon,\infty) {\rm~for~all~} k=1\ldots n\right\}~~,
\end{equation}
Here $z_k$ are the insertion points of $\phi_{i_k}^{(2)}$, which of course
are  integrated over. 

\begin{figure}[t]
  \begin{center}
     \epsfysize=4.5cm\centerline{\epsffile{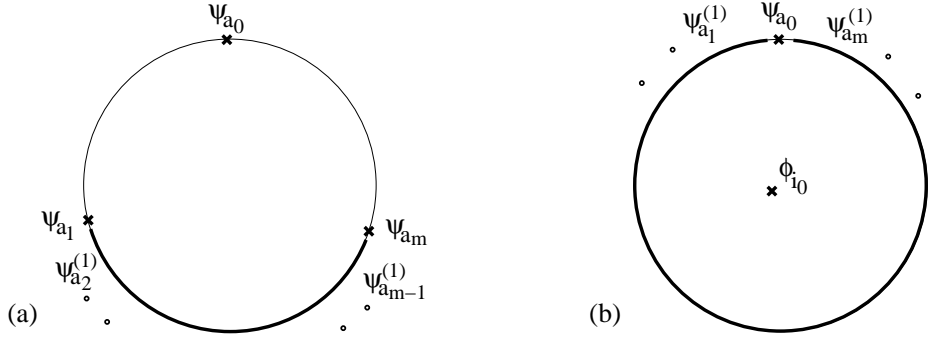}}
    \parbox{12truecm}{
      \caption{Boundary and bulk insertions for disk amplitudes. 
      (a) Three boundary fields $\psi_{a_0}$, $\psi_{a_1}$ and
      $\psi_{a_m}$ are at fixed positions, the others are integrated
      in a path ordered way between $\psi_{a_1}$ and $\psi_{a_m}$.
      (b) One bulk and one boundary field are fixed.
      In both cases additional bulk operators may be present, which are integrated over the whole 
      disk.
        \label{fig:corr}}
    }
  \end{center}
\end{figure}

We next consider boundary insertions. Using $\PSL(2,\bbR)$-invariance, 
three of them can be fixed while the others are
integrated (see fig. \ref{fig:corr}(a)). A typical disk amplitude has
the form: 
\begin{equation}
  \label{eq:corr1}
  \bra~ \psi_{a_1} \psi_{a_2} ~
  P \int\psi^{(1)}_{a_3}\ldots \int\psi^{(1)}_{a_{m-1}} ~
  \psi_{a_m} \int \phi^{(2)}_{i_1}\ldots \int \phi^{(2)}_{i_n}~
  \ket~~,
\end{equation}
where we fixed the positions of 
  $\psi_{a_1}$, $\psi_{a_2}$ and $\psi_{a_m}$ to the points 
  $\tau_1, \tau_2, \tau_m \in \bbR$, with
  the restriction $\tau_1 < \tau_2 < \tau_m$. The path-ordering symbol $P$
  means that the integral over 
  $\tau_{3} \ldots \tau_{m-1}$ runs between
  $\tau_{2}$ and $\tau_{m}$ with the constraint
  $\tau_2 < \tau_3 < \ldots < \tau_{m-1} < \tau_m$. 
  Including a regulator, the exact integration domain will be chosen as follows:
\begin{equation}
  \label{eq:bounddomain1}
  \bbS_m(\tau_2,\tau_m) = \bigl\{(\tau_3,\ldots,\tau_{m-1})\in
  \bbR^{m-3} | \nn \tau_k-\tau_j> [2(k-j)-1]\epsilon \mathrm{~~for}~~2\leq j<k\leq m  \bigr\}~~.
\end{equation}

\paragraph{\bf Remark}
Notice that we are requiring slightly increased separations for non-consecutive
  boundary insertions, rather than working with the naive point-splitting
  constraint $|\tau_k-\tau_j|\geq |k-j|\epsilon$.
This somewhat unusual choice is made for the following reason.
The factorization procedure of the following sections makes
  use of the descent equation $[Q,G] = \frac{d}{d\tau}$, which implies that 
  acting with $Q$ on an integrated boundary insertion produces terms
  involving the associated zero-form
  operator evaluated at the boundaries of its integration interval, generally 
  with some integrated insertions squeezed in. The
  increased separations chosen in
  (\ref{eq:bounddomain1}) ensure the presence of non-void
  integration domains for the squeezed-in operators. For instance,
  if we consider
  the BRST operator acting on $\int \psi_{a_4}^{(1)}$, then our choice for the
  integration domain  $\bbS_m(\tau_2,\tau_m)$ leads to a term of the form: 
$$
\psi_{a_2}(\tau_2) \int_{\tau_2+\e}^{\tau_4-\e}
  \!\!d\tau_3~ \psi_{a_3}^{(1)}(\tau_3)~ \psi_{a_4}(\tau_4)
  \Bigr|_{\tau_4 = \tau_2+3\e}= 
\psi_{a_2}(\tau_2) \int_{\tau_2+\e}^{\tau_2+2\e}
  \!\!d\tau_3~ \psi_{a_3}^{(1)}(\tau_3)~ \psi_{a_4}(\tau_2+3\e)~,
$$
which involves integration over a non-void interval.
Had we used the naive condition
$|\tau_k-\tau_l|>|k-l|\epsilon$, the integral in the last equation would
have been $\int_{\tau_2+\e}^{\tau_2+\e}\psi_{a_3}^{(1)}=0$.

Besides (\ref{eq:corr1}) one can also consider amplitudes
in which $\PSL(2,\bbR)$-invariance is used to fix the positions of 
one bulk and one boundary insertion (see fig. \ref{fig:corr}(b)):
\begin{equation}
  \label{eq:corr2}
  \bra~ \phi_{i_1} \psi_{a_1} ~
  P \int\psi^{(1)}_{a_2}\ldots \int\psi^{(1)}_{a_m} ~
  \int \phi^{(2)}_{i_2}\ldots \int \phi^{(2)}_{i_n}~
  \ket~~.  
\end{equation}
Naively, the integration domain is obtained from $\bbS_m(\tau_2,\tau_m)$ by
replacing both $\tau_2$ and $\tau_m$ by $\tau_1$, where we 
integrate over the real line and identify $-\infty$ and $\infty$.
However, the integrals approach $\psi_{a_1}$ from both sides, so we have to
introduce a further cut-off. We will choose the following integration domain:
\begin{eqnarray}
  \label{eq:bounddomain2}
  \bbS_{m}(\tau_1) = \{ (\tau_2,\ldots,\tau_m)\in \bbR^{m-3}~&|&~
  (\tau_2\dots \tau_m) \mathrm{~is~cyclically~ordered~and}\\
  &&
  \tau_k-\tau_l > [2(k\!\!-\!\!l)\!\!-\!\!1]\epsilon ~
  \mathrm{for}~
  \tau_k>\tau_l>\tau_1~ \mathrm{or}~\tau_1>\tau_k>\tau_l~~,~~\nonumber\\
 && \tau_k-\tau_l > [2(k\!\!-\!\!l\!\!+\!\!m)\!\!-\!\!1]\epsilon~
  \mathrm{for}~ \tau_k>\tau_1>\tau_l
   \}~~.\nonumber
\end{eqnarray}
\del
In a certain sense, the correlation function (\ref{eq:corr1}) is naturally
considered when one has few bulk operators, whereas (\ref{eq:corr2})
is natural for few boundary operator insertions. 
\enddel
We will see in a moment that (after removing the regulator) 
the two kinds amplitudes are equal up to sign,
as has been argued before in \cite{HofmanMaA}.

Having defined the regularized amplitudes, we shall explore the
implications of the conformal Ward identities and of the Ward identities
for $G$. Using the doubling trick, one easily proves the relation: 
\begin{eqnarray}
  \label{eq:Gward}
  \nn
  \oint \xi(z) \bra G(z) 
 \psi_{a_0} \ldots \psi_{a_m} \phi_{i_1} \ldots \phi_{i_n} \ket
  &=&
  \sum\limits_{k=0}^{m} \pm ~\xi(\tau_k) 
  \bra \psi_{a_0} \ldots \psi^{(1)}_{a_k} \ldots \psi_{a_m} 
  \phi_{i_1} \ldots \phi_{i_n} \ket \\[5pt]
  &\pm&
  \sum\limits_{k=0}^{n} \xi(w_k)
  \bra \psi_{a_0} \ldots \psi_{a_m} 
  \phi_{i_1} \ldots \phi^{(1,0)}_{i_k} \ldots \phi_{i_n} \ket \\[5pt]
  &\pm&
  \sum\limits_{k=0}^{n} \bar \xi(\bar w_k)
  \bra \psi_{a_0} \ldots \psi_{a_m} 
  \phi_{i_1} \ldots \phi^{(0,1)}_{i_k} \ldots \phi_{i_n} \ket \nn\\[15pt]
 &=& 0~~,\nn~
\end{eqnarray}
where $\xi(z)=a z^2 + b z + c$ with $a,b,c \in \R$ is a globally-defined
holomorphic vector field on the upper half plane and the signs account
for the grading on boundary fields. By the doubling trick, the contour
  integral on the left hand side encircles all fields and their images with
  respect to the real axis in the complex
  plane (which is viewed as a double cover of the upper half plane). 
  In the right hand side we evaluated the residue at every
  insertion, including the images. The terms containing $\phi_i^{(1,0)}$ arise 
  from the residue at $\phi_i$, while the terms
  containing  $\phi_i^{(0,1)}$ arise from the residues at the the
  images of these insertions.

In the bulk sector, a similar identity implies constancy of the bulk
topological metric along the moduli space and integrability of 
the deformed amplitudes. Below, we will study the consequences of 
(\ref{eq:Gward}).


\subsection{Equivalence of the two types of amplitudes}


We start by explaining the relation between the two kinds of disk amplitudes
(\ref{eq:corr1}) and (\ref{eq:corr2}). We will show that these a priori
different quantities are in fact equal up to sign factors.  This was already
discussed in \cite{HofmanMaA} and we shall review the argument below in order
to extract the correct signs for the case of boundary fields with different
degrees. The derivation uses the Ward identities of $G$ to relate integration
over a bulk descendant with two integrations over boundary descendants.

As an example, consider the amplitudes $\bra\psi_a\psi_b\psi_c\int\phi_i^{(2)}\ket$ 
and $\bra\phi_i\psi_a~P\int\psi_b\int\psi_c\ket$.
We use the Ward identities:
\begin{equation}
  \nonumber
  \oint\xi_3 \bra G \oint\xi_2 G ~
   \psi_a(\tau_1)~\psi_b(\tau_2)~\psi_c(\tau_3)~\phi_i(z,\bar z)\ket
  = 0~~,
\end{equation}
and:
\begin{equation}
\oint \xi_3 \bra G  \psi_a\psi_b\psi_c^{(1)}\phi_i\ket=0~~,~~ 
\oint \xi_2 \bra G  \psi_a\psi_b^{(1)}\psi_c\phi_i\ket=0~~,\nn
\end{equation}
with the following choice for the global holomorphic vector fields: 
\begin{equation}
\xi_2(z)=(z-\tau_1)(z-\tau_3)~{\rm~and~}~
\xi_3(z)=(z-\tau_1)(z-\tau_2)~~.\nn
\end{equation}
We assume the ordering $\tau_1 < \tau_2 < \tau_3$. Using equation~(\ref{eq:Gward}), we obtain:
\begin{equation}
  \label{eq:Xirel}
  \frac{\xi_2(z)\bar\xi_3(\bar z) - \bar\xi_2(\bar z)\xi_3(z)}
       {\xi_2(\tau_2)\xi_3(\tau_3)}~
  \bra \psi_a\psi_b\psi_c\phi_i^{(2)} \ket~=
  (-1)^{\td b}~ 
  \bra \psi_a\psi_b^{(1)}\psi_c^{(1)}\phi_i \ket~~.
\end{equation}
The conformal Ward identities ensure that
both sides of equation~(\ref{eq:Xirel}) depend only on 
the cross-ratio $\zeta = \frac{(z-\tau_3)(\tau_2-\tau_1)}{(z-\tau_2)(\tau_3-\tau_1)}$
and its complex conjugate. Using the relations:
\[  \xi_i(\tau_i)\frac{\6 \zeta}{\6\tau_i} + 
  \xi_i(z)\frac{\6 \zeta}{\6 z} =0~~, \quad\textrm{for}~i=2,3~~,
\]
we find:
\[  \frac{\xi_2(z)\bar\xi_3(\bar z) - \bar\xi_2(\bar z)\xi_3(z)}
       {\xi_2(\tau_2)\xi_3(\tau_3)} =
  \left(\frac{\6\zeta}{\6 z}\frac{\6\bar\zeta}{\6\bar z}\right)^{-1}
  \left(\frac{\6\zeta}{\6\tau_2}\frac{\6\bar\zeta}{\6\tau_3}-
  \frac{\6\zeta}{\6\tau_3}\frac{\6\bar\zeta}{\6\tau_2}\right)^{-1}~~.
\]
Hence the prefactor in equation~(\ref{eq:Xirel}) is 
the Jacobian of the coordinate transformation from $(z, \bar z)$ to
$(\tau_2,\tau_3)$. Notice that we are free to rescale bulk
and boundary descendants via
$\phi^{(2)}\rightarrow\lambda_{bulk}\phi^{(2)}$ and 
$ \psi^{(1)}\rightarrow\lambda_{bound}\psi^{(1)}$.
\del
independently in order to define the amplitudes
$\bra \psi_a\psi_b\psi_c\int\phi_i^{(2)} \ket$ and 
$\bra \phi_i~\psi_a~P\!\int\!\psi_b^{(1)}\!\int\!\psi_c^{(1)}
\ket$. 
\enddel
Taking this into account, we find:  
\begin{equation}
  \label{eq:HofmanWard}
  (-1)^{\td b}~\bra \psi_a\psi_b\psi_c\int\phi_i^{(2)} \ket =
  - \bra \phi_i~\psi_a~P\!\int\!\psi_b^{(1)}\!\int\!\psi_c^{(1)} \ket~~,
\end{equation}
where we chose the relative normalization factor to be $-1$. Of
  course, this locks the rescalings together through the relation
  $\lambda_{bulk}\propto \lambda_{bound}^2$.

One can easily generalize the analysis to arbitrary numbers of bulk
and boundary insertions. This gives:\footnote{Notice that for $B_{a_0 a_1 a_2}$ as well as $B_{a_0;i_1}$ and
$B_{a_0 a_1;i_1}$ such a relation does not exist for obvious reasons.}
\begin{eqnarray}
  \label{B}
  B_{a_0\ldots a_m;i_1\ldots i_n} &:=&
  (-1)^{\td a_1+\ldots +\td a_{m-1}}\bra~ \psi_{a_0} \psi_{a_1} ~
  P \int\psi^{(1)}_{a_2}\ldots \int\psi^{(1)}_{a_{m-1}} ~
  \psi_{a_m} \int \phi^{(2)}_{i_1}\ldots \int \phi^{(2)}_{i_n}~
  \ket
  \nn\\[5pt]
  &=&
  - ~\bra~ \phi_{i_1} \psi_{a_0} ~
  P \int\psi^{(1)}_{a_1}\ldots \int\psi^{(1)}_{a_m} ~
  \int \phi^{(2)}_{i_2}\ldots \int \phi^{(2)}_{i_n}~
  \ket~~.
\end{eqnarray}
Thus (\ref{eq:corr1}) and (\ref{eq:corr2}) are equal up to sign, 
and they determine the single object $B_{a_0\ldots a_m;i_1\ldots i_n}$ defined
  by the expression above. These amplitudes vanish unless:
\begin{equation}
\label{B_selrule}
\td a_0 + \ldots + \td a_m = {\tilde \omega}~~.
\end{equation}
As we shall see below, it is notationally convenient to define:
\begin{equation}
\label{low_B}
B_{a_0a_1}=B_{a_0}=B_{i}=0~~.
\end{equation}

We make one final remark about equation (\ref{B}). The first line 
is manifestly symmetric in the
bulk indices, but this is not obvious for the second line. As in the pure
bulk theory, there exists a Ward identity \cite{WDVV,HofmanMaA}
which switches fixed and integrated bulk insertions. This can be used to show
directly that the 
second line in (\ref{B}) is also totally symmetric in the
bulk insertions.

For later reference, let us translate the topological sewing constraints of Subsection
\ref{sewing_bdry} in terms of the amplitudes $B_{a_0\ldots a_m;i_1\ldots i_n}$. From equation (\ref{B}), we have:
\begin{equation}
B_{abc}=(-1)^b D_{abc}~~,~~B_{a; i}=-e_{ia}=-f_{ai}~~.\nn
\end{equation}
Introducing the quantities:
\begin{equation}
  \label{eq:pullindex}
  B^a{}_{a_1\ldots a_m;i_1\dots i_n} := \omega^{ab} B_{b a_1\ldots a_m; i_1\ldots i_n}~~,~~
  B^i{}_{a_0\ldots a_m;i_2\dots i_n}:=\eta^{ij}B_{a_0\ldots a_m;j, i_2\ldots i_n}~~,
\end{equation}
we find the relations:
\begin{equation}
B_{0ab}=(-1)^{\tilde a}\omega_{ab}~~,
~~B^a_{bc}=(-1)^{\tilde b} D^a_{bc}~~,~~B_i^a=-e_i^a~~,~~B_a^i=-f_a^i~~.\nn
\end{equation}
Thus equations (\ref{assoc}), (\ref{bb1}), (\ref{bb2}) and (\ref{cardy}) take
  the form:
\begin{eqnarray}
B^{a_0}_{c a_3} B^c_{a_1 a_2} &=&-(-1)^{\td a_1} 
B^{a_0}{}_{a_1 c} B^c{}_{a_2a_3} 
\label{assoc_B}\\
B_{ab}^cB_i^b&=&-B_i^bB_{ba}^c
\label{bb1_B}\\
B_{ab}^cB_i^aB_j^b &=& C_{ij}^k \,B_k^c
\label{bb2_B}\\
\label{cardy_B}\eta^{ij}B_{ia}B_{jb} &=& (-1)^{s(c,d)}B_{ad}^cB_{cb}^d
~~,
\end{eqnarray}
where the sign factor $s$ depends on $c,d$.


\subsection{ Two point correlation functions are not deformed}
\label{sec:2point}

In this subsection, we show that the two-point boundary correlators are constant under bulk and boundary
deformations. Let us start with the Ward identity for $G$ in the presence of  
two fixed boundary insertions:
\begin{equation}
  \nonumber
  \oint \xi(z) \bra G(z) ~
   \psi_{a_1}(\tau_1) \psi_{a_2}(\tau_2) \psi_{a_3}(\tau_3) \ket  = 0~~.
\end{equation}
Choosing $\xi(z)= (z-\tau_1)(z-\tau_2)$, we find:
\begin{equation}
  \label{eq:2pointG}
\bra \psi_{a_1}~\psi_{a_2}~\psi_{a_3}^{(1)} \ket =0~~. 
\end{equation}
The analogous relation for a bulk perturbation:
\begin{equation}
  \label{eq:2pointGbulk}
  \bra \psi_{a_1}~\psi_{a_2}~\phi_{i}^{(2)} \ket =0~~,
\end{equation}
requires a bit more work. For this, consider the Ward identity:
\begin{equation}
\nn
\oint \xi_2 \bra G \oint \xi_1 G 
 \psi_{a_1}(\tau_1)\psi_{a_2}(\tau_2)\phi_{i}(w,\bar w)\ket = 0~~,
\end{equation}
where $\xi_1(z)=(z-\tau_1)(z-\tau_2)$ and $\xi_2(z)=(z-\tau_2)(z-\Re w)$. 
Combining this with the relation:
\begin{equation}
\nn
\oint \xi_1 \bra G  \psi_{a_1} \psi_{a_2}^{(1)}\phi_{i}\ket = 0~~,
\end{equation}
leads to equation (\ref{eq:2pointGbulk}).
 
Since the supercharge $G$ does not act on additional descendants
$\int \psi_a^{(1)}$ and $\int \phi_i^{(2)}$, we easily infer the
generalization:
\begin{equation}
  \label{eq:full2point1}
  \bra \psi_{a_1}~\psi_{a_2}~P\int\psi_{a_3}^{(1)}\ldots\int\psi_{a_m}^{(1)}~
  \int \phi^{(2)}_{i_1}\ldots \int \phi^{(2)}_{i_n} \ket
  = 0~~,
  ~~{\rm~for}~m\geq 3~{\rm or}~~n \geq 1~~.
\end{equation}
In similar manner, one shows:
\begin{eqnarray}
  \label{eq:full2point2}
  \bra \phi_{i_0}^{(1,0)}~\psi_{a_1}~~P\int\psi_{a_2}^{(1)}\ldots\int\psi_{a_m}^{(1)}~
  \int \phi^{(2)}_{i_1}\ldots \int \phi^{(2)}_{i_n} \ket
  &=& 0~~, \nn \\
  \bra \phi_{i_0}^{(0,1)}~\psi_{a_1}~~P\int\psi_{a_2}^{(1)}\ldots\int\psi_{a_m}^{(1)}~
  \int \phi^{(2)}_{i_1}\ldots \int \phi^{(2)}_{i_n} \ket
  &=& 0~~,
\end{eqnarray}
In terms of the quantities defined in equation (\ref{B}), relation
(\ref{eq:full2point1}) takes the form: 
\begin{eqnarray}
\label{B_vanish}
B_{0a_1\dots a_m;i_1\dots i_n}=0&&
{\rm~for~}~m\geq 3~{\rm or}~ n\geq 1 ~~.
\end{eqnarray}

The identities discussed in this subsection will be important for subsequent
arguments. As we shall see, they are relevant for proving 
independence of the amplitudes of the positions of unintegrated insertions and
more generally of the worldsheet  metric. Moreover, they relate 
to special properties of the boundary algebra and topological metric.

\subsection{Independence of the positions of unintegrated insertions}

We will now show that the fundamental amplitudes (\ref{B})
are independent of the positions of unintegrated insertions. 
As an example, consider the 4-point boundary amplitude.
Differentiating it  with respect to $\tau_1$ and using the
descent equations, we find:
\begin{eqnarray}
  \nonumber
  \frac{\6}{\6 \tau_{1}} \bra \psi_{a_0} \psi_{a_1}
  \int\limits_{\tau_1}^{\tau_3}\!\!\psi_{a_2}^{(1)}
  \psi_{a_3} \ket 
  &=& \bra \psi_{a_0} [Q,\psi_{a_1}^{(1)}]
  \int\limits_{\tau_1}^{\tau_3}\!\!\psi_{a_2}^{(1)} \psi_{a_3}\ket -
  \bra \psi_{a_0} \psi_{a_1}
  \psi_{a_2}^{(1)}|_{\tau_1}
  \psi_{a_3}\ket \\
  \nonumber
&=&   (-1)^{\td a_1} \bra \psi_{a_0} \psi_{a_1}^{(1)}
  (\psi_{a_2}|_{\tau_1}-\psi_{a_2}|_{\tau_3})
  \psi_{a_3}\ket -
  \bra \psi_{a_0} \psi_{a_1}
  \psi_{a_2}^{(1)}|_{\tau_1}
  \psi_{a_3}\ket \\[15pt]
  \nonumber
&=&  (-1)^{\td a_1} \Bigl(
  \bra \psi_{a_0} (\psi_{a_1} \psi_{a_2})^{(1)}
  \psi_{a_3} \ket - \bra \psi_{a_0} \psi_{a_1}^{(1)}
  (\psi_{a_2} \psi_{a_3}) \ket \Bigr)
  = 0~~.
\end{eqnarray}
In the last line we used relation (\ref{eq:full2point1}). Generalizing this
argument, it is not hard to show that all amplitudes (\ref{B}) are
independent on the positions of unintegrated insertions.

\subsection{Independence of the worldsheet metric}

Due to the nontrivial terms in the right hand side of equation 
(\ref{eq:Qvardescbound}), it is not immediately clear that the
amplitudes (\ref{B}) are independent of the worldsheet metric. 
The usual recipe of topological field theory does not work: the variation of the
correlation function with respect to the metric produces an insertion of
the energy-momentum tensor, which can be written as
$[Q,G_{\mu\nu}]$. When pulling the BRST operator through the integrated
boundary insertions, one obtains nontrivial terms induced by equation
(\ref{eq:Qvardescbound}), so one cannot immediately conclude that integrated
correlators are independent of the worldsheet metric. 
However, conformal invariance comes to the rescue, through the conformal Ward identity:
\begin{eqnarray}
  \bra T(z)~ \phi_{i_1} \ldots \phi_{i_n}~ 
            \psi_{a_1} \ldots \psi_{a_m} \ket &=&
  \sum\limits_{k=1}^n 
  \left(\frac{h_k}{(z-z_k)^2}+\frac{1}{z-z_k}\frac{\6}{\6 z_k}\right)
  \bra \phi_{i_1} \ldots \phi_{i_n} 
       \psi_{a_1} \ldots \psi_{a_m} \ket~ +\nn\\[10pt]
  &+& \sum\limits_{k=1}^n 
  \left(\frac{\bar h_k}{(z-\bar z_k)^2}+
        \frac{1}{z-\bar z_k}\frac{\6}{\6\bar z_k}\right)
  \bra \phi_{i_1} \ldots \phi_{i_n} 
       \psi_{a_1} \ldots \psi_{a_m} \ket~ + \nn\\[10pt]
  &+& \sum\limits_{l=1}^m 
  \left(\frac{h_l}{(z-\tau_l)^2}+
        \frac{1}{z-\tau_l}\frac{\6}{\6\tau_l}\right)
  \bra \phi_{i_1} \ldots \phi_{i_n} 
       \psi_{a_1} \ldots \psi_{a_m} \ket \nn ~~,
\end{eqnarray}
where $\phi_{i_k}=\phi_{i_k}(z_k,\bar z_k)$ and 
$\psi_{a_l}=\psi_{a_l}(\tau_l)$ are bulk and boundary conformal primaries.
In the case of interest, the conformal weights of zero-form operators 
are $h_l = h_k = \bar h_k = 0$, while for descendants one has 
$h_l = h_k = \bar h_k = 1$. 

Let us first consider the simplest case, namely the boundary 4-point amplitude:
\begin{eqnarray}
  \int\limits_{\tau_2}^{\tau_4} d\tau_3 ~\bra T(z)~ \psi_{a_1} \psi_{a_2} 
  \psi_{a_3}^{(1)} \psi_{a_4} \ket &=&
  \sum\limits_{l=1,2,4} \int\limits_{\tau_2}^{\tau_4} d\tau_3
  \frac{1}{z-\tau_l}\frac{\6}{\6\tau_l}
  \bra \psi_{a_1} \psi_{a_2} 
  \psi_{a_3}^{(1)} \psi_{a_4} \ket ~+ \nn\\[10pt]
  &+& \int\limits_{\tau_2}^{\tau_4} d\tau_3~ \frac{\6}{\6\tau_3} 
  \left(\frac{1}{z-\tau_3} \bra \psi_{a_1} \psi_{a_2} 
  \psi_{a_3}^{(1)} \psi_{a_4} \ket \right) \nn ~~.
\end{eqnarray}
Using the descent relations (\ref{eq:bounddesc}) in the first line
and evaluating the integral in the second line, we find:
\begin{eqnarray}
  \nn
  \int\limits_{\tau_2}^{\tau_4} d\tau_3 ~\bra T(z)~ \psi_{a_1} \psi_{a_2} 
  \psi_{a_3}^{(1)} \psi_{a_4} \ket &=&
  (-1)^{\td a_1 + \td a_2}\frac{1}{z-\tau_1}
  \left( \bra \psi_{a_1}^{(1)}\psi_{a_2}(\psi_{a_3}\psi_{a_4})\ket - 
  \bra \psi_{a_1}^{(1)}(\psi_{a_2}\psi_{a_3})\psi_{a_4}\ket \right) +\\
  \nn
  &+& (-1)^{\td a_2}\frac{1}{z-\tau_2}
  \left( \bra \psi_{a_1}(\psi_{a_2}\psi_{a_3})^{(1)}\psi_{a_4}\ket - 
  \bra \psi_{a_1}\psi_{a_2}^{(1)}(\psi_{a_3}\psi_{a_4})\ket \right) +\\
  \nn
  &+& \frac{1}{z-\tau_3}
  \left( \bra \psi_{a_1}\psi_{a_2}(\psi_{a_3}\psi_{a_4})^{(1)}\ket - 
  \bra \psi_{a_1}(\psi_{a_2}\psi_{a_3})\psi_{a_4}^{(1)}\ket \right)\\
  \nn
  &=&0~~.
\end{eqnarray}
In the last step, we used again equation (\ref{eq:full2point1}). In the same
manner one can show that all amplitudes (\ref{B}) are independent of the
worldsheet 
metric.

\subsection{Cyclicity and bulk permutation invariance}

We shall now prove that disk correlation functions are (graded) cyclically
symmetric with respect to boundary insertions and symmetric under arbitrary 
permutations of bulk insertions. 

Let us illustrate this with the boundary 4-point amplitude:
\begin{equation}
  \label{eq:4point}
  \oint \xi(z) \bra G(z) ~
  \psi_{a_1}(\tau_1) \psi_{a_2}(\tau_2) 
  \psi_{a_3}(\tau_3) \psi_{a_4}(\tau_4)\ket 
  = 0~~,
\end{equation}
where $\tau_4 > \ldots > \tau_1$.
Taking $\xi(z) = (z-\tau_4)(z-\tau_1)$ in equation~(\ref{eq:4point}) and
  using relation (\ref{eq:Gward}), we obtain 
$\xi(\tau_2) \bra \psi_{a} \psi^{(1)}_{b} \psi_{c} \psi_{d}\ket = 
(-1)^{\td b} \xi(\tau_3) \bra \psi_{a} \psi_{b} \psi^{(1)}_{c} \psi_{d}\ket$.
From the conformal Ward identities we know that the unintegrated
  4-point function depends only on the cross-ratio
$\zeta=\frac{(\tau_4-\tau_3)(\tau_2-\tau_1)}{(\tau_4-\tau_2)(\tau_3-\tau_1)}$~,
which satisfies the relation:
\[
  \xi(\tau_2) \frac{\6 \zeta}{\6 \tau_2} + 
  \xi(\tau_3) \frac{\6 \zeta}{\6 \tau_3} = 0~~.
\]
Hence the Ward identity (\ref{eq:4point}) implies:
\[ \left(\frac{\6 \zeta}{\6\tau_2}\right)^{-1}
 \bra\psi_a\psi^{(1)}_b\psi_c\psi_d\ket
  = - (-1)^{\td b}\left(\frac{\6 \zeta}{\6 \tau_3}\right)^{-1}
  \bra\psi_a\psi_b\psi^{(1)}_c\psi_d\ket~~.
\]

Let us integrate this equation over   $\zeta$, 
taking  into account that on the right-hand side the
integration runs in the `wrong' direction, i.e. 
$\int_0^1 d\zeta (\frac{\6 \zeta}{\6\tau_2})^{-1} =
\int_{\tau_1}^{\tau_3} d\tau_2$, 
but $\int_0^1 d\zeta(\frac{\6 \zeta}{\6 \tau_3})^{-1} =
- \int_{\tau_2}^{\tau_4} d\tau_3$. This gives the relation: 
\begin{equation}
\langle \psi_a P \int\psi^{(1)}_b \psi_c \psi_d\ket = 
(-1)^{\td b} \bra \psi_a \psi_b P \int\psi^{(1)}_c \psi_d
\rangle~~.\nn
\end{equation}

Generalizing the argument to more integrated insertions, one finds the following identities: 
\begin{equation}
  \label{eq:cyclic1}
  \bra \psi_{a_0} \psi_{a_1}
  ~P \int\!\!\psi_{a_2}^{(1)} \ldots \int\!\!\psi_{a_{m-1}}^{(1)} 
  \psi_{a_m}\ket =
  (-1)^{\td a_1+\ldots+\td a_{m-2}}\bra \psi_{a_0}
  ~P \int\!\!\psi_{a_1}^{(1)} \ldots \int\!\!\psi_{a_{m-2}}^{(1)} 
  \psi_{a_{m-1}}\psi_{a_m}\ket~~
\end{equation}
and:
\begin{equation}
  \label{eq:cyclic2}
  \bra \phi_i \psi_{a_0}
  ~P \int\!\!\psi_{a_1}^{(1)} \ldots \int\!\!\psi_{a_m}^{(1)} \ket =
  (-1)^{\td a_0+\ldots+\td a_{m-1}}
  \bra \phi_i
  ~P \int\!\!\psi_{a_0}^{(1)} \ldots \int\!\!\psi_{a_{m-1}}^{(1)}
  \psi_{a_m}\ket~~.
\end{equation}

Additional bulk perturbations do not change these results. 
We conclude that the fundamental disk amplitudes
$B_{a_0\ldots a_m;i_1\ldots i_n}$ with $m,n\geq 0$ and $2n+m>1$ are
cyclically symmetric in the boundary indices:
\begin{equation}
  \label{eq:cyclic}
  B_{a_0\ldots a_m;i_1\ldots i_n} = 
  (-1)^{\td a_m(\td a_0+\ldots +\td a_{m-1})}
  B_{a_ma_0\ldots a_{m-1};i_1\ldots i_n}~~.
\end{equation}
Moreover, all such amplitudes are totally symmetric in the bulk indices (the
  argument is the same as for the pure bulk case \cite{WDVV}).

\section{The effective superpotential}
\label{sec:superpot}

In this section, we explain how one can package open-closed disk amplitudes
into a generating function, and how this relates to the effective
superpotential mentioned in the introduction. 

\subsection{Deformed amplitudes on the disk}
\label{sec:deformed_amplitudes}

The last statement of the previous subsection implies that 
we can integrate all bulk perturbations to produce generating functions:
\begin{equation}
  \label{eq:intAinfty}
  \cF_{a_0\ldots a_m} (t)~~\quad \textrm{for}\quad m\geq 0~~
\end{equation}
with the following property:
\begin{equation}
\label{eq:diffAinfty}
B_{a_0\ldots a_m;i_1\ldots i_n}=\partial_{i_1}\dots\partial_{i_n}
{\cal F}_{a_0\dots a_m}(t)|_{t=0}~~.
\end{equation}
For $m\geq 2$, the generating functions are given by the
  expressions:
\begin{equation}
\nn
{\cal F}_{a_0\dots a_m}(t)=(-1)^{{\tilde a}_1+\dots +{\tilde a}_{m-1}}
\langle \psi_{a_0}\psi_{a_1}P\int{\psi_{a_2}}\dots \int{\psi_{a_{m-1}}}\psi_{a_m}
 e^{\sum_{p}{t_p\int_{D^2}{\phi_p^{(2)}}}}\rangle ~~,
\end{equation}
which are understood as the formal power series:
\begin{equation}
\label{eq:formalpower}
{\cal F}_{a_0\dots a_m}(t)=(-1)^{{\tilde a}_1+\dots +{\tilde a}_{m-1}}
\sum_{N_0\dots N_{h_c-1}=0}^\infty
\prod_{p=0}^{h_c-1}{\frac{t_p^{N_p}}{N_p!}}
\langle 
\psi_{a_0}\psi_{a_1}P\int{\psi_{a_2}}\dots \int{\psi_{a_{m-1}}}\psi_{a_m}
{\left[\int \phi^{(2)}_{p}\right]^{N_p}} \rangle~~.
\end{equation}
The cases $m=0$ and $m=1$ of (\ref{eq:corr2}) are special, because one
bulk operator is not integrated. However, through the Ward identity for
$G$, such  correlators are again totally symmetric in the bulk indices. 
Thus one can define $\cBB_a(t)$ and $\cBB_{ab}(t)$ through the relations:
\begin{eqnarray}
\6_{i} \cBB_a(t) &=&-
\langle \phi_i~ \psi_{a}~
 e^{\sum_{p}{t_p\int_{D^2}{\phi_p^{(2)}}}}\rangle~~,  \nn\\[5pt]
\6_{i} \cBB_{ab}(t) &=&-
\langle \phi_i~ \psi_{a} P\int \!\psi^{(1)}_{b}~
e^{\sum_{p}{t_p\int_{D^2}{\phi_p^{(2)}}}}\rangle\nn~~, 
\end{eqnarray}
which determine these quantities up to $t$-independent terms. 

Cyclicity of disk amplitudes with respect to boundary insertions 
(equation (\ref{eq:cyclic})) implies:
\begin{equation}
\label{F_cyc}
{\cal F}_{a_0\dots a_m}(t)=(-1)^{\td a_m(\td a_0+\ldots +\td a_{m-1})}
{\cal F}_{a_ma_0\dots a_{m-1}}(t)~~,
\end{equation}
while equations (\ref{B_vanish}) give:
\begin{equation}
\label{F_vanish}
{\cal F}_{0a_1\dots a_m}(t)=0~~{\rm~for~}~m\neq 2~~.
\end{equation}
and:
\begin{equation}
\label{F_constant}
{\cal F}_{0a_1 a_2}(t)=\omega_{a_1a_2}={\rm~independent~of~}t~~.
\end{equation}

Mimicking the closed string case reviewed in Section \ref{sec:preliminaries}, we
define deformed amplitudes by:
\begin{equation}
\label{B_deformed}
B_{a_0\ldots a_m;i_1\ldots i_n}(t):=\partial_{i_1}\dots\partial_{i_n}{\cal
  F}_{a_0\dots a_m}(t)~~,
\end{equation}
i.e.:
\begin{equation}
\nn
B_{a_0\ldots a_m;i_1\ldots i_n}(t)=(-1)^{{\tilde a}_1+\dots +{\tilde a}_{m-1}}
\langle \psi_{a_0}\psi_{a_1}P\int{\psi_{a_2}^{(1)}}\dots \int{\psi_{a_{m-1}}^{(1)}}\psi_{a_m} 
P\int{\phi_{i_1}^{(2)}} \dots\int{\phi_{i_n}^{(2)}} 
e^{\sum_{p}{t_p\int_{D^2}{\phi_p^{(2)}}}}\rangle~~.
\end{equation}

Notice that $B_{a}(t)={\cal F}_a(t)$ and $B_{ab}(t)={\cal F}_{ab}(t)$ need not vanish, though 
  they must be of order at least one in $t_i$ 
  (cf. equations (\ref{low_B})). In
  particular, this means that deformations of the closed string background
  will generally induce tadpoles:
\begin{equation}
B_a(t):=\langle \psi_a \rangle_t~~, \nn
\end{equation}
where $\langle \dots \rangle_t$ stands for the expectation value on the disk
  taken in the deformed theory. Such tadpoles must of course be canceled
  (for example by performing a shift of the boundary topological vacuum) if
  the deformed theory is to be conformal (and generally a meaningful string background). 
This means that deformations of the bulk and boundary
  sectors must be locked together in order to solve the obstructions, a
  phenomenon well-known from joint deformation theory. We shall further discuss this
  phenomenon in Subsection \ref{subsubsec:tadpole}, and 
  exemplify it for concrete physical models in Section \ref{sec:TLG}.

\subsection{The formal generating function and the effective superpotential}

\label{sec:generating_function}

It is possible to package the cyclic amplitudes $\cF_{a_0\ldots a_m}$
defined in (\ref{eq:intAinfty}) into a single generating function as
follows. Consider the noncommutative and associative superalgebra of formal power series 
${\hat \A}=\C[[{\hat s}_a]]$ in the variables ${\hat s}_a$ of degrees 
${\tilde a}\in \Z_2$ , where $a$ runs from $0$ to $h_o-1$. 
We define the {\em formal generating function} ${\hat {\cal W}}$ through
the expression:
\begin{equation}
\label{Wformal}
{\hat {\cal W}}=\sum_{m\geq 1}{\frac{1}{m}{\hat s}_{a_m}\dots
  {\hat s}_{a_1}{\cal F}_{a_1\dots a_m}({\hat t}) }~~,
\end{equation}
where ${\cal F}_{a_1\dots a_m}({\hat t})$ are viewed as formal power
series. 
This quantity is an element of the associative superalgebra
${\hat {\cal B}}:=\C[[{\hat t}]]\otimes {\hat \A}$, where $\C[[{\hat t}]]:=\C[[{\hat t}_0\dots
{\hat t}_{h_c-1}]]$ is the algebra 
of formal power series in the even and commuting variables ${\hat t}_i$. 

Since ${\hat s}_a$ are non-commuting, the quantity ${\hat {\cal W}}$ 
has no obvious physical interpretation, so the reader might
wonder what is the use of considering non-commuting parameters in the first
place.  To understand this, notice that we can evaluate (\ref{Wformal}) on
supercommuting variables $s_a$ of degrees ${\tilde a}$ 
(so that $s_as_b=(-1)^{{\tilde a}{\tilde b}}s_bs_a$). More precisely, consider a
morphism of unital superalgebras $\pi:{\hat {\cal B}}\rightarrow {\cal B}$,
where ${\cal B}$ is a
unital Banach {\em commutative} superalgebra and let $s_a=\pi({\hat
  s}_a)$ and $t_a:=\pi({\hat t}_a)$.  Then we define the evaluation of ${\hat
  {\cal W}}$ at $(s,t)$ through: 
\be
\label{W}
{\cal W}(s,t):=\sum_{m\geq 1}{\frac{1}{m}s_{a_m}\dots
  s_{a_1}{\cal F}_{a_1\dots a_m}(t)}\in {\cal B}~~,
\ee
where we assume that the series in the right hand side is absolutely
convergent. Here ${\cal F}_{a_1\dots a_m}(t)$ is the evaluation of 
${\cal F}_{a_1\dots a_m}$ at $t$. Formally, we
have ${\cal W}(s,t)=\pi({\hat {\cal W}})$.

Since $s_a$ super-commute and have the same $\Z_2$-degree as the boundary
descendants $\psi_a^{(1)}$, they can be can be viewed as honest boundary
deformation parameters of the world-sheet theory.  For appropriate choices of
$\pi$ and ${\cal B}$, the quantity ${\cal 
W}$ can be viewed as the space-time effective superpotential of
the untwisted $N=2$ model, when such an interpretation of the world-sheet
theory is available.

Because $s_a$ super-commute, it follows that 
monomials in these variables differing by a permutation are related through:
\begin{equation}
s_{a_{\sigma(m)}}\dots s_{a_{\sigma(1)}}=\eta(\sigma; a_1\dots
a_m)s_{a_m}\dots s_{a_1}~~.\nn
\end{equation}
Here $\sigma$ is a permutation on $n$ elements and $\eta(\sigma;a_1\dots
a_m)$  is defined as the sign produced when
permuting $s_a$ to relate the left and right hand
sides.  Using this relation, ${\cW}(s,t)$ reduces to:
\begin{equation}
\label{W_asym}
{\cal W}(s,t)\ =\ \sum_{m\geq 1}{\frac{1}{m!}s_{a_m}\dots s_{a_1}{\cal
    A}_{(a_1\dots a_m)}(t)}~~,
\end{equation}
where:
\begin{equation}
{\cal A}_{a_1\dots a_m}(t):= (m-1)!{\cal F}_{(a_1\dots a_m)}(t)\ :=\ \frac{1}{m}
\sum_{\sigma \in S_m}{\eta(\sigma;a_1\dots a_m){\cal F}_{a_{\sigma(1)}\dots
    a_{\sigma(m)}}}(t)\nn
\end{equation}
are (super-)symmetrized combinations of the cyclic amplitudes 
${\cal F}_{a_1\dots a_m}(t)$ and
$S_m$ is the group of permutations of $m$ objects. These are the relevant,
physically observable quantities, because tree-level scattering amplitudes are
summed over permutations of indistinguishable incoming states. By construction, these functions
are integrable with respect to the boundary deformation parameters, namely they are given
by partial derivatives of ${\cal W}$:
\begin{equation}
\label{FWf}
{\cal A}_{a_1\dots a_m}=(\partial_{a_1}\dots \partial_{a_m}{\cal
 {\hat  W}})(s,t){\Big |}_{s=0}~~,
\end{equation}
where $\partial_a:=\frac {\vec \partial}{\partial {\hat s}_a}$ are the
canonical left derivations of ${\hat \A} $ and the right hand side is evaluated
at $(s,t)$. 

It is clear that ${\cal W}(s,t)$ carries less information than
the full set of disk amplitudes. In other words, one cannot package
the entire information of the topological string theory in this
quantity alone. As explained above,  one way to encode tree-level
world-sheet data without loosing any information is to consider the
formal generating function ${\hat \cW}$ in (\ref{Wformal}). 
In practical applications (for example in Section \ref{sec:TLG} below), we shall 
choose the evaluation map $\pi$ such that $t_i=\pi({\hat t}_i)\in \C \cdot
1_{\cal B}$ for all
$i$ and $s_a=\pi({\hat s}_a) \in 1\in \C \cdot 1_{\cal B}$ for all ${\hat
  s}_a$ 
of even degree (here $1_{\cal B}$ is the unit of ${\cal B}$). Then the restriction
$W(s,t)|_{s^{odd}=0}$, where $s^{odd}=(s_a)_{{\tilde a}=odd}$ defines
a function of the complex variables $t_i$ and $(s_a)_{{\tilde a}=even}$.


\section{Homotopy associativity constraints on boundary amplitudes on the disk}
\label{sec:Ainfty}


In this section, we discuss a countable set of algebraic constraints on tree-level boundary
amplitudes on the disk, which can be viewed as the Ward identities of the BRST 
symmetry. These constraints arise from the relations \cite{HofmanMaA}:
\begin{equation}
  \label{eq:AinftyWard}
  \langle [Q,\psi_{a_0} \psi_{a_1} P \!\int\! \psi_{a_2}^{(1)}
  \ldots \!\int\! \psi_{a_{m-1}}^{(1)} 
  \psi_{a_m}] \rangle = 0~~(m\geq 2)~~,
\end{equation}
which encode BRST invariance of the topological vacuum. They are due to
equation (\ref{eq:Qvardescbound}), which induces nontrivial contributions when
  taking the commutator with the BRST operator in the left
  hand side of (\ref{eq:AinftyWard}). From 
  (\ref{eq:Qvardescbound}), it is clear that the resulting terms will involve
  amplitudes in which two boundary insertions approach each other in the
  limit when the regulator $\epsilon$ is removed.
  Therefore, the contribution on the left hand side of (\ref{eq:AinftyWard})
  is due entirely to contact singularities, and hence it can be factorized into
  amplitudes with lower numbers of insertions. Performing the
  computation, one finds that the Ward identities of the BRST symmetry can be
  brought to a form known in the mathematics literature 
  as a ''minimal $A_\infty$ algebra''.

  Before proceeding with the computation, we briefly 
  mention another, and perhaps more fundamental, point of view.  
  Since the contributions in the left hand side of (\ref{eq:AinftyWard}) arise
  entirely from contact terms, it is clear that their factorization in the
  limit $\epsilon\rightarrow 0$ is intimately connected with an appropriate
  choice of compactification of the moduli space of disks with boundary markings. 
  As in the closed string case, the appropriate
  compactification is provided by so-called "stable disks", which describe the
  allowed degenerations of such geometric objects. In the limit
  $\epsilon\rightarrow 0$, factorization of the terms produced on the right
  hand side of (\ref{eq:AinftyWard}) corresponds to a contribution to the disk
  amplitude coming from the boundary of this compactified moduli
  space. Writing the amplitude as the integral of a closed differential form over
  this space, equation (\ref{eq:AinftyWard}) amounts to the statement that 
  this boundary contribution must vanish. Hence the $A_\infty$ structure
  can be viewed as a consequence of the
  topology of this boundary. In abstract terms, it arises because the 
  strata of the stable compactification obey the defining rules of the
  so-called ``little intervals operad'' with respect to the composition law
  induced by sewing of stable disks at their boundary punctures. This
  point of view on the origin of the $A_\infty$ constraints is intimately
  connected with open string field theory in its general formulation given by
  Zwiebach (see \cite{Zwiebachoc} and references therein). In fact, the 
  string field theory perspective provides maybe the most elegant derivation of
  such constraints, but it lies outside the scope of the present paper, so we 
  shall give the more elementary derivation based on conformal field theory arguments.

As sketched above, acting explicitly with the BRST operator on the left hand side of 
equation (\ref{eq:AinftyWard}) and using the descent relation
$[Q,\psi^{(1)}_{a_k}]=[Q,[G,\psi_{a_k}]] = \6_{\tau_k} \psi_{a_k}$ 
produces an integration over the boundary of the stable compactification of
the moduli space of the boundary-punctured disk, where two or more punctures get together very
closely. The discussion of the resulting terms involves the regularization 
(\ref{eq:bounddomain1}) in an essential manner. 

For clarity, we first discuss the case  $m=4$. The regularized
configuration space and its boundary components are 
shown in Figure \ref{fig:modulispace4}. The left-hand side of equation
(\ref{eq:AinftyWard}) becomes:
  \begin{figure}[t]
  \begin{center}
 \epsfysize=7cm\centerline{\epsffile{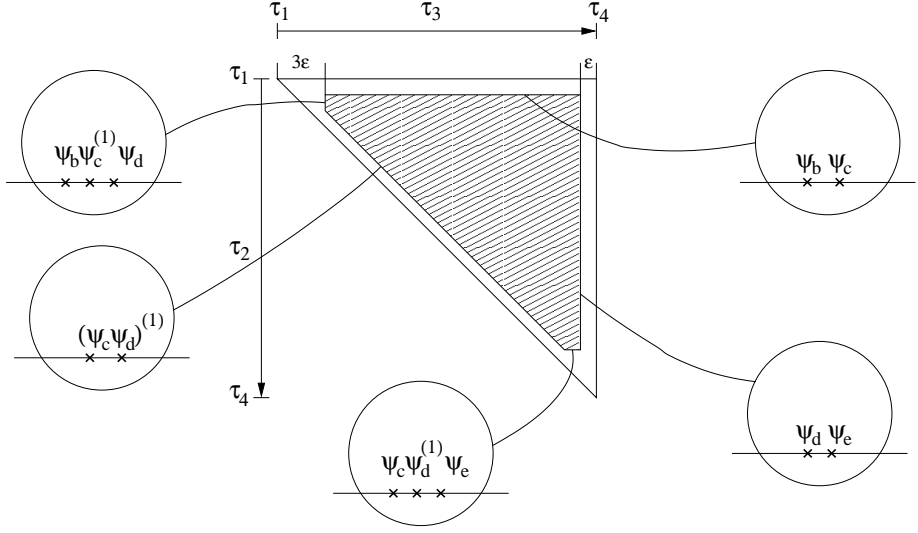}}
    \parbox{12truecm}{
      \caption{
        \label{fig:modulispace4}
        The integration domain $\bbS_5(\tau_2,\tau_5)$ and its
        boundary components 
        (through a magnifying glass) for the correlation function 
        $\bra \psi_a(\tau_0) \psi_b(\tau_1)~
        P\!\int\!\psi_c^{(1)}(\tau_2)\!\int\!\psi_d^{(1)}(\tau_3)~
        \psi_e(\tau_4)\ket$.
         }
    }
  \end{center}
  \end{figure}
\begin{eqnarray}
  \label{eq:Ainftyder}
  \sum\limits_{k=2}^{m-1} (-1)^{s_k} \hspace*{-14pt}&&\hspace*{-14pt}
  \bra \psi_{a_0} \psi_{a_1} P \!\int\! \psi_{a_2}^{(1)} \ldots
  \!\int \6_{\tau_k}\!\psi_{a_k} \ldots \!\int\! \psi_{a_{m-1}}^{(1)} 
  \psi_{a_m}\ket  \\ \nonumber
  =
  \sum\limits_{k=2}^{m-1} (-1)^{s_k} 
  \int\limits_{\tau_1}^{\tau_m} \!\!d\tau_k 
  \hspace*{-10pt}&\biggl(&\hspace*{-10pt}
  \6_{\tau_k}\!
  \bra \psi_{a_0} \psi_{a_1} 
  \!\int\limits_{\tau_1}^{\tau_k}\!\psi_{a_2}^{(1)} 
  \!\int\limits_{\tau_2}^{\tau_k}\! \psi_{a_3}^{(1)} \ldots
  \!\int\limits_{\tau_{k-2}}^{\tau_k}\!\!\psi_{a_{k-1}}^{(1)}
  \psi_{a_k}
  \!\int\limits_{\tau_k}^{\tau_{k+2}}\!\!\psi_{a_{k+1}}^{(1)}  
  \ldots 
  \!\int\limits_{\tau_k}^{\tau_{m}}\! \psi_{a_{m-1}}^{(1)} 
  \psi_{a_m}\ket\\ \nonumber
  \hspace*{-10pt}&-&\hspace*{-10pt}
  \sum\limits_{l=2}^{k-1}
  \bra \psi_{a_0} \psi_{a_1} 
  \!\int\limits_{\tau_1}^{\tau_k}\!\psi_{a_2}^{(1)} \ldots
  \Bigl[\psi_{a_l}^{(1)}\bigr|_{\tau_l \rightarrow \tau_k}
  \!\int\limits_{\tau_{l}}^{\tau_k}\!\!\psi_{a_{l+1}}^{(1)} \dots
  \psi_{a_k}\Bigr]
  \!\int\limits_{\tau_k}^{\tau_{k+2}}\!\!\psi_{a_{k+1}}^{(1)}  
  \ldots 
  \!\int\limits_{\tau_k}^{\tau_{m}}\! \psi_{a_{m-1}}^{(1)} 
  \psi_{a_m}\ket\\ \nonumber
  \hspace*{-10pt}&+&\hspace*{-10pt}
  \sum\limits_{l=k+1}^{m-1}
  \bra \psi_{a_0} \psi_{a_1} 
  \!\int\limits_{\tau_1}^{\tau_k}\!\psi_{a_2}^{(1)} \ldots
  \Bigl[\psi_{a_k} \ldots
  \!\int\limits_{\tau_k}^{\tau_l}\!\!\psi_{a_{l-1}}^{(1)}
  \psi_{a_l}^{(1)}\bigr|_{\tau_k \leftarrow \tau_l}  \Bigr] \ldots
  \!\int\limits_{\tau_k}^{\tau_{m}}\! \psi_{a_{m-1}}^{(1)} 
  \psi_{a_m}\ket  
  \biggr)\\ \nonumber
  = 
  \sum\limits_{k=2}^{m-1} (-1)^{s_k} 
  \hspace*{-10pt}&\biggl(&\hspace*{-10pt}
  \bra \psi_{a_0} \psi_{a_1} P \int \psi_{a_1}^{(1)} \ldots 
  \int \psi_{a_{k-1}}^{(1)} 
  \Bigl[ \psi_{a_k}\bigr|_{\tau_k \rightarrow \tau_m} 
  P \int \psi_{a_{k+1}}^{(1)} \ldots 
  \int \psi_{a_{m-1}}^{(1)} \psi_{a_m} \Bigr]
  \ket \\ \nonumber
  \hspace*{-10pt}&-&\hspace*{-10pt}
  \bra \psi_{a_0} 
  \Bigl[ \psi_{a_1} P \int \psi_{a_2}^{(1)} \ldots 
  \int \psi_{a_{k-1}}^{(1)} \psi_{a_k}\bigr|_{\tau_1\leftarrow\tau_k}
  \Bigr]
  P \int \psi_{a_{k+1}}^{(1)} \ldots 
  \int \psi_{a_{m-1}}^{(1)}\psi_{a_m} \ket \\ \nonumber
  \hspace*{-10pt}&-&\hspace*{-10pt}
  \sum\limits_{l=2}^{k-1}
  \bra \psi_{a_0} \psi_{a_1} 
  P \int\psi_{a_2}^{(1)} \ldots
  \int \Bigl[\psi_{a_l}\bigr|_{\tau_l \rightarrow \tau_k}
  P \int\psi_{a_{l+1}}^{(1)} \dots
  \psi_{a_k}\Bigr]^{(1)}
  \ldots 
  \int\psi_{a_{m-1}}^{(1)} 
  \psi_{a_m}\ket
\biggr)  ~~,
\end{eqnarray}
where the sign is given by $s_k = \td a_0+\ldots +\td a_{k-1}$. In the
second step we used the fact that the regularized configuration space is a 
simplex, which means that we have nested integration domains
\footnote{For sake of easier reading, the nested integrals over
          $\tau_{k+1}$ to $\tau_{m-1}$ are partly
          written in the `wrong' order.}.
For notational simplicity, we do not indicate the cut-off $\epsilon$ in
the integrals.

In the last form of (\ref{eq:Ainftyder}), the terms in square
brackets are products of boundary operators. In the limit
$\epsilon \rightarrow 0$, we can factorize the result by pulling these terms
out while inserting the sum $\sum_{a,b}\psi_c\omega^{cd}\psi_d$ 
over a basis of the on-shell space of boundary observables. This gives:
\begin{eqnarray}
  \label{eq:Ainftyfact} \nonumber
  \sum\limits_{k=2}^{m-1} (-1)^{s_k} 
  \hspace*{-10pt}&\biggl(&\hspace*{-10pt}
  \bra \psi_{a_0} \psi_{a_1} P \int \psi_{a_1}^{(1)} \ldots 
  \int \psi_{a_{k-1}}^{(1)} \psi_c \ket ~\omega^{cd}~
  \bra \psi_d \psi_{a_k}
  P \int \psi_{a_{k+1}}^{(1)} \ldots 
  \int \psi_{a_{m-1}}^{(1)} \psi_{a_m}\ket \\
  \hspace*{-10pt}&-&\hspace*{-10pt}
  \bra \psi_{a_0} \psi_c P \int \psi_{a_{k+1}}^{(1)} \ldots 
  \int \psi_{a_{m-1}}^{(1)}\psi_{a_m} \ket
  ~\omega^{cd}~ \bra \psi_d \psi_{a_1} P \int \psi_{a_2}^{(1)} \ldots 
  \int \psi_{a_{k-1}}^{(1)} \psi_{a_k} \ket \\ \nonumber
  - \sum\limits_{l=2}^{k-1}
  \hspace*{-10pt}& &\hspace*{-10pt}
  \bra \psi_{a_0} \psi_{a_1} 
  P \int\psi_{a_2}^{(1)} \ldots
  \int \psi_c^{(1)}
  \ldots 
  \int\psi_{a_{m-1}}^{(1)} \psi_{a_m}\ket ~\omega^{cd}~
  \bra \psi_d \psi_{a_l}
  P \int\psi_{a_{l+1}}^{(1)} \dots
  \psi_{a_k}\ket
  \biggr) = 0~~.
\end{eqnarray}
We next re-write this equation in terms of the quantities defined in equation (\ref{eq:pullindex}).
Using (\ref{B}), we find:
\begin{equation}
  \label{eq:Ainfty}
  \sum\limits_{\tiny
    \begin{array}{c} k,l=2\\ 
      k\!\!-\!\!m\!\!+\!\!2\!<\!l\!\leq\! k
   \end{array}}^{m}
  \hspace*{-10pt}
  (-1)^{\td a_1 + \ldots + \td a_{l-2}}
  B^b{}_{a_1\ldots a_{l-2}ca_{k+1}\ldots a_m} B^c{}_{a_{l-1}\ldots a_k}
  = 0~~{\rm~for~}m\geq 2~~.
\end{equation}
In deriving (\ref{eq:Ainfty}) we used the selection rules
$\td b = \td a_1+\ldots +\td a_m + 1$ for $B^b{}_{a_1\ldots a_m}$ and 
$\grd = \td a_0+\ldots +\td a_m$ for $B_{a_0\ldots a_m}$.
The restrictions in the sum account for the fact that the amplitudes
$B_{a_0\ldots a_m}$ are considered only for $m \geq 2$ (alternatively, one can
remove these constraints and use definitions (\ref{low_B})).
The first equation in (\ref{eq:Ainfty}) is obtained for $m=2$, and coincides
with the associativity condition (\ref{assoc_B}) for the boundary product.

\subsection{Algebraic description}

To make contact with expressions found in the mathematics literature, 
let us bring (\ref{eq:Ainfty}) to a more familiar form. For this, we 
define tree-level {\em boundary scattering products} $r_m:H_o^{\otimes
  m}\rightarrow H_o$ to be the multilinear maps determined by the
equations:
\begin{equation}
\label{sproducts}
r_m(\psi_{a_1}\dots \psi_{a_m})=B^{a_0}_{a_1\dots a_m}\psi_{a_0}~~,
\end{equation}
where, as usual, we use implicit summation over repeated indices. 
The selection rule for $B^{a_0}_{a_1\dots a_m}$ gives:
\begin{equation}
\nn
\deg~r_m(\psi_{a_1}\dots \psi_{a_m})=1+\sum_{j=1}^m{\td a_j}~~,
\end{equation}
so all maps $r_m$ have degree one when $H_o$ is endowed with the
suspended grading.
Equation (\ref{eq:Ainfty}) takes the form:
\begin{equation}
\label{Ainf_std}
\sum\limits_{\tiny
    \begin{array}{c} k+l=m+1\\ 
      j=0\dots k-1
   \end{array}}^{m}
  \hspace*{-10pt}
  (-1)^{\td a_1 + \ldots + \td a_j}
  r_k(\psi_{a_1}\ldots \psi_{a_j}, r_l(\psi_{a_{j+1}}\ldots \psi_{a_{j+l}}),
  \psi_{a_{j+l+1}}\ldots \psi_{a_m})  = 0~~,
\end{equation}
where we set $r_0=r_1=0$. Relations (\ref{Ainf_std}) define an 
$A_\infty$ algebra \cite{StasheffA,StasheffB}, in conventions in which all products
  have degree one. For reader's convenience, we summarize
  the standard terminology concerning such algebras:

(1) A collection of multilinear maps $r_m:H_o^{\otimes m}\rightarrow 
    H_o$ of degree $+1$ satisfying (\ref{Ainf_std}) is called a {\em weak}
    $A_\infty$ algebra if $m$ is allowed to run from $0$ to $\infty$.

(2) Such a collection is called a {\em strong} $A_\infty$ algebra (or simply
    an $A_\infty$ algebra) if $m$ runs from $1$ to infinity.

(3) Such a collection is a {\em minimal} $A_\infty$ algebra if $m$ runs from
    $2$ to infinity. 

Thus a (strong) $A_\infty$ algebra is a weak $A_\infty$ algebra for which
$r_0=0$, while a minimal $A_\infty$ algebra is a (strong) $A_\infty$ algebra 
for which $r_1=0$. The algebra obtained above is a minimal $A_\infty$
algebra. As we shall see below, bulk perturbations will generically deform this to 
a {\em weak} $A_\infty$ algebra. This corresponds to the appearance of a
tadpole induced by deformations of the closed string background.

Due to the cyclicity property (\ref{eq:cyclic}) of disk amplitudes, 
our minimal $A_\infty$ algebra is in fact {\em cyclic} with respect to the
bilinear form on $H_o$ defined by the boundary topological metric. 
Writing:
\begin{equation}
B_{a_0\dots a_m}=\omega(\psi_{a_0}, r_m(\psi_{a_1}\dots \psi_{a_m}))~~,\nn
\end{equation}
this is simply condition (\ref{eq:cyclic}) expressed in terms of string
scattering products:
\begin{equation}
\label{r_cyc}
\omega(\psi_{a_0}, r_m(\psi_{a_1}\dots \psi_{a_m}))=(-1)^{{\tilde a}_m({\tilde
    a}_0+\dots +{\tilde a}_{m-1})}\omega(\psi_{a_m}, r_m(\psi_{a_0}\dots \psi_{a_{m-1}}))~~.
\end{equation}
A further constraint follows from equations (\ref{B_vanish}), which imply:
\begin{equation}
B^c_{a_1\dots a_{i-1} 0 a_{i+1}\dots a_m}=0~~{\rm~for~}~~m\geq
2~~{\rm~and~all}~i=1\dots m~~,\nn
\end{equation}
i.e.:
\begin{equation}
\label{unit1}
r_m(\psi_{a_1}\dots \psi_{a_{i-1}}, 1_o, \psi_{a_{i+1}}\dots \psi_{a_{m-1}})=0~~{\rm~for~}~~m\geq
3~~{\rm~and~all}~i=1\dots m-1~~.
\end{equation}
On the other hand, we have:
\begin{equation}
r_2(\psi_a, \psi_b)=B_{ab}^c\psi_c=(-1)^{\tilde
    a}D_{ab}^c\psi_c~~.\nn
\end{equation}
Using the fact that $1_o$ is a unit for the boundary algebra, this gives:
\begin{equation}
\label{unit2}
r_2(1_o,\psi_a)=(-1)^{\tilde a}r_2(\psi_a,1_o)=\psi_a~~.
\end{equation}
Equations (\ref{unit1}) and (\ref{unit2}) mean that $(H_o,r_*)$ is a {\em
    unital} $A_\infty$ algebra (see, for example, \cite{FukayamirrorB}). 

\paragraph{\bf Observation} When considering boundary condition changing
sectors, the $A_\infty$ algebra discussed above generalizes to an $A_\infty$
category \cite{Fukayainfty}.

\

The relevance of $A_\infty$ algebras was originally
pointed out in \cite{Gaberdiel} in the context of open string field theory
in the general, non-polynomial formulation given by Zwiebach 
(see \cite{Zwiebachoc} and references therein). 
In this approach, one obtains $A_\infty$ constraints on 
open string products. Such products are associated with geometric vertices whose 
construction  depends on a positive parameter $l$, 
which measures the length of their external strips. 
The scattering products considered above can be viewed as the limit
$l\rightarrow +\infty$ of the string products of  \cite{Gaberdiel}, while the
limit $l\rightarrow 0^+$ recovers the better known formulation of
\cite{WittenSFT}, in which only the cubic vertex survives. 
$A_\infty$ algebras were originally introduced by J. Stasheff
\cite{StasheffA,StasheffB}, while $A_\infty$ categories were first discussed by 
K. Fukaya \cite{Fukayainfty}. They play a central role in the homological mirror symmetry program 
\cite{KontsevichHMS,Fukayainfty,Fukayabook,Fukayarev,Fukayactg,FukayamirrorA,FukayamirrorB}, 
where they arise via topological string field theory (see \cite{CILf} and references therein).

\section{Constraints on deformed amplitudes}
\label{sec:def_constraints}

We are now ready to discuss the consistency constraints for mixed bulk-boundary
amplitudes on the disk, and
derive the generalization of the homotopy associativity constraints of
Section \ref{sec:Ainfty}. 
As we shall see below, the relevant consistency conditions take the form of a 
weak, cyclic and unital $A_\infty$ algebra, which can be viewed as an
all-order deformation of the minimal $A_\infty$ algebra of Section
\ref{sec:Ainfty}. The appearance of a weak $A_\infty$ algebra under
deformations of the closed string background is due to the generation of an
open string tadpole, which must be canceled by a shift of the open string
vacuum. This encodes interlocking of open and closed string deformation parameters
when solving the joint deformation problem for the bulk and boundary sectors. After 
discussing the algebraic and physical interpretation of this phenomenon, we 
investigate the remaining constraints, which encode the 
stringy generalization of the second bulk-boundary sewing condition and of the 
Cardy relation. This completes the set of consistency conditions
constraining open-closed amplitudes on the disk.


\subsection{Weak $A_\infty$ constraints for mixed amplitudes on the disk}


In the present subsection, we extend the discussion of $A_\infty$ constraints
to general open-closed amplitudes on the disk. We shall show that the
$A_\infty$ structure exhibited in Section \ref{sec:Ainfty} is promoted to a so-called
{\em weak} $A_\infty$ algebra, which is again cyclic and unital. For
simplicity we start by discussing the case of a single boundary insertion. 
As we shall see below, these amplitudes can be used to define a
first order deformation of the $A_\infty$ algebra of Section \ref{sec:Ainfty}, 
a deformation which preserves cyclicity and unitality but 
need not preserve minimality. We shall also discuss the general case of multiple insertions,
which defines an all-order (formal) deformation in the bulk parameters $t_i$.

\subsubsection{Disk amplitudes with a single bulk insertion}
\label{sec:lindef}

Insertions of bulk operators perturb the minimal $A_\infty$ algebra
extracted in Section \ref{sec:Ainfty}. We first consider linear
perturbations, which amount to inserting just one bulk operator in the disk amplitudes:
\begin{equation}
  \label{eq:linbulkWard}
  \bra [Q,\phi_i \psi_{a_0} P \!\int\! \psi_{a_1}^{(1)}
  \ldots \!\int\! \psi_{a_m}^{(1)}] \ket = 0~~.
\end{equation}
As in Section \ref{sec:Ainfty}, acting with the BRST commutator on the
integrated descendants on the left hand side
produces terms in which several boundary fields approach each
other. In the limit $\epsilon\rightarrow 0$, we can factorize the result
by inserting complete systems of open string observables. Notice that the
integration domain
for equation (\ref{eq:linbulkWard}) differs from that of equation
(\ref{eq:AinftyWard}), 
because we have only one fixed boundary operator. This makes the computation
more involved.

\begin{figure}[t]
  \begin{center}
  \epsfysize=7cm\centerline{\epsffile{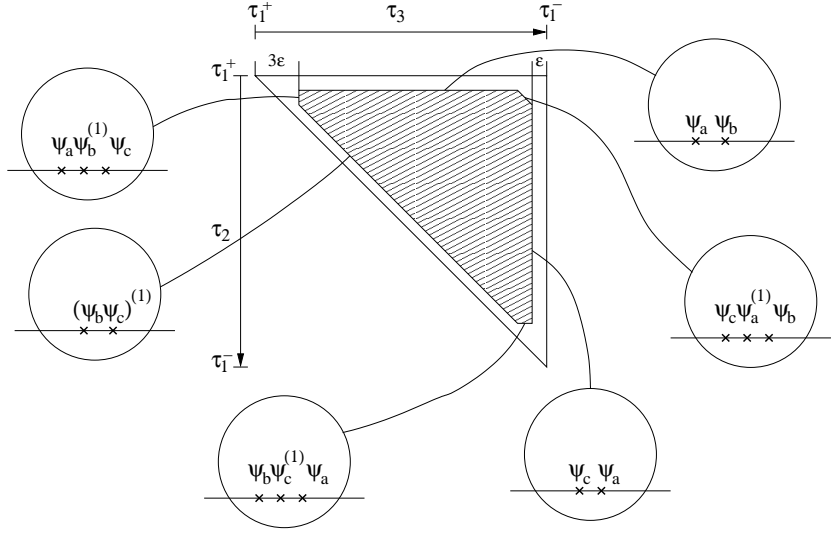}}
    \parbox{12truecm}{
      \caption{
        The integration domain $\bbS_3(\tau_1)$ and its boundary
        components (through a magnifying glass) for the correlation
        function $\bra \phi_i(w,\bar w)~ \psi_a(\tau_1)~ 
        P\!\int\!\psi_b(\tau_2)\!\int\!\psi_c(\tau_3) \ket$. The real
        line, as boundary of the disk, was compactified to a circle
        by identifying $\tau_1^+$ and $\tau_1^-$.
        \label{fig:modulispace3}}
    }
  \end{center}
\end{figure}

Let us illustrate this with the simplest
non-trivial case, namely $m=2$. The integration domain and its
boundary components are shown in Figure~\ref{fig:modulispace3}.
Using the descent equation (\ref{eq:bounddesc}), the left-hand side of
(\ref{eq:linbulkWard}) becomes: 
\begin{equation}
  \label{eq:linbulkder}
  (-1)^{a_0} \bra \phi_i \psi_{a_0} 
  \int\limits_{\tau_0^+ +\e}^{\tau_0^- -3\e}\!\! \6_{\tau_1} 
  \psi_{a_1} 
  \int\limits_{L_2(\tau_0,\tau_1)}^{R_2(\tau_0,\tau_1)}\!\! 
  \psi_{a_2}^{(1)}\ket
  +(-1)^{a_0 + a_1+1}\bra \phi_i \psi_{a_0} 
  \int\limits_{L_1(\tau_0,\tau_2)}^{R_1(\tau_0,\tau_2)}\!\!
  \psi_{a_1}^{(1)} 
  \int\limits_{\tau_0^+ +3\e}^{\tau_0^- -\e}\!\!\6_{\tau_2}
  \psi_{a_2}\ket~~.
\end{equation}
The boundary of the integration domain can be inferred from 
(\ref{eq:bounddomain2}) and is shown in Figure
\ref{fig:modulispace3}. Its components are given by: 
\begin{eqnarray}
&&R_2(\tau_0,\tau_1) = 
\Big\{\begin{array}{lll} 
  \tau_0^- - \e &
  \textrm{for}&
  \tau_1 > \tau_0^+ + 2\e \\
  \tau_0^- - 3\e + (\tau_1 -\tau_0^+)  &
  \textrm{for}&
  \tau_1 < \tau_0^+ + 2\e 
\end{array} \nn\\
&&L_2(\tau_0,\tau_1) = 
\Big\{\begin{array}{lll} 
  \tau_1 + \e &
  \textrm{for}&
  \tau_1 > \tau_0^+ + 2\e \\
  \tau_0^+ + 3\e &
  \textrm{for}&
  \tau_1 < \tau_0^+ + 2\e 
\end{array} ~~,\nn
\end{eqnarray}
with similar expressions for $R_1$ and $L_1$.
As in the derivation of Section \ref{sec:Ainfty},  we use partial
integration taking into account all boundary contributions. 
Compared to Section \ref{sec:Ainfty}, we have an additional contribution from the upper
right corner of the regularized configuration space in Figure \ref{fig:modulispace3},
which comes from the boundary components $R_2$ for 
$\tau_0^+ + \e<\tau_1 < \tau_0^+ + 2\e$ and  $L_1$ for $\tau_0^- -
\e>\tau_2>\tau_0^- - 2\e$. This contribution takes 
the form:
\begin{eqnarray}
  \nonumber
  (-1)^{a_0+1 + (a_2+1)(a_0+a_1)} \bra \phi_i 
  \!\!\int\limits_{\tau_0^++\e}^{\tau_0^++2\e}\!\! d\tau_1~
  \Bigl(&&\hspace*{-15pt} \psi_{a_2}^{(1)}(\tau_1\!-\!3\e)~
  \psi_{a_0}(\tau_0)~\psi_{a_1}(\tau_1) +\\ \nonumber
  +~ (-1)^{a_0+a_2}&&\hspace*{-15pt}
  \psi_{a_2}(\tau_1\!-\!3\e)~ 
  \psi_{a_0}(\tau_0)~\psi_{a_1}^{(1)}(\tau_1)
  \Bigr) \ket \\ \nonumber
  =~ -(-1)^{a_0+(a_2+1)(a_0+a_1+1)}&&\hspace*{-15pt}
 \bra  \phi_i~ \Bigl(\psi_{a_2}(\tau_1\!-\!3\e)
 \!\!\int\limits_{\tau_1-2\e}^{\tau_1-\e}\!\!d\tau_0 
 \psi_{a_0}^{(1)}(\tau_0)~
 \psi_{a_1}(\tau_1) \Bigr) \ket~~,
\end{eqnarray}
where we used a Ward identity corresponding to the current $G$ to
`move' the integral from $\tau_1$ to $\tau_0$
\footnote{More precisely, we used the relation:
$$
  \oint \xi(w) \bra G(w)~ \phi_i(z,\bar z)~ 
  \psi_{a_2}(\tau_2)~ \psi_{a_0}(\tau_0)~ \psi_{a_1}(\tau_1) \ket = 0~,
$$
with $\xi(w)=(w-z)(w-\bar z)$ as well as the fact that correlators depend only
on the cross ratio $\zeta_l=\frac{(z-\bar z)(\tau_l-\tau_0)}{(z-\tau_l)(\bar z -
  \tau_0)}$ for $l=1,2$. In the limit  $\tau_2 \rightarrow \tau_1-3\e$, we
have $\zeta_2 = \zeta_1 + \cO(\e)$ and we obtain the Jacobian
$\left|\frac{\6\tau_1}{\6\tau_0}\right|$ up to terms $\cO(\e)$, which
vanish in our limit.
}.
Collecting all terms and factorizing as in Section \ref{sec:Ainfty}, 
we find that relation (\ref{eq:linbulkWard}) reduces to:
\begin{eqnarray}
  \label{eq:linbulkfact}
  & &
  (-1)^{\td a_0} 
  \bra \phi_i \psi_c \int \psi_{a_2}^{(1)}\ket
  \omega^{cd} \bra \psi_d \psi_{a_0} \psi_{a_1} \ket 
  \nonumber \\ &+&
  (-1)^{\td a_0(\td a_1+\td a_2)+\td a_1+\td a_2}
  \bra \phi_i \psi_c \ket \omega^{cd} 
  \bra \psi_d \psi_{a_1} \int \psi_{a_2}^{(1)} \psi_{a_0} \ket 
  \nonumber \\ &+&
  (-1)^{\td a_2(\td a_0+\td a_1)+\td a_2} 
  \bra \phi_i \psi_c \int \psi_{a_1}^{(1)}\ket
  \omega^{cd} \bra \psi_d \psi_{a_2} \psi_{a_0} \ket 
            \\ &+&
  (-1)^{\td a_0+\td a_1}
  \bra \phi_i \psi_c \ket \omega^{cd} 
  \bra \psi_d \psi_{a_0} \int \psi_{a_1}^{(1)} \psi_{a_2} \ket 
  \nonumber \\ &+&
  (-1)^{\td a_0+\td a_2 +\td a_2(\td a_0+\td a_1)}
  \bra \phi_i \psi_c \ket \omega^{cd} 
  \bra \psi_d \psi_{a_2} \int \psi_{a_0}^{(1)} \psi_{a_1} \ket 
  \nonumber \\ &+&
  (-1)^{\td a_0+\td a_1} 
  \bra \phi_i \psi_{a_0} \int \psi_c^{(1)}\ket
  \omega^{cd} \bra \psi_d \psi_{a_1} \psi_{a_2} \ket ~=~0~~.  
  \nonumber
\end{eqnarray}
\begin{figure}[t]
  \begin{center}
    \epsfysize=4.0cm\centerline{\epsffile{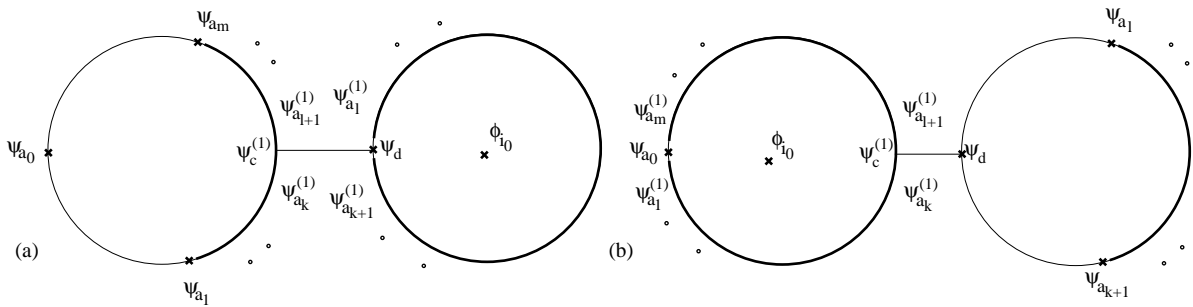}}
    \parbox{12truecm}{
      \caption{
        The two contributions to the factorizations leading to equation
        (\ref{eq:linpert}). Figure (a) shows a summand of the first
        term in this equation, while Figure (b) shows a summand
        of the second term.
        \label{fig:Hochfact}}
    }
  \end{center}
\end{figure}
When expressed in terms of the quantities defined in (\ref{B}), this equation takes the form:
\begin{eqnarray}
  \label{eq:linpertm2}
  & &
  B^{a_0}{}_{d;i} B^d{}_{a_1 a_2}
  + B^{a_0}{}_{d a_2} B^d{}_{a_1;i}
  + (-1)^{\td a_1} B^{a_0}{}_{a_1 d} B^d{}_{a_2;i} \\[10pt]
  &+& \nn
  B^{a_0}{}_{d a_1 a_2} B^d{}_{i}
  + (-1)^{\td a_1} B^{a_0}{}_{a_1 d a_2} B^d{}_{i}
  + (-1)^{\td a_1+\td a_2} B^{a_0}{}_{a_1 a_2 d} B^d{}_{i}~=~0~~.
\end{eqnarray}

The general case is a straightforward generalization,
but the computations are 
much more tedious. Therefore, we shall give the result without presenting the details
of its proof. It is the natural generalization of (\ref{eq:linpertm2}) for $m\geq 1$:
\begin{eqnarray}
  \label{eq:linpert}
  \sum\limits_{j=0}^{m}\sum\limits_{k=0}^{j}
  (-1)^{\td a_1+ \ldots +\td a_k}~
  B^{a_0}{}_{a_1\ldots a_kca_{j+1}\ldots a_m}~
  B^{c}{}_{a_{k+1}\ldots a_j;i} &+& \\ \nonumber
  +~ \sum\limits_{j=2}^{m}\sum\limits_{k=0}^{j}
  (-1)^{\td a_1+ \ldots +\td a_k}~
  B^{a_0}{}_{a_1\ldots a_kca_{j+1}\ldots a_m;i}~
  B^{c}{}_{a_{k+1}\ldots a_j}&=&0~~,
\end{eqnarray}
where we explicitly set $B^c{}_{a} = B^c = 0$, consistent with definitions (\ref{low_B}). 
The first of equations (\ref{eq:linpert}) 
is obtained for $m=1$ and coincides with the first bulk-boundary sewing
  constraint (\ref{bb1_B}) of TFT.


\subsubsection{General disk amplitudes}


We now turn to the general case, allowing for an arbitrary
number of bulk insertions. 
Extending the argument of the previous subsection, 
we will extract a series of constraints 
which amount to the statement that the deformed
disk amplitudes $B_{a_0 \dots a_m}(t)={\cal F}_{a_0\dots a_m}(t)$ satisfy the
defining relations of a weak $A_\infty$ algebra.

Consider a general disk amplitude written in the form:
\begin{equation}
  \label{eq:linbulkWard_gen}
  \bra [Q,\phi_{i_0} \psi_{a_0} P \!\int\! \psi_{a_1}^{(1)}
  \ldots \!\int\! \psi_{a_m}^{(1)} 
  \int\phi_{i_1}^{(2)}\ldots\int\phi_{i_n}^{(2)}] \ket = 0~~.
\end{equation}
Acting with the $Q$-commutator on the integrated boundary insertions produces a sum 
  over the terms appearing in equation (\ref{eq:linpert}). Additionally, we
  also have all contributions from integrated bulk descendants:
\begin{eqnarray}
  \label{eq:partofall}
  \sum\limits_{I \subseteq I_{0,n}}
  \sum\limits_{k\leq j}^{m}
  (-1)^{\td a_1+ \ldots +\td a_k}~
  B^{a_0}{}_{a_1\ldots a_kca_{j+1}\ldots a_m;I_{0,n}\backslash I}~
  B^{c}{}_{a_{k+1}\ldots a_j;I}~~,
\end{eqnarray}
where $I_{p,q}=\{i_p, i_{p+1}, \ldots,i_{q-1}, i_q\}$. Note
  that the BRST variation of boundary fields does not produce terms containing
  $B^{a}{}_{I\backslash i_0}$ and $B^{a}{}_{b;I\backslash i_0}$. Instead,
  these missing  terms arise from the $Q$-variation of the integrated bulk
  insertion:
\begin{equation}
  \label{eq:bulkloop}
  [Q,\int \phi_{i_k}^{(2)}] =
  \lim\limits_{\e \rightarrow 0}
  \oint\limits_{k\e} \phi_{i_k}^{(1)}~~. 
\end{equation}
The integral runs along a loop, which follows the boundary
(namely the real axis) at a distance $k\e$. Using equation
  (\ref{eq:bulkloop}), 
we obtain contributions from a bulk operator approaching the
boundary far away from any boundary operator, and from a bulk
operator approaching a boundary insertion.
Due to our regularization (\ref{eq:bulkdomain}), the
loop (\ref{eq:bulkloop}) cuts the integration domains of the operators
$\phi^{(2)}_{i_1}\dots \phi^{(2)}_{i_{k-1}}$  into a part near the
boundary and a bulk part. On the other hand, the 
operators with $\phi^{(2)}_{i_{k+1}}\dots \phi^{(2)}_{i_n}$ are inside the loop
and hence they don't produce more contact terms. 
In the limit $\epsilon\rightarrow 0$, factorization 
proceeds by distributing the former operators in all possible ways on the two
emerging disks. 

As an example, consider the piece of the boundary sitting between
$\psi_{a_l}$ and $\psi_{a_{l+1}}$. A typical term produced by the process
above has the form:
\begin{equation}
\bra \ldots \int\psi_{a_l} 
  \int\limits_{\tau_l}^{\tau_{l+1}}\phi_{i_k}^{(1)}
  \int\psi_{a_{l+1}} \ldots
  \prod_{j\neq k} \int\phi_{i_j}^{(2)}\ket~~.\nn
\end{equation}
Its factorization produces the contributions
$\pm \sum_{I \subseteq I_{1,k-1}} B_{\ldots a_l c a_{l+1}\ldots ;i_0 
  I_{1,n}\backslash \{i_k I\} } 
B^c{}_{i_k I}$, where we used the notation $\{i_kI\}=\{i_k\}\cup I$. 
Summing over $k$ leads to a total
contribution:
\begin{equation}
\pm \sum_k \sum_{I \subseteq I_{1,k-1}} 
  B_{\ldots a_l c a_{l+1}\ldots;i_0 
    I_{1,n}\backslash \{i_k I \}} B^c{}_{i_k I} =
  \pm \sum_{I\subseteq I_{1,n}}B_{\ldots a_l c a_{l+1}\ldots;i_0 
    I_{1,n}\backslash I} B^c{}_{I}~~.\nn
\end{equation}
Similarly, the factorization of a bulk operator approaching an
\emph{integrated} boundary field gives rise to the terms:
\begin{equation}
\pm \sum_{I\subseteq I_{1,n}}B_{\ldots a_{l-1} c a_{l+1}\ldots;i_0 
    I_{1,n}\backslash I} B^c{}_{a_l;I}~~. \nn
\end{equation}
Finally, bulk operators approaching the fixed insertion $\psi_{a_0}$ produce:
\begin{equation}
\pm \sum_{I\subseteq I_{1,n}}  
B^{a_0}{}_{c;I_{1,n}\backslash I} B^c{}_{a_1 \ldots a_n;i_0 I}~~.\nn
\end{equation}
This completes the list of contributions from the boundary of the
configuration space.

Gathering all terms, we find that equation (\ref{eq:linbulkWard_gen})
can be written in the following form:
\begin{equation}
  \label{eq:genearlpert}
  \sum\limits_{I \subseteq I_{0,n}}
  \sum\limits_{\tiny\begin{array}{c}k,j=0\\ k\leq j\end{array}}^{m}
  (-1)^{\td a_1+ \ldots +\td a_k}~
  B^{a_0}{}_{a_1\ldots a_kca_{j+1}\ldots a_m;I_{0,n}\backslash I}~
  B^{c}{}_{a_{k+1}\ldots a_j;I}~=~0~~.
\end{equation}
For $n=0$, this reduces to equation (\ref{eq:linpert}) extracted in the
  previous subsection.

Notice that indices distribute in the same manner as 
would derivatives with respect to the formal parameters $t^j$. 
This means that we can concisely write relations (\ref{eq:genearlpert}) as weak $A_\infty$ 
constraints for the perturbed boundary amplitude $B_{a_0\dots a_m}(t):=\cF_{a_0\ldots a_m} (t)$:
\begin{equation}
  \label{eq:weakAinfty}
  \sum\limits_{\tiny\begin{array}{c}k,j=0\\ k\leq j\end{array}}^{m}
  (-1)^{\td a_1+ \ldots +\td a_k}~
  \cF^{a_0}{}_{a_1\ldots a_kca_{j+1}\ldots a_m}(t)~
  \cF^{c}{}_{a_{k+1}\ldots a_j}(t)~=~0~~.
\end{equation}
Expanding this as a power series in $t$ reproduces equations
(\ref{eq:genearlpert}). The first two deformed amplitudes:
$\cF_a$ and $\cF_{ab}$ are of order at least one in $t_i$, since
$B_a$ and $B_{ab}$ vanish.  The presence of these terms for $t\neq 0$ promotes
the minimal $A_\infty$  algebra of Section \ref{sec:Ainfty} to a weak $A_\infty$ algebra,
  and corresponds to the generation of non-vanishing tadpoles, as discussed
  in Subsection \ref{sec:deformed_amplitudes}.

\subsubsection{Algebraic formulation}

Extending the discussion of Section \ref{sec:Ainfty}, let us 
define {\em deformed open string scattering products}
$r_m^t:H_o^{\otimes m}\rightarrow H_o$ through the relations:
\begin{equation}
\label{r_t}
r_m^t(\psi_{a_1}\dots \psi_{a_m})={\cal F}_{a_1\dots
  a_m}^a(t)\psi_a~{\rm~for~}~m\geq 1~~.
\end{equation}
and:
\begin{equation}
r_o^t(1):={\cal F}^{a}(t)\psi_a~~,\nn
\end{equation}
where ${\cal F}^a(t)=\omega^{ab}{\cal F}_b(t)$ and we used the fact  
that the product $r_o:H_o^{\otimes
  0}\approx \C\rightarrow H_o$ is determined by its value at the complex unit
$1\in \C$~. As in Section \ref{sec:Ainfty}, equations (\ref{eq:weakAinfty})
become:
\begin{equation}
\label{weakAinf_std}
\sum\limits_{\tiny
    \begin{array}{c} k+l=m+1\\ 
      j=0\dots k-1
   \end{array}}^{m}
  \hspace*{-10pt}
  (-1)^{\td a_1 + \ldots + \td a_j}
  r^t_k(\psi_{a_1}\ldots \psi_{a_j}, r^t_l(\psi_{a_{j+1}}\ldots \psi_{a_{j+l}}),
  \psi_{a_{j+l+1}}\ldots \psi_{a_m})  = 0~~,
\end{equation}
which are the standard relations defining a weak $A_\infty$ algebra.

Remembering equation (\ref{F_cyc}), we find that this weak $A_\infty$ algebra is cyclic:
\begin{equation}
\label{weak_r_cyc}
\omega(\psi_{a_0}, r^t_m(\psi_{a_1}\dots \psi_{a_m}))=(-1)^{{\tilde a}_m({\tilde
    a}_0+\dots +{\tilde a}_{m-1})}\omega(\psi_{a_m}, r^t_m(\psi_{a_0}\dots
    \psi_{a_{m-1}}))~~{\rm for}~~m\geq 1~~.
\end{equation}
Moreover, equations (\ref{F_vanish}) and (\ref{F_constant}) show that $(H_o,r^t_*)$ 
is unital:
\begin{equation}
\label{weak_unit1}
r^t_m(\psi_{a_1}\dots \psi_{a_{i-1}}, 1_o, 
      \psi_{a_{i+1}}\dots \psi_{a_{m-1}})=0~~{\rm~for~}~~
      (m=1 {\rm~or~} m\geq 3)~~{\rm~and~all}~i=1\dots m-1~~
\end{equation}
and:
\begin{equation}
\label{weak_unit2}
r^t_2(1_o,\psi_a)=(-1)^{\tilde a}r^t_2(\psi_a,1_o)=\psi_a~~.
\end{equation}
To arrive at the last equation, we used 
relation (\ref{F_constant}) and non-degeneracy of $\omega$.

\subsubsection{Interpretation in terms of open string field theory}

The algebraic formulation given above allows us to 
give an alternate description of the effective superpotential, which makes contact
with open string field theory as formulated by Zwiebach (see \cite{Zwiebachoc}
and references therein). Let us consider the object:
\begin{equation}
\label{psi}
\psi:=\sum_{a}{s_a\psi_a}~~,
\end{equation}
viewed as an element of the graded vector space $H_o^e:=\A\otimes H_o$, which is 
naturally a super bi-module over the commutative superalgebra $\A=\pi({\hat
  \A})\subset {\cal B}$ (the notation here follows Subsection \ref{sec:generating_function}). 
When $H_o$
is endowed with the suspended degree $\deg$, the quantity $\psi$ has even
degree as an element of this module.  
Using definition (\ref{r_t}), we find the following expression for
the deformed boundary amplitudes:
\begin{equation}
{\cal F}_{a_0\dots a_m}(t)=\omega(\psi_{a_0}, r^t_m(\psi_{a_1}\dots \psi_{a_m}))~~.\nn
\end{equation}

We would like to express this in terms of $\psi$. For this, consider the 
natural extension of $\omega$ to $H_o^e$, which we shall denote by the same
letter. This is an $\A$-valued bilinear form on
$H_o^e$ given as follows on decomposable elements:
\begin{equation}
\label{omega_e}
\omega(f\otimes \psi_a, g\otimes \psi_b)=(-1)^{{\tilde a}{\tilde g}}fg\omega_{ab}~~,
\end{equation}
where $f,g$ are homogeneous elements of $\A$ of degrees ${\tilde f}$ and
${\tilde g}$. We also extend $r^t_m$ to multilinear products on $H_o^e$ (again
denoted by the same symbol) through:
\begin{equation}
\nn
r_m^t(f_1\psi_{a_1}\dots f_m\psi_{a_m})=(-1)^{\sum_{j=2}^{m}{({\tilde
      a}_1+\dots +{\tilde a}_{j-1}){\tilde f}_j}+{\tilde f}_1+\dots +{\tilde
    f}_m } f_1\dots f_mr_m^t(\psi_{a_1}\dots \psi_{a_m})~~.
\end{equation}
With these definitions, we have:
\begin{equation}
r_m^t(\psi\dots \psi)=s_{a_m}\dots s_{a_1}r_m^t(\psi_{a_1}\dots \psi_{a_m})~~\nn
\end{equation}
and:
\begin{equation}
s_{a_m}\dots s_{a_0}{\cal F}_{a_0\dots a_m}(t)=\omega(\psi,r_m(\psi\dots \psi))~~.\nn
\end{equation}
Thus equation (\ref{W}) becomes:
\begin{equation}
{\cal W}(s,t)=\sum_{m\geq 0}{\frac{1}{m+1}\omega(\psi, r^t_m(\psi^{\otimes m}))}~~.\nn
\end{equation}
This is the standard form \cite{Gaberdiel} of an open string field action, though
built around a background which need not satisfy the open string equations of
motion (as reflected by the presence of the product $r_0^t$). 
In this interpretation, the object $\psi$ is identified with the
string field. As expected, the parameters $t$ encode a deformation of this action, 
parameterized by a choice of the closed string background. The fact that the
effective superpotential of a topological open string theory can be
viewed as a string field action follows from the observation that the renormalization
group flow\footnote{The RG flow in open string field theory was studied from
  the algebraic point of view in \cite{Nakatsu,KajiuraA}. It corresponds to changing
  the parameter $l$ giving the length of external strips used in the
  construction of open string vertices in the non-polynomial formulation 
(see, for example, \cite{Zwiebachoc}). } 
in the "target space" (=string field) formulation
of such models is a semigroup of homotopy equivalences -- thus no information is lost
when passing from the microscopic to the long wavelength description. 

Fixing the closed string background (i.e. treating $t_i$ as fixed parameters),
the open string equations of motion take the form:
\begin{equation}
\label{MC}
(\partial_a {\cal {\hat W}})(s,t)=0~~{\rm for~all}~~ a~~
\Longleftrightarrow \sum_{m=0}^\infty{r^t_m(\psi^{\otimes m})}=0~~,
\end{equation}
where we assume that ${\cal B}$ is chosen such that the extended bilinear form 
(\ref{omega_e}) is non-degenerate.

This algebraic condition is known as the Maurer-Cartan equation for a weak
$A_\infty$ algebra. The presence of $r_0^t$ signals the fact
that the reference point $s=0$ does not satisfy this equation. 
Indeed, the left hand side of (\ref{MC}) at $s=0$ equals
$r_0(\psi^{\otimes 0}):=r_0(1)$.

\subsubsection{Canceling the tadpole}
\label{subsubsec:tadpole}
As mentioned above, deformations of the closed string background will
generally produce a tadpole which must be canceled if the deformed theory
is to have a chance of being conformal. In this subsection, we explain how
this can be achieved by shifting the open string vacuum, thereby making contact with
previous mathematical work.

Consider a shift:
\begin{equation}
\label{psi_shift}
s_a\rightarrow s_a+\sigma_a~~,
\end{equation}
with $\sigma_a\in \A$. 
In terms of the string field (\ref{psi}), this operation amounts to:
\begin{equation}
\psi\rightarrow \psi+\alpha~~,\nn
\end{equation}
where $\alpha:=\sum_{a}{\sigma_a\psi_a}$ is an even element of $H_o^e$~~.

It is not hard to check that under such
a transformation the deformed scattering products 
change as:
\begin{equation}
r_m^t\rightarrow r_m^{t;\sigma}\nn
\end{equation}
where:
\begin{equation}
\label{r_a}
r_m^{t;\sigma}(u_1\dots u_m)=r^t(e^\alpha, u_1, e^\alpha,
u_2,\dots ,e^\alpha, u_m, e^\alpha)~~
\end{equation}
for all $u_1\dots u_m\in H_o^e$. 

In equation (\ref{r_a}), we used the notations:
\begin{equation}
e^\alpha:=\sum_{k=0}^\infty{\alpha^{\otimes k}}\nn
\end{equation}
and:
\begin{equation}
r^t:=\sum_{m=0}^{\infty}{r_m^t}~~.\nn
\end{equation}
Notice that $r^t$ is a map from $\oplus_{m=0}^\infty{(H_o^e)^{\otimes m}}$ to
$H_o^e$. 

In particular, the product $r_0^t$ becomes:
\begin{equation}
r_0^{t;\sigma}=r^t(e^\alpha)=\sum_{m=0}^\infty{r_m^t(\alpha\dots \alpha)}~~.\nn
\end{equation}
Hence the tadpole amplitude $B_a(t)={\cal F}_a(t)=\omega( \psi_a, r_0^t(1))$ 
vanishes if and only if:
\begin{equation}
\label{MCa}
\sum_{m=0}^\infty{r_m^t(\alpha\dots \alpha)}=0\Longleftrightarrow
(\partial_a {\cal {\hat W}})(\sigma,t)=0~~.
\end{equation}
This amounts to the well-known fact that the equations of motion for (open) string
theory amount to the tadpole cancellation condition. 
It is not hard to check by direct computation 
that the products $r_m^{t;\sigma}$
with $m\geq 1$ form a {\em strong} $A_\infty$ algebra provided that this
equation is satisfied.  Hence the Maurer-Cartan condition (\ref{MCa}) 
describes possible transformations of a weak $A_\infty$
algebra into a (strong) $A_\infty$ algebra obtained by shifts of the form (\ref{psi_shift}).

Given a solution $\sigma$ of (\ref{MCa}), the expansion of ${\cal W}$ around
the new open string vacuum takes the form:
\begin{equation}
{\cal W}(s,t)=\sum_{m\geq 2}{\frac{1}{m+1}\omega(\psi,
  r_m^{t,\sigma}(\psi^{\otimes m}))}+{\cal W}(\sigma,t)~~.\nn
\end{equation}
Up to the last term (which is $s$-independent), this is the standard form of
the open string field action in the formulation of \cite{Gaberdiel}. 

We mention that condition (\ref{MCa}) plays a crucial role in the work of 
\cite{Fukayainfty,Fukayabook, FukayamirrorA, FukayamirrorB}, where it originates in a
  very similar manner (see \cite{CILf} for a detailed discussion).

\subsubsection{Relation to the deformation theory of cyclic
  $A_\infty$ algebras}

The results deduced in this subsection are intimately related to the deformation theory 
of cyclic $A_\infty$ algebras as developed in \cite{Penkava}. This
interpretation is quite obvious, so we can be brief. 

It is clear from the discussion above that insertion of bulk operators
realizes an all-order deformation of the $A_\infty$ algebra of Section 
\ref{sec:Ainfty}, viewed as a weak $A_\infty$ algebra which happens to be 
strong and minimal for the particular value $t=0$ of the deformation
parameters. Moreover, such deformations preserve cyclicity
and unitality. 

To make contact with the work of \cite{Penkava}, let us consider the case of
infinitesimal deformations discussed in Subsection \ref{sec:lindef}. This can
be recovered from the more general results of the previous subsection by
expanding the products $r_m^t$ to first oder in $t$. Writing:
\begin{equation}
r_m^t=r_m+t_i\Phi^i_m+{\cal O}(t^2)~~,\nn
\end{equation}
we extract morphisms:
\begin{equation}
\label{Phi_i}
\Phi^i_m=\frac{\partial r_m^t}{\partial t_i}\Big{|}_{t=0}:H_o^{\otimes m}\rightarrow H_o~~.
\end{equation}
The objects $\Phi^i:=\sum_{m=0}^\infty{\Phi_m^i}$ belong to the so-called (weak) Hochschild complex
$C=\oplus_{m=0}^\infty{C^m(H_o)}$ of $H_o$, whose graded subspaces are defined through:
\begin{equation}
C^m(H_o):=Hom(H_o^{\otimes m}, H_o)~~\nn
\end{equation}
and whose differential is given by the first order variation of the weak 
$A_\infty$ constraints (\ref{weakAinf_std}):
\begin{equation}
\label{delta}
(\delta \Phi^i)_m(\psi_{a_1}\dots \psi_{a_m})=(\partial_i A_m)\Big{|}_{t=0}~~,
\end{equation}
where $A_m(t)$ is the left hand side of (\ref{weakAinf_std}):
\begin{equation}
A_m(t):=\sum_{\tiny
    \begin{array}{c} k+l=m+1\\ 
      j=0\dots k-1
   \end{array}}^m{
  (-1)^{\td a_1 + \ldots + \td a_j}
  r^t_k(\psi_{a_1}\ldots \psi_{a_j}, r^t_l(\psi_{a_{j+1}}\ldots \psi_{a_{j+l}}),
  \psi_{a_{j+l+1}}\ldots \psi_{a_m})}~~.\nn
\end{equation}
In equation (\ref{delta}), it is understood that we replace 
$\frac{\partial r_m^t}{\partial t_i}\Big{|}_{t=0}$ by $\Phi_m^i$ through
  relation (\ref{Phi_i}) and view the result $\delta\Phi^i$ as the action of
  an algebraic operator\footnote{Strictly speaking, this specifies $\delta$ only
    for elements $\Phi$ of $C(H_o)$ such that each $\Phi_m$ has degree one as
    a map from $H_o^{\otimes m}$ to $H_o$. However the definition generalizes
    to the entire Hochschild complex.} $\delta$ on $\Phi^i$. The fact that $\delta$ squares
  to zero follows from the $A_\infty$ constraints. 

Because our $A_\infty$ algebras are cyclic, one has further restrictions on
$\Phi^i$ which amount to the statement that they are elements of a certain
subcomplex $CC(H_o)$ known as the {\em cyclic complex}. This can be defined as 
the set of elements $\Phi=\sum_{m}{\Phi_m}$ in $C(H_o)$ with the property that the
quantities $\omega (\psi_{a_0}, \Phi_m(\psi_{a_1}\dots \psi_{a_m})) $ obey the
cyclicity constraints: 
\begin{equation}
\omega(\psi_{a_0}, \Phi_m(\psi_{a_1}\dots \psi_{a_m}))=(-1)^{{\tilde a}_m({\tilde
    a}_0+\dots +{\tilde a}_{m-1})}\omega(\psi_{a_m}, \Phi_m(\psi_{a_1}\dots \psi_{a_{m-1}}))~~.\nn
\end{equation}
For our maps $\Phi^i$, these conditions follow by differentiating
(\ref{weak_r_cyc}) with respect to $t_i$ at $t=0$. The Hochschild differential
$\delta$ preserves the subspace $CC(H_o)$. Denoting its restriction by
the same letter, one obtains the cyclic complex $(CC(H_o),\delta)$ considered
\footnote{Our sign conventions differ from those of \cite{Penkava} by
  suspension. Moreover, we allow for the subspace $C^0(H_0)=CC(H_0)=\C$ in the
  Hochschild and cyclic complexes, since we consider deformations of {\em
    weak} and cyclic $A_\infty$ algebras.} in \cite{Penkava}.

Since the deformed products (\ref{r_t}) obey weak $A_\infty$ constraints for
all $t$, differentiation of (\ref{weakAinf_std}) at $t=0$ shows that $\Phi^i$
are $\delta$-closed:
\begin{equation}
\delta \Phi^i=0~~.\nn
\end{equation}
Thus $\Phi^i$ define elements $[\Phi^i]$ of the cohomology of $(CC(H_o), \delta)$, known
as the (weak) cyclic cohomology of the $A_\infty$ algebra $(H_o,
r_*)$. Comparing with Subsection \ref{sec:lindef}, it is easy to see that
$\Phi^i$ can be written as:
\begin{equation}
\Phi^i_m(\psi_{a_1}\dots \psi_{a_m})=B^a_{a_1\dots a_m;i}\psi_a~~.\nn
\end{equation}
This shows that they are completely determined by the disk amplitudes 
$B_{a_0\dots a_m;i}$ with a single bulk insertion. Thus one has a map:
\begin{equation}
\phi_i\rightarrow [\Phi^i] \nn
\end{equation}
from BRST-closed bulk zero-form observables to the cyclic
cohomology of the $A_\infty$ algebra $(H_o, r_*)$. A similar statement was
proposed in \cite{HofmanMaA} in a particular case.


\subsection{Bulk-boundary crossing symmetry}
\label{sec:bb_crossing}

The second bulk-boundary crossing constraint (\ref{bb2}) of two-dimensional 
topological field theory states that the bulk-boundary map is a morphism from the 
bulk to the boundary algebra~\cite{CILtop}. In this
section, we discuss the `stringy' generalization of this constraint.

It is clear that the relevant property arises from factorization of the
amplitude:
\begin{equation}
  \label{eq:cupcorr}
  \bra\phi_i \phi_j ~\psi_{a_0} P \int \psi_{a_1}^{(1)}
  \ldots \int \psi_{a_m}^{(1)} \ket \;,
\end{equation}
into the channel where the two bulk fields approach each other and
the channel where both bulk fields approach the boundary.
In contrast to the $A_\infty$ constraints, this factorization follows from
explicit movement of the bulk operators rather than from the Ward identities
  of the BRST symmetry.
This is similar to the mechanism leading to the WDVV equations
(\ref{eq:WDVV}). In the 
case at hand we have to deal with a subtlety which requires closer
examination: we know from Section~\ref{sec:corrfunc} that only
the fundamental amplitudes (\ref{B}) are independent of the positions of
  unintegrated insertions. This is not the case for the amplitude (\ref{eq:cupcorr}),
  since it contains two bulk and one boundary unintegrated insertions. 
  Therefore, it is not immediately clear that
  factorizing (\ref{eq:cupcorr}) makes sense. 
The naive guess for the factorization takes the form:
\begin{eqnarray}
  \label{eq:factbubo}
  & & 
  {C_{ij}}^l\,
  \bra\phi_l~\psi_{a_0} P \!\int\!\! \psi_{a_1}^{(1)}
  \ldots \!\!\int\!\! \psi_{a_m}^{(1)} \ket_\ws = \nonumber\\
  &=& \hspace*{-5mm}\sum\limits_{0\leq m_1\leq \ldots m_4\leq m}
  \hspace*{-5mm}
  \bra \psi_{a_0} P \!\int\!\! \psi_{a_1}^{(1)}
  \ldots \!\!\int\!\! \psi_{a_{m_1}}^{(1)}~\psi_{b}~ P \!\int\!\! \psi_{a_{m_2+1}}^{(1)}
  \ldots \!\!\int\!\! \psi_{a_{m_3}}^{(1)}~\psi_{c}~ P \!\int\!\! \psi_{a_{m_4+1}}^{(1)}
  \ldots \!\!\int\!\! \psi_{a_m}^{(1)}\ket_\ws ~~~~\\
  & & \omega^{bd}~\omega^{ce}~
  \bra\phi_i~\psi_{d} P \!\int\!\! \psi_{a_{m_1+1}}^{(1)}
  \ldots \!\!\int\!\! \psi_{a_{m_2}}^{(1)} \ket_\ws~
  \bra\phi_j~\psi_{e} P \!\int\!\! \psi_{a_{m_3+1}}^{(1)}
  \ldots \!\!\int\!\! \psi_{a_{m_4}}^{(1)} \ket_\ws ~\nn~.
\end{eqnarray}
Since the correlation function (\ref{eq:cupcorr}) is not
independent of the positions of the fixed insertions, we shall give an
independent argument for why this relation holds.
In the following, we shall denote the left hand side of
(\ref{eq:factbubo}) simply by $\lhs$, and the right hand side by
$\rhs$. 

\begin{figure}[t]
  \begin{center}
     \epsfysize=7cm\centerline{\epsffile{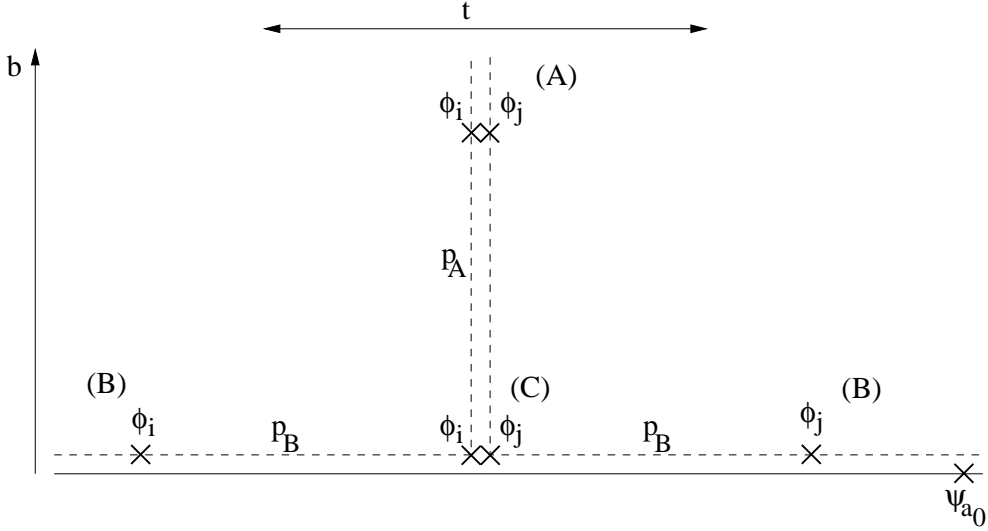}}
    \parbox{12truecm}{
      \caption{
        The factorization associated with the stringy version of the second
        bulk-boundary crossing constraint. Configuration
        $(A)$ corresponds to the topological bulk product and $(B)$ to
        the factorization at the boundary. Configuration $(C)$ connects these
        channels. The quantity $b$ is the equal distance of the bulk fields from
        the boundary, while $t$ is the horizontal separation of the bulk
        fields.
        \label{fig:cupfactorization}}
    }
  \end{center}
\end{figure}

To establish equation (\ref{eq:factbubo}),  we consider the amplitude
(\ref{eq:cupcorr}) for the configurations $(A)$ and $(B)$ of bulk
operators on the upper half plane shown in
Figure \ref{fig:cupfactorization}. Let us denote the distance between
the bulk operators by $t \in \bbR$ and assume that the two bulk operators 
sit on a line parallel at a distance $b$ to the boundary. In the limit $t
\rightarrow 0$, configuration $(A)$ corresponds to the left hand side of 
equation (\ref{eq:cupcorr}), while the right hand side of this equation 
arises in the limit $b \rightarrow 0$ of configuration $(B)$.

For configuration $(A)$,  we have $t=t_0$ with
$|t_0| \ll 1$ and we can perform a bulk operator product expansion in
$t_0$, so that (\ref{eq:cupcorr}) becomes $\lhs + g_1(t_0,b)$, where
$g_1(t_0,b) = \cO(t_0)$ denotes contributions from higher terms in the OPE.
Moving along the path $p_A$ down toward configuration $(C)$, the
function $g_1(t_0,b)$ changes with $b$ and becomes $g_1(t_0,b_0)$.
Configuration $(B)$ shows the bulk operators at the distance
$b=b_0 \ll 1$ from the boundary. According to the bulk-boundary
operator product the amplitude (\ref{eq:cupcorr})
takes the form $\rhs + g_2(t,b_0)$, where 
$g_2(t,b_0) = \cO(b_0)$. Following the path $p_B$ we 
reach again the point $(C)$.
At $(C)$ we have $\lhs + g_1(t_0,b_0) = \rhs + g_2(t_0,b_0)$, which implies
that $g_1(t_0,b_0)$ and $g_2(t_0,b_0)$ are non-singular for $b_0\rightarrow 0$
and $t_0\rightarrow 0$, respectively. 
Hence we can safely take the factorization limit
$t_0,b_0\rightarrow 0$, in which $g_1$ and $g_2$ vanish, so that
$\lhs = \rhs$. This shows that equation (\ref{eq:factbubo}) holds.

Using the Ward identity (\ref{eq:cyclic1}) to move
the integral contours, and taking into account definition
(\ref{B}), equation (\ref{eq:factbubo}) gives:
\begin{equation}
  \label{eq:cupprod}
  B^{a_0}{}_{a_1 \ldots a_m;l} C^l{}_{ij} =
  \hspace*{-5mm}\sum\limits_{0\leq m_1\leq \ldots m_4\leq m}
 \hspace*{-5mm}(-1)^{\tilde a_{m_1+1}+\ldots +\tilde a_{m_3}}
  B^{a_0}{}_{a_1\ldots a_{m_1}ba_{m_2+1}\ldots a_{m_3}ca_{m_4+1}\ldots a_m}
  B^b{}_{a_{m_1+1}\ldots a_{m_2};i}B^c{}_{a_{m_3+1}\ldots a_{m_4};j}~~,
\end{equation}
where $C^l{}_{ij}$ are the usual bulk ring structure constants. Note
  that the left-hand side is manifestly symmetric in $i$ and $j$ whereas this
  symmetry is not manifest in the
  right-hand side. This reflects the fact that one can also
  accomplish the factorization of Figure~\ref{fig:cupfactorization}
  after exchanging $i$ and $j$.

Additional integrated bulk insertions spread in the usual way when we
factorize, so we can treat them as derivatives and combine all
relations into a single equation involving the quantities
$\cF_{a_0\ldots a_m} (t)$ for $m\geq 0$ and the bulk WDVV potential $\cF(t)$:
\begin{eqnarray}
  \label{eq:boundWDVV}
  &&\6_i \6_j \6_k \cF(t) ~\eta^{kl}~ \6_l \cF_{a_0 a_1 \ldots a_m}(t) =
  \\[5pt]
  &=&\hspace*{-7mm}\sum\limits_{0\leq m_1\leq \ldots m_4\leq m}
  \hspace*{-5mm}(-1)^{\tilde a_{m_1+1}+\ldots +\tilde a_{m_3}}
  \cF_{a_0\ldots a_{m_1}ba_{m_2+1}\ldots a_{m_3}ca_{m_4+1}\ldots a_m}(t)
  ~\6_i \cF^b{}_{a_{m_1+1}\ldots a_{m_2}}(t)
  ~\6_j \cF^c{}_{a_{m_3+1}\ldots a_{m_4}}(t)~~.
   \nonumber
\end{eqnarray}
For  $m=0$ and $m=1$, these equations take the form:
\begin{eqnarray}
  \label{eq:casem0}
  \6_i \6_j \6_k \cF~ \eta^{kl}~ \6_l \cF_{a_0} &=&
  \cF_{a_0bc}~ \6_i\cF^b~ \6_j\cF^c~\;,
\\[10pt]
  \label{eq:casem1}
  \6_i \6_j \6_k \cF~ \eta^{kl}~ \6_l \cF_{a_0a_1} &=&
  \cF_{a_0bca_1}~ \6_i\cF^b~ \6_j\cF^c +
  \cF_{a_0bc}~ \6_i\cF^b~ \6_j\cF^c{}_{a_1} 
\nonumber\\[5pt] &+&
  (-1)^{\td a_1}~ \cF_{a_0ba_1c}~ \6_i\cF^b~ \6_j\cF^c +
  (-1)^{\td a_1}~ \cF_{a_0bc}~\6_i\cF^b{}_{a_1}~ \6_j\cF^c 
\nonumber\\[5pt] &+&
  \cF_{a_0a_1bc~} \6_i\cF^b~ \6_j\cF^c \;.  
\end{eqnarray}
The undeformed version of (\ref{eq:casem0}) coincides with the second 
bulk-boundary crossing constraint (\ref{bb2_B}) of two-dimensional TFT.


\subsection{Cardy conditions}
\label{sec:Cardy}

The Cardy condition is probably the most interesting sewing constraint
of 2d TFT \cite{MooreSegal,Moore,CILtop}, since it connects the exchange of closed strings
between D-branes at the tree level with a one-loop open string amplitude. 
Allowing for insertions of both bulk and boundary fields in the
corresponding cylinder amplitude leads to the following factorization:
\begin{eqnarray}
  \label{eq:Cardycorr}
  & & \bra\phi_i\psi_{a_0} P \int\psi_{a_1}\ldots\int\psi_{a_n} \ket 
      ~\eta^{ij}~
      \bra\phi_j\psi_{b_0} P \int\psi_{b_1}\ldots\int\psi_{b_m} \ket = \\
  &=&\hspace*{-5mm}
  \sum\limits_{\tiny \begin{array}{c}0\leq n_1 \leq n_2 \leq n \\
      0\leq m_1 \leq m_2 \leq m \end{array}}\hspace*{-5mm}  
  (-1)^{s}~\omega^{c_1d_1}\omega^{c_2d_2}~
  \nonumber\\
  &&\bra \psi_{a_0}~ P \int\psi_{a_1}\ldots\int\psi_{a_{n_1}}
       \psi_{d_1} P \int\psi_{b_{m_1+1}}\ldots\int\psi_{b_{m_2}}
       \psi_{c_2} P \int\psi_{a_{n_2+1}}\ldots\int\psi_{a_n} \ket 
  \nonumber\\
  &&\bra \psi_{b_0}~ P \int\psi_{b_1}\ldots\int\psi_{b_{m_1}}
       \psi_{c_1} P \int\psi_{a_{n_1+1}}\ldots\int\psi_{a_{n_2}}
       \psi_{d_2} P \int\psi_{b_{m_2+1}}\ldots\int\psi_{b_m} \ket 
  ~~,\nonumber
\end{eqnarray}
where the sign $s$ accounts for reshuffling of the boundary fields. The
left hand side of (\ref{eq:Cardycorr}) is the factorization in
the closed string channel, in which the cylinder becomes infinitely long. The
right hand side corresponds to the generalization of the double-twist
diagram \cite{Moore} of the open string channel.

Taking into account further integrated bulk insertions, equations (\ref{eq:Cardycorr})
become:
\begin{eqnarray}
  \label{eq:Cardy}
  &&\6_i \cF_{a_0 \ldots a_n} \eta^{ij}
  \6_j \cF_{b_0 \ldots b_m} = \\[5pt]
  &=&\hspace*{-9mm}
  \sum\limits_{\tiny \begin{array}{c}0\leq n_1 \leq n_2 \leq n \\
      0\leq m_1 \leq m_2 \leq m \end{array}}\hspace*{-7mm}
  (-1)^{s+\td c_1 +\td c_2}~
  \omega^{c_1d_1}~ \omega^{c_2d_2}~
  \cF_{a_0 \ldots a_{n_1} d_1 b_{m_1\!\!+\!1} \ldots b_{m_2} c_2 a_{n_2\!+\!1}
  \ldots a_{n}}~
  \cF_{b_0 \ldots b_{m_1} c_1 a_{n_1\!\!+\!1} \ldots a_{n_2} d_2 b_{m_2\!+\!1}
  \ldots b_{m}}
  \;.\nonumber
\end{eqnarray}
The first relations in this hierarchy of constraints take the form:
\begin{eqnarray}
  \label{eq:firstCardy}
  \6_i \cF_{a_0} \eta^{ij} \6_j \cF_{b_0} &=&
  (-1)^{s+\td c_1 +\td c_2}~
  \omega^{c_1d_1}~ \omega^{c_2d_2}~
  \cF_{a_0 d_1 c_2}~\cF_{b_0 c_1 d_2}\;,
  \nonumber \\[5pt]
  \6_i \cF_{a_0 a_1} \eta^{ij} \6_j \cF_{b_0} &=&
  (-1)^{s+\td c_1 +\td c_2}~
  \omega^{c_1d_1}~ \omega^{c_2d_2}~
  \cF_{a_0 d_1 c_2 a_1}~\cF_{b_0 c_1 d_2} 
  \\ &+&
  (-1)^{s+\td c_1 +\td c_2}~
  \omega^{c_1d_1}~ \omega^{c_2d_2}~
  \cF_{a_0 d_1 c_2}~\cF_{b_0 c_1 a_1 d_2} 
 \nonumber\\ &+&
  (-1)^{s+\td c_1 +\td c_2}~
  \omega^{c_1d_1}~ \omega^{c_2d_2}~
  \cF_{a_0 a_1 d_1 c_2}~\cF_{b_0 c_1 d_2}~~.
  \nonumber
\end{eqnarray}
Taking the limit $t=0$ in the first equation recovers the Cardy
constraint (\ref{cardy_B}) of two-dimensional TFT. 
Notice that the left hand side of the first equation 
vanishes identically if we consider insertions of the identity operator, and
if the suspended degree ${\tilde \omega}$ of the symplectic
structure vanishes; this reflects vanishing of the Witten index in that case.

It is worth pointing out that the arguments of Section 
\ref{sec:corrfunc} cannot be used to show that the annulus amplitude is independent 
on the world-sheet metric and of the positions of unintegrated boundary insertions. 
In fact, experience with the bulk theory \cite{BCOV} suggests 
that there are BRST anomalies in open string correlators beyond tree level, so there is
indeed no {\it a priori} reason why the annulus amplitude should be 
metric-independent.  However, we will take the point of view that when {\it imposing}
the Cardy condition (\ref{eq:Cardy}), one focuses by definition on the
topological part of the amplitude. It is not clear to us whether the Cardy
relation is satisfied by the complete amplitude, which potentially involves
supplementary anomalous contributions.

In the present paper we will be concerned only with the topological part of
the annulus amplitude.\footnote{For the Landau-Ginzburg examples 
described in Section \ref{sec:superpot}, we shall find that imposing the
generalized Cardy condition as written above agrees with independently known 
results (namely, with the factorization property of the Landau-Ginzburg
potential).} To capture the full amplitude including possible
holomorphic anomalies would require the analog of $t-t^*$
equations \cite{BCOV} for open strings, a subject which is beyond the scope of the
present paper.

\section{Application: Superpotentials for D-branes in topological minimal models}
\label{sec:TLG}

In this section, we demonstrate the power of the consistency conditions
derived in this paper (namely cyclicity (\ref{eq:cyclic})), weak
$A_\infty$ structure (\ref{eq:weakAinfty}), bulk-boundary 
sewing (\ref{eq:boundWDVV}) and Cardy relations (\ref{eq:Cardy})) by
applying them to certain families of D-branes in B-type 
topological minimal models. In the examples considered below, we shall find 
that the totality of these constraints suffices to determine the effective superpotential. 

Let us recall some facts \cite{KapA,BHLS,KapB,KapC} 
about D-branes in B-type topological minimal models. As usual, the bulk sector
is  characterized by the level $k$, while D-brane
boundary sectors are labeled by $\bl=0,1,...[k/2]$.
It is convenient to switch to the Landau-Ginzburg realization of
these models. Then the bulk sector is described by a univariate polynomial 
$W_{LG}^{(k+2)}(\vp)$ of degree $k+2$ in the complex variable $\vp$, which
gives the worldsheet superpotential.
On the other hand, `non-multiple' B-type D-branes in the boundary sector $\bl$ correspond to
factorizations of the bulk superpotential: 
\begin{equation}
\label{eq:Wfact}
W_{LG}^{(k+2)}(\vp)\ =\ J\ssc{\bl+1}(\vp)\,E\ssc{k+1-\bl}(\vp)\ ,
\qquad \bl=0,1,...[k/2]~~
\end{equation}
where  $J\ssc{\bl+1}(\vp)$ is a polynomial of degree $\bl+1$ \cite{KapA,BHLS}.

The open string spectrum consists of boundary
changing and boundary preserving sectors. We focus first on
boundary preserving sectors, each of which corresponds to
a degree label $\bl$. The on-shell boundary algebra $H_o$ is isomorphic 
with a supercommutative ring $\cR_\dl$ with even and odd generators $\vp$
and $\omega$, subject to relations which can be described as follows. 
Let $H$ denote the greatest common denominator of $J$ and $E$, ie.,
$J\ssc{\bl+1}(\vp)=p(\vp)H\ssc{\ell+1}(\vp)$ and
$E\ssc{k+1-\bl}(\vp)=q(\vp)H\ssc{\ell+1}(\vp)$ for some polynomials
$p,q$.\footnote{In the unperturbed theory, about which we will
expand, we can take both $p$ and $q$ to be constant, and we 
normalize them by setting $pq=-\frac{\varphi^{k-2\kappa}}{k+2}$.}
Then the relations in the boundary ring are \cite{BHLS}:
\begin{equation}
\label{eq:boundideal}
{{\mathcal I}}:\ \
\big\{H\ssc{\ell+1}(\vp)\,=\,0,\ \omega^2\,=\,p(\vp) q(\vp)\big\}~~.
\end{equation}
The $U(1)$ charges of the generators are given by:
$$
q(\phi)\ =\ 1,\qquad q(\omega)\ =\ k/2-\ell~~.
$$
When viewed as a complex vector space, the boundary algebra $H_o$ admits the basis:
\begin{equation}
\label{eq:ringbasis}
\cR_\dl\ \equiv\ \big\{\,\psi_a\,\big\}\ =\ \big\{\vp^\al,\omega\vp^\al\big\},
\qquad a=0,...,2\ell+1,\ \ \al=0,...,\ell\ .
\end{equation}

Recall that the bulk algebra is given by the Newton ring
${\cal R}=\C[\phi]/\langle\6_\vp W_{LG}^{(k+2)}(\vp)\rangle$, which admits the
following basis when viewed as a complex vector space:
$$
\cR\ \equiv\ \big\{\,\phi_i\,\big\}\ =\ \big\{\,\vp^i\,\big\},
\qquad i=0,...,k .
$$

When suitably integrated, each of the fields can be used to deform
the theory. In the bulk sector we have the deformation $\delta S=\sum_{i=0}^k
t_{k+2-i}\int d^2z [G_{-1},[{\bar G}_{-1},\phi_i]$, while in the boundary sector
we have:
\begin{equation}
\label{deltaS_boundary}
\delta S_\dl\ =\ 
\left(\sum_{\al=0}^\ell u_{\ell+1-\al}\int dx\, G(\omega\vp^\al)\right)
+
\left(\sum_{\al=0}^\ell v_{k/2+1-\al}\int dx\, G\vp^\al\right)~~.
\end{equation}
In this equation, we divided the boundary deformation parameters $s_a$ into even and 
odd variables $u_\al$ and $v_\al$. These parameters can be formally assigned $U(1)$
charges, which can be used as labels; this is the convention
employed in equation (\ref{deltaS_boundary}). 
Notice that super-integration over the moduli of boundary punctures flips the
$\Z_2$ degree, so that odd ring elements (=topological tachyon excitations $\omega\phi^\alpha$) are
associated with the bosonic deformation parameters $u$, and vice versa.

We are now ready to present some computations. 
We first consider a few explicit examples and determine their effective
superpotentials. As a rule,  we shall find that the generalized WDVV equations lead to a
unique solution, but only once {\em all} constraints
are imposed on the open-closed amplitudes. For example, fixing $t=0$ 
and imposing only the $A_\infty$ conditions 
(\ref {eq:Ainfty}) leaves some parameters undetermined in the effective
superpotential ${\cal W}(0,s)={\cal W}(0,u,v)$. It is only after considering
both open and closed deformations and imposing the 
bulk-boundary constraints (\ref {eq:boundWDVV}) and Cardy conditions 
(\ref{eq:Cardy}) that all coefficients of ${\cal W}(s,t)$ become uniquely
determined\footnote{Strictly speaking, this is true only up to choosing the
normalization of the 3-point boundary and 2-point bulk-boundary correlators. 
In the computations below, we  normalized these correlators in a manner 
which is natural in the LG description. Notice that
the sign in the Cardy condition (\ref{eq:Cardy}) 
is given by: $(-1)^s=(-1)^{(\tilde c_1+\tilde a_0)(\tilde c_2+\tilde b_0)}$ in
the present case.}.

For the example $(k,\ell)=(3,1)$, we find the following expressions
for the perturbed boundary correlators (\ref{eq:formalpower}) on the disk:
\begin{eqnarray}
\label{eq:Fk3l1}
\BB{021} = -\BB{003} = -\BB{012} = -\BB{1213} &=& 1,
 \nonumber \\
  \BB{222} =
  \BB{2233} = 
  \BB{2323} = 
  \BB{23333} = 
  \BB{333333} &=&\!\! - 1/5  ,
  \nonumber \\  
  \BB{22} = \BB{233} = \BB{3333} &=& {t_2},
   \\   
  \BB{23} = \BB{333} &=& {t_3},
   \nonumber \\     
  \BB{2} = \BB{33} &=& t_4 -{{t_2}}^2,
   \nonumber \\   
  \BB{3} &=& {t_5} -{t_2}\,{t_3}. \nonumber
\end{eqnarray}
Our notation was explained after equations (\ref{eq:ringbasis}), namely $a=0,1$
and $a=2,3$  label even and odd boundary ring elements respectively.  
Moreover, we listed one representative per cyclic
orbit, and only the non-vanishing amplitudes. The value of $-1/5$ for
the correlators which contain three unintegrated fermionic insertions arises
from our normalization, which is $\omega^2=-\frac{\varphi}{5}$. Notice that ordering
of boundary indices is indeed important; for example $\BB{1123}=0$ while
$\BB{1213}=-1$.  

In this example, the effective superpotential takes the form:
\begin{eqnarray}
\label{eq:W31}
-\cW(t,u) &=&     \frac15\Big({\frac{{{u_1}}^6}{6} + {{u_1}}^4{u_2} 
      + \frac{3}2{{u_1}}^2{{u_2}}^2 + \frac{{{u_2}}^3}{3}}\Big)
       + 
  {t_2}\Big( \frac{-{{u_1}}^4}{4} - {{u_1}}^2{u_2} - \frac{{{u_2}}^2}{2} \Big)
  - 
  {t_3}\Big( \frac{{{u_1}}^3}{3} + {u_1}{u_2} \Big)  
        \nonumber\\ &&\
 + 
  \Big(  {t_4} -{{t_2}}^2 \Big) \Big( \frac{-{{u_1}}^2}{2} - {u_2} \Big) 
-\Big(    {t_5}- {t_2}{t_3} \Big) {u_1} \ .
\end{eqnarray}
Since the parameters $v$ are odd while appearing only in anti-commutators,
they drop out from the effective superpotential, even though the corresponding
non-symmetrized amplitudes are non-zero and have to be taken into account when
solving the constraint equations. 

The effective superpotentials for some other examples are as follows.  For
$(k,\ell)=(4,2)$ we find:
\begin{eqnarray}
\label{eq:W42}
-\cW(t,u)\ &=& \
  \frac16\Big({\frac{{{u_1}}^7}{7} + {{u_1}}^5{u_2} + 2{{u_1}}^3{{u_2}}^2 + 
     {u_1}{{u_2}}^3 + {{u_1}}^4{u_3} + 3{{u_1}}^2{u_2}{u_3} + 
     {{u_2}}^2{u_3} + {u_1}{{u_3}}^2}\Big)
            \nonumber\\ &&\!
     - {t_2}
   \Big( \frac{{{u_1}}^5}{5} + {{u_1}}^3{u_2} + {u_1}{{u_2}}^2 + 
     {{u_1}}^2{u_3} + {u_2}{u_3} \Big)   
   + 
  {t_3}\Big( \frac{-{{u_1}}^4}{4} - {{u_1}}^2{u_2} - \frac{{{u_2}}^2}{2} - 
     {u_1}{u_3} \Big) 
                \nonumber\\ &&\!
      + 
  \Big( \frac{  {t_4} -3{{t_2}}^2}{2}\Big) 
   \Big( \frac{-{{u_1}}^3}{3} - {u_1}{u_2} - {u_3} \Big) 
    + \Big(   {t_5}-2{t_2}{t_3} \Big) 
   \Big( \frac{-{{u_1}}^2}{2} - {u_2} \Big)  
      \nonumber\\ &&\!
        -\Big(  {t_6}+ \frac{{{t_2}}^3}{3} - \frac{{{t_3}}^2}{2} - {t_2}{t_4} 
       \Big) {u_1}  \ ,
\nonumber 
\end{eqnarray}
while for $(k,\ell)=(5,2)$ we obtain:
\begin{eqnarray}
\label{eq:W52}
&&-\cW(t,u)=
 \frac17\Big({\frac{{{u_1}}^8}{8} \!+\! {{u_1}}^6{u_2} \!+\! \frac{5{{u_1}}^4{{u_2}}^2}{2} \!+\! 
     2{{u_1}}^2{{u_2}}^3 \!+\! \frac{{{u_2}}^4}{4} \!+\! {{u_1}}^5{u_3} \!+\! 
     4{{u_1}}^3{u_2}{u_3} \!+\! 3{u_1}{{u_2}}^2{u_3} \!+\! 
     \frac{3{{u_1}}^2{{u_3}}^2}{2} \!+\! {u_2}
     {{u_3}}^2}\Big)
       \nonumber\\ &&\qquad\qquad
      - 
  {t_2}\Big( \frac{{{u_1}}^6}{6} + {{u_1}}^4{u_2} + 
     \frac{3{{u_1}}^2{{u_2}}^2}{2} + \frac{{{u_2}}^3}{3} + {{u_1}}^3{u_3} + 
     2{u_1}{u_2}{u_3} + \frac{{{u_3}}^2}{2} \Big)  
         \nonumber\\ &&\qquad\qquad
       + {t_3}
   \Big( \frac{-{{u_1}}^5}{5} - {{u_1}}^3{u_2} - {u_1}{{u_2}}^2 - 
     {{u_1}}^2{u_3} - {u_2}{u_3} \Big)    
        + 
  \Big(  {t_4} -2{{t_2}}^2 \Big) 
   \Big( \frac{-{{u_1}}^4}{4} - {{u_1}}^2{u_2} - \frac{{{u_2}}^2}{2} - 
     {u_1}{u_3} \Big) 
       \nonumber\\ &&\qquad\qquad
    + 
  \Big(   {t_5} -3{t_2}{t_3}\Big) 
   \Big( \frac{-{{u_1}}^3}{3} - {u_1}{u_2} - {u_3} \Big)
  + \Big( t_6+ {{t_2}}^3 - {{t_3}}^2 - 2{t_2}{t_4}  \Big)
     \Big( \frac{-{{u_1}}^2}{2} - {u_2} \Big) 
       \nonumber\\ &&\qquad\qquad
   - \Big(t_7+ {{t_2}}^2{t_3} - {t_3}{t_4} - {t_2}{t_5} \Big) 
     {u_1}  \ .
 \nonumber 
 \end{eqnarray}
Notice that $\cW(t,u)$ has $U(1)$ charge equal to $k+3$, which is one-half of the charge
of the effective prepotential $\cF(t)$ of the bulk sector.

These results, obtained by painstakingly solving the generalized WDVV constraints, 
suggest the following closed formula for the effective superpotential
in the general boundary preserving sector labeled by $(k,\ell)$:\footnote{We
  plan to discus this in more detail elsewhere.}
\be
\label{eq:closedW}
\cW(t,u)\ =\ -\sum_{i=0}^{k+2} g_{k+2-i}^{(k)}(t)\,h_{i+1}^{(\ell)}(u)~~,
\ee
where $h_i^{(\ell)}(u)$ are defined by:
\be
\label{eq:schurfct}
\log\big[1-\sum_{n=1}^{\ell+1}u_n\, y^n\big]\ :=\
\sum_{i=1}^\infty h_{i}^{(\ell)}(u)\,y^i
\ee
and $g_{k+2-i}^{(k)}(t)$ are the coefficients of $\varphi^i$ in the
bulk LG superpotential: 
$$
W_{LG}^{(k+2)}(t)\ =\ -\sum_{i=0}^{k+2}\, g_{k+2-i}^{(k)}(t)\,\varphi^{i}\ ,
$$
whose explicit form can be found for example in \cite{WDVV}
 (here $g_0^{(k)}=-1/(k+2)$ and $g_1^{(k)}=0$).
Equation (\ref {eq:closedW}) implies the following expression for 
the deformed one-point correlators on the disk:
$$
\cF_{\ell+1-\al}(t)\ \equiv\ \6_{u_\al}\cW(t,u)|_{u=0}\ 
=\ g_{k+3-\al}(t)\ .
$$

As explained in Subsection
\ref{subsubsec:tadpole}, in the presence of
deformations the tadpoles must be canceled by shifting to a new vacuum for which $\6_{u_\al}\cW(t,u)=0$.
A reassuring consistency check is provided by solving this
condition,  where $t_i$ are treated as parameters.
These equations give a set of constraints relating $u$ and $t$, 
thereby determining an affine algebraic variety ${\cal Z}_{k,l}$.
This can be parameterized by solving for $\{t_\bullet\}=\{t_{k+2-\ell},...,t_{k+2}\}$ in terms of
$u_\al$ and $\{t_\circ\}=\{t_{2},...,t_{k+1-\ell}\}$. 
We find that the locus ${\cal Z}_{k,l}$ has the property that the bulk superpotential
factorizes along it as follows:
\begin{equation}
    \label{eq:Weffact}
    W_{LG}^{(k+2)}(\vp,u,t_\circ,t_\bullet(t_\circ,u))\ =\ 
    J\ssc{\ell+1}(\vp,u)\,E\ssc{k+1-\ell}(\vp,u,t_\circ)~~,
\end{equation}
with
\begin{equation}
    \label{eq:Jeffact}
    J\ssc{\ell+1}(\vp,u)\ =\ \vp^{\ell+1} - 
    \sum_{\al=0}^\ell u_{\ell+1-\al}\,\vp^\al\ ,
\end{equation}
and
\begin{equation}
    \label{eq:Eeffact}
    E\ssc{k+1-\ell}(\vp,u,t_\circ)\ =\
     -\sum_{i=\ell+1}^{k+2}g_{k+2-i}(t_\circ)\,\Big(\sum_{n=0}^{i-\ell-1}
    \, \varphi^n\,f_{i -\ell-n-1}^{(\ell)}(u)\Big)~~,
\end{equation}
where the coefficients $f^{(\ell)}_i$ are determined by the relation
$\frac{1}{1-\sum_{n=1}^{l+1} u_n y^n} := - \sum_{i=0}^{\infty} {f_i^{(l)} y^i}$~.

In the untwisted model, the physical interpretation is as follows. Generic bulk ($t$) and
boundary ($u$) perturbations break supersymmetry, a phenomenon which can be traced
back to the boundary terms (\ref{eq:Qvardesc}) and (\ref{eq:Qvardescbound}) in
    the BRST variation of integrated descendants. Thus $t$ and $u$ `feel' a potential which
represents an obstruction against such  deformations.  
The effects of the boundary terms cancel and supersymmetry is maintained
only when bulk and boundary deformations are locked together
through the relation $W_{LG}=J\,E$~~--~ this cancellation was indeed precisely why one had
to introduce a boundary potential in the first place
\cite{WarnerProblem,KapA,BHLS,KapC,CILLG}. Thus it is
no surprise, though a welcome check on our computations, that the
critical set ${\cal Z}$ of $\cW(t,u)$ with respect to the boundary
    deformation parameters $u$ 
corresponds to the factorization locus of the worldsheet LG superpotential in the
combined, bulk and boundary parameter space. As expected, this is precisely
the locus along which the boundary data preserve half of the supersymmetry of the
worldsheet action, thereby allowing for a meaningful coupling to $B$-type D-branes. This is similar
in spirit to ref.~\cite{CaVa}, where, in a different
physical context,  critical points of effective superpotentials
were associated with factorization loci in the target space geometry.

If $E(\vp,u,t_\circ)$ is generic, its greatest common denominator with
$J(\vp,u)$ is trivial, hence according to (\ref{eq:boundideal})
 no physical open
string states survive after turning on bulk and boundary deformations by
allowing for general $t,u$. This
reflects tachyon condensation of the $D2\bar{D2}$ system \cite{HoriTC,KapA}, leading
to the trivial open string vacuum.  Only upon appropriately
specializing $E(\vp,u,t_\circ)$ such that it has 
a non-trivial common factor $H(\vp,u)$ with $J(\vp,u)$, does one find that 
some open string states remain in the
physical spectrum. Such sub-loci of the factorization variety ${\cal Z}_{k,l}$ correspond
to (a topological model of) tachyon condensation with non-trivial endpoint. 
In this version of tachyon condensation, 
the open string spectrum gets truncated while moving between different strata 
of the supersymmetry preserving locus of the effective superpotential. 
A very similar picture was found in  
\cite{CILd} for the case of the open A model close to a large radius
point of a Calabi-Yau compactification (see figure 2 of that paper).
The topological version of tachyon condensation was discussed in detail in 
\cite{CILa,CILb,CILg,CILh,CILd,CILe} in the context of open string field
theory. It also plays a central role in the work of \cite{Fukayabook,FukayamirrorB}.

Finally, let us give an example of effective superpotentials for the boundary
changing sector of minimal models. For simplicity we will not turn
on bulk deformations. In this situation we can study the
formation of D-brane composites in a fixed conformal bulk
background. Let us consider the minimal
model at level $k=3$ with the D-brane configuration
$\BS_{\ell=0}\oplus\BS_{\ell=1}$. In this setting, one has fermionic boundary operators,
$(\omega^{(00)},\omega^{(01)},\omega^{(10)},\omega^{(11)},\vp\omega^{(11)})$,
associated with the deformation parameters 
$(u^{(00)}_1,u^{(01)}_{3/2},u^{(10)}_{3/2},u^{(11)}_2,u^{(11)}_1)$ (see \cite{BHLS}).
The generalized WDVV equations again determine all amplitudes, giving the
following effective superpotential for the bosonic deformation parameters:
\begin{eqnarray}
  \label{eq:W3}
  \cW(t\!\!=\!\!0,u)&=&
  -\frac{1}{30} u^{(00)}_1 {}^6 - \frac{1}{15}   ~u^{(11)}_2 {}^3 -
  \frac{3}{10} 
  ~u^{(11)}_2 {}^2  ~u^{(11)}_1 {}^2 -  
  \frac{1}{5}   ~u^{(11)}_2   ~u^{(11)}_1 {}^4 -\frac{1}{30}
  ~u{}^{(11)}_1 {}^6- 
  \nn\\[15pt]
  && - \frac{1}{5}~u^{(00)}_1 {}^3  ~u^{(01)}_{3/2}   ~u^{(10)}_{3/2}  -  
  \frac{1}{5}  ~u^{(00)}_1    ~u^{(11)}_2   ~u^{(01)}_{3/2}   ~u^{(10)}_{3/2}-
  \\[15pt]
  &&- \frac{1}{5}  ~u^{(00)}_1 {}^2  ~u^{(11)}_1   ~u^{(01)}_{3/2}
  ~u^{(10)}_{3/2}  -  
  \frac{2}{5}   ~u^{(11)}_2   ~u^{(11)}_1   ~u^{(01)}_{3/2}   ~u^{(10)}_{3/2}-
  \nn\\[15pt]
  && - \frac{1}{5}  ~u^{(00)}_1   ~u^{(11)}_1 {}^2  ~u^{(01)}_{3/2}
  ~u^{(10)}_{3/2}  -  
  \frac{1}{5}  ~u^{(11)}_1 {}^3  ~u^{(01)}_{3/2}   ~u^{(10)}_{3/2}  -
  \frac{1}{10} 
  ~u^{(01)}_{3/2} {}^2  ~u^{(10)}_{3/2} {}^2 \nn ~~.
\end{eqnarray}
In view of the results of this paper, one expects the situation to be similar
to that found  for boundary preserving sectors, 
namely the critical set of $\cW(t\!\!=\!\!0,u)$ should
parameterize deformations compatible with a factorized  bulk potential:
$W_{LG}^{(k+2)} \unit = J\ssc{01} E\ssc{01}=E\ssc{01} J\ssc{01}$, where
$J\ssc{01}$ and $E\ssc{01}$ are now matrices comprising both the boundary
 changing and boundary preserving sectors. Making an appropriate ansatz for
the dependence of $J\ssc{01}$ and $E\ssc{01}$ of the parameters $u^{(\ell\ell')}_a$
(namely an ansatz compatible with the $U(1)$ charges), one indeed finds that 
the critical locus of $\cW(t\!\!=\!\!0,u)$, characterized by:
\begin{eqnarray}
  \label{eq:minW}
  u_1^{(11)} &=& - u_1^{(00)}~~\nn\\[10pt]
  u_2^{(11)} &=& - u_1^{(00)}{}^2~~\\[10pt]
  u_{3/2}^{(01)} u_{3/2}^{(10)} &=& - u_1^{(00)}{}^3 \nn ~~,
\end{eqnarray}
implies factorization of $W_{LG}\ssc{k+2} \unit$. The
matrices $J\ssc{01}$ and $E\ssc{01}$ are uniquely determined by
(\ref{eq:minW}) and take the form:
\begin{eqnarray}
  \label{eq:JEbc}
  J\ssc{01} &=& \left(\begin{array}{cc}
    \vp - u_1^{(00)} & -u_{3/2}^{(01)} \\[5pt]
    - u_{3/2}^{(10)} & \vp^2 + u_1^{(00)} \vp + u_1^{(00)}{}^2
  \end{array}\right)~~,\\ \nonumber
  E\ssc{01} &=& \frac{1}{5}\left(\begin{array}{cc}
    \vp^4 + u_1^{(00)} \vp^3 + u_1^{(00)}{}^2 \vp^2 &
    u_{3/2}^{(01)} \vp^2\\[5pt]
    u_{3/2}^{(10)} \vp^2 &
    \vp^3 - u_1^{(00)} \vp^2
  \end{array}\right)~~.
\end{eqnarray}
Although the deformations (\ref{eq:minW}) preserve 
half of bulk supersymmetry, the open string spectrum is
truncated for generic values of the boundary deformation parameters.
This is different from the boundary flows studied recently in
\cite{HoriLG}, for which the open string spectrum does not truncate
along the deformation locus considered there.

\section{Outlook}

Our work brings up a number of interesting questions. One is how to make
contact with the Landau-Ginzburg formulation of B-type topological open
strings. This would amount to investigating the $A_\infty$ structure of the
contact terms through the methods of open string field theory, as proposed in
a general context in \cite{CILc}. We intend to present our findings in this
direction in a subsequent paper.

Another important problem is to apply the generalized WDVV equations to
theories allowing for exactly marginal bulk deformations, like Calabi-Yau
manifolds with D-branes, and use them to learn about D-brane
superpotentials. This was one of the main motivations for the present paper,
and should eventually allow one to make contact with geometric computations
based on mirror symmetry, such as those performed in \cite{Vafa,Klemm,LMWb}.

Just as for the bulk theory, one expects that there is a connection of open-closed
topological minimal models with matrix models and integrable systems. It would
be especially interesting to understand the relation with the recent work of
\cite{ADKMV}. 
A related question concerns open string gravitational descendants in these
models, which requires a systematic study of topological gravity on bordered
Riemann surfaces. In particular, open string gravitational descendants should
lead lead to interesting generalizations of the Virasoro and
$W$-constraints. Some work in this direction, though from a different
perspective, was recently carried out in ~\cite{rastelli}.

\section{Acknowledgments}

The authors thank Ilka Brunner for collaboration in the early stages of
this project. W.L. thanks Peter Mayr and Nick Warner for
discussions. M.H. thanks Tristan Maillard for valuable 
communication. C.I.L thanks Albrecht Klemm for support
and interest in his work, as well as Anton Kapustin and Kentaro Hori for
stimulating conversations. W.L. and C.I.L. thank the
Kavli Institute for Theoretical Physics and University of California at Santa Barbara for a
pleasant stay, during which the present collaboration was formed. 
This
research was supported in part by the National Science Foundation
under Grant No. PHY99-07949.



\bib{\Segal}{G. Segal, ''Two-dimensional conformal field theories and modular
functors,'' IX-th International Congress on Mathematical Physics (Swansea,
1988), pp 22--37, Hilger, Bristol, 1989.}
\bib{\GGMV}{L.~Alvarez-Gaume, C.~Gomez, G.~W.~Moore and C.~Vafa,
``Strings In The Operator Formalism,''
Nucl.\ Phys.\ B {\bf 303}(1988)455.}

\bib{\WittenSFT}{E.~Witten,
``Noncommutative Geometry And String Field Theory,''
Nucl.\ Phys.\ B {\bf 268}(1986)253.}

\bib{\WittenNLSM}{E.~Witten,
``Topological Sigma Models,''
Commun.\ Math.\ Phys.  {\bf 118}(1988)411.}
\bib{\Wittenmirror}{E.~Witten,
``Mirror manifolds and topological field theory,''
arXiv:hep-th/9112056.}
\bib{\WittenTG}{J.~M.~F.~Labastida, M.~Pernici and E.~Witten,
``Topological Gravity In Two-Dimensions,''
Nucl.\ Phys.\ B {\bf 310}(1988)611.}
\bib{\Labastida}{J.~M.~F.~Labastida and P.~M.~Llatas,
``Topological matter in two-dimensions,''
Nucl.\ Phys.\ B {\bf 379}(1992)220 [arXiv:hep-th/9112051].}
\bib{\BCOV}{ M.~Bershadsky, S.~Cecotti, H.~Ooguri and C.~Vafa,
``Kodaira-Spencer theory of gravity and exact results for quantum string
amplitudes,'' Commun.\ Math.\ Phys.\ {\bf 165}(1994)311 [arXiv:hep-th/9309140].}
\bib{\BK}{S.~Barannikov, M.~Kontsevich, 
{\em Frobenius Manifolds and Formality of Lie Algebras of Polyvector Fields},  
Internat. Math. Res. Notices  {\bf 4}(1998) 201--215, alg-geom/9710032.} 
\bib{\JS}{J.~S.~Park, 
``Topological open p-branes,'' in {\em Symplectic geometry and mirror symmetry} (Seoul,
2000) pp. 311--384, World Sci. Publishing, River Edge, NJ, 2001 [arXiv:hep-th/0012141].}


\bib{\Candelas}{P.~Candelas, X.~C.~De la Ossa, P.~S.~Green and L.~Parkes,
``An Exactly Soluble Superconformal Theory From A Mirror Pair Of Calabi-Yau
Manifolds,'' Phys.\ Lett.\ B {\bf 258}(1991)118.}
\bib{\KlemmA}{
S.~Hosono, A.~Klemm, S.~Theisen and S.~T.~Yau,
``Mirror symmetry, mirror map and applications to Calabi-Yau hypersurfaces,''
Commun.\ Math.\ Phys.\  {\bf 167}(1995)301
[arXiv:hep-th/9308122].}
\bib{\KlemmB}{
S.~Hosono, A.~Klemm, S.~Theisen and S.~T.~Yau,
``Mirror symmetry, mirror map and applications to complete intersection
Calabi-Yau spaces,'' Nucl.\ Phys.\ B {\bf 433}(1995)501
[arXiv:hep-th/9406055].}
\bib{\KlemmC}{P.~Berglund, S.~Katz and A.~Klemm, 
``Mirror symmetry and the moduli space for generic hypersurfaces in toric
varieties,'' Nucl.\ Phys.\ B {\bf 456}(1995)153 [arXiv:hep-th/9506091].}
\bib{\Greene}{
P.~S.~Aspinwall, B.~R.~Greene and D.~R.~Morrison,
``Calabi-Yau moduli space, mirror manifolds and spacetime topology  change in
string theory,'' Nucl.\ Phys.\ B {\bf 416}(1994)414 [arXiv:hep-th/9309097].}


\bib{\WDVV}{
R.~Dijkgraaf, H.~Verlinde and E.~Verlinde,
``Topological Strings In D $<$ 1,''
Nucl.\ Phys.\ B {\bf 352}(1991)59.}
\bib{\Dubrovin}{
B.~Dubrovin, ``Geometry Of 2-D Topological Field Theories,''
Lecture Notes in Math. {\bf 1620}, Springer, Berlin, 1996 [arXiv:hep-th/9407018].}
\bib{\KlemmTheisen}{A.~Klemm, S.~Theisen and M.~G.~Schmidt,
``Correlation Functions For Topological Landau-Ginzburg Models With $C<=3$,''
Int.\ J.\ Mod.\ Phys.\ A {\bf 7}(1992)6215.}


\bib{\KachruA}{
S.~Kachru, S.~Katz, A.~E.~Lawrence and J.~McGreevy,
``Open string instantons and superpotentials,''
Phys.\ Rev.\ D {\bf 62}(2000)026001 [arXiv:hep-th/9912151].}
\bib{\KachruB}{S.~Kachru, S.~Katz, A.~E.~Lawrence and J.~McGreevy,
``Mirror symmetry for open strings,''
Phys.\ Rev.\ D {\bf 62}(2000)126005[arXiv:hep-th/0006047].}
\bib{\Vafa}{M.~Aganagic and C.~Vafa,
``Mirror symmetry, D-branes and counting holomorphic discs,''
arXiv:hep-th/0012041.}
\bib{\Klemm}{M.~Aganagic, A.~Klemm and C.~Vafa,
``Disk instantons, mirror symmetry and the duality web,''
Z.\ Naturforsch.\ A {\bf 57}(2002)1 [arXiv:hep-th/0105045].}
\bib{\LMWa}{
W.~Lerche, P.~Mayr and N.~Warner,
``Holomorphic N = 1 special geometry of open-closed type II strings,''
arXiv:hep-th/0207259.}
\bib{\LMWb}{
W.~Lerche, P.~Mayr and N.~Warner,
``N = 1 special geometry, mixed Hodge variations and toric geometry,''
arXiv:hep-th/0208039.}


\bib{\Zwiebachclosed}{
B.~Zwiebach, ``Closed string field theory: Quantum action and the B-V master equation,''
Nucl.\ Phys.\ B {\bf 390}(1993)33 [arXiv:hep-th/9206084].}
\bib{\Zwiebachoc}{B.~Zwiebach,
``Oriented open-closed string theory revisited,'' Annals Phys.\  {\bf 267},
(1998)193 [arXiv:hep-th/9705241].}
\bib{\Zwiebachareaclosed}{B.~Zwiebach, 
``How covariant closed string theory solves a minimal area problem'',  
Comm. Math. Phys. {\bf 136}(1991),  no. 1, 83--118.}
\bib{\Zwiebachproof}{B.~Zwiebach, ``A proof that Witten's open string
theory gives a single cover of moduli space'', Comm. Math. Phys.  {\bf 142} (1991),
no. 1, 193--216.}
\bib{\Zwiebachareaopen}{B.~Zwiebach, ``Minimal area problems and quantum open strings''.
Comm. Math. Phys.  {\bf 141} (1991), no. 3, 577--592.}
\bib{\WolfZwiebach}{M.~Wolf, B.~Zwiebach,
``The plumbing of minimal area surfaces,'' arXiv:hep-th/9202062.}
\bib{\WittenZwiebach}{E.~Witten and B.~Zwiebach, ``Algebraic structures
and differential geometry in $2-D$ string theory,'' Nucl.\ Phys.\ B {\bf 377},
(1992)55 [arXiv:hep-th/9201056].}


\bib{\WittenCS}{E.~Witten,
``Chern-Simons gauge theory as a string theory,''
Prog.\ Math.\  {\bf 133}(1995)637 [arXiv:hep-th/9207094].}
\bib{\Gaberdiel}{M.~R.~Gaberdiel and B.~Zwiebach,
``Tensor constructions of open string theories I: Foundations,''
Nucl.\ Phys.\ B {\bf 505}(1997)569 [arXiv:hep-th/9705038].}
\bib{\CILa}{C.~I.~Lazaroiu,
``Generalized complexes and string field theory,''
JHEP {\bf 0106}(2001)052 [arXiv:hep-th/0102122].}
\bib{\CILb}{C.~I.~Lazaroiu,
``Unitarity, D-brane dynamics and D-brane categories,''
JHEP {\bf 0112}(2001)031 [arXiv:hep-th/0102183].}
\bib{\CILc}{C.~I.~Lazaroiu,
``String field theory and brane superpotentials,'' JHEP {\bf 0110}(2001)018 
[arXiv:hep-th/0107162].}
\bib{\CILd}{C.~I.~Lazaroiu and R.~Roiban,
``Holomorphic potentials for graded D-branes,''
JHEP {\bf 0202}(2002)038 [arXiv:hep-th/0110288].}
\bib{\CILe}{C.~I.~Lazaroiu and R.~Roiban,
``Gauge-fixing, semiclassical approximation and potentials for graded
Chern-Simons theories,'' JHEP {\bf 0203}(2002)022 [arXiv:hep-th/0112029].}
\bib{\CILf}{C.~I.~Lazaroiu,
``D-brane categories,'' Int.J.Mod.Phys. A18 (2003) 5299-5335 [arXiv:hep-th/0305095].}
\bib{\Diaconescu}{D.~E.~Diaconescu,
``Enhanced D-brane categories from string field theory,''
JHEP {\bf 0106}(2001)016 [arXiv:hep-th/0104200].}
\bib{\CILg}{C.~I.~Lazaroiu,
``Graded Lagrangians, exotic topological D-branes and enhanced  triangulated
categories,'' JHEP {\bf 0106}(2001)064 [arXiv:hep-th/0105063].}
\bib{\CILh}{C.~I.~Lazaroiu, R.~Roiban and D.~Vaman,
``Graded Chern-Simons field theory and graded topological D-branes,''
JHEP {\bf 0204}(2002)023 [arXiv:hep-th/0107063].}

\bib{\KajiuraA}{
H.~Kajiura, ``Homotopy algebra morphism and geometry of classical string field  theory,''
Nucl.\ Phys.\ B {\bf 630}(2002)361 [arXiv:hep-th/0112228].}
\bib{\KajiuraB}{H.~Kajiura, 
``Noncommutative homotopy algebras associated with open strings,'' arXiv:math.QA/0306332.}
\bib{\HofmanMaA}{
C.~Hofman and W.~K.~Ma, ``Deformations of topological open strings,'' JHEP {\bf 0101}(2001)035
[arXiv:hep-th/0006120].}
\bib{\HofmanMaB}{
C.~M.~Hofman and W.~K.~Ma, ``Deformations of closed strings and topological open membranes,''
JHEP {\bf 0106}(2001)033 [arXiv:hep-th/0102201].}
\bib{\HofmanB}{C.~Hofman, ``On the open-closed B-model,''
JHEP {\bf 0311}(2003)069 [arXiv:hep-th/0204157].}
\bib{\Nakatsu}{
T.~Nakatsu,``Classical open string field theory: A(infinity)-algebra,  renormalization
group and boundary states,'' Nucl.\ Phys.\ B {\bf 642}, 13 (2002) [arXiv:hep-th/0105272].}

\bib{\VafaLG}{C.~Vafa,
``Topological Landau-Ginzburg Models",
Mod.\ Phys.\ Lett.\ A {\bf 6}(1991)337.}
\bib{\LercheWarner}
{W.~Lerche, C.~Vafa and N.~P.~Warner,
``Chiral Rings In N=2 Superconformal Theories,''
Nucl.\ Phys.\ B {\bf 324}(1989)427.}
\bib{\BLNMW}{
M.~Bershadsky, W.~Lerche, D.~Nemeschansky and N.~P.~Warner,
 ``Extended N=2 superconformal structure of gravity and W gravity coupled to matter,''
Nucl.\ Phys.\ B {\bf 401}(1993)304 [arXiv:hep-th/9211040].}
\bib{\LSW}{W.~Lerche, D.~J.~Smit and N.~P.~Warner,
 ``Differential equations for periods and flat coordinates in two-dimensional
topological matter theories,'' Nucl.\ Phys.\ B {\bf 372}(1992)87 [arXiv:hep-th/9108013].}


\bib{\Hori}{K.~Hori, A.~Iqbal and C.~Vafa,
``D-branes and mirror symmetry,''arXiv:hep-th/0005247.}

\bib{\WarnerProblem}{N.~P.~Warner,
``Supersymmetry in boundary integrable models,''
Nucl.\ Phys.\ B {\bf 450}(1995)663 [arXiv:hep-th/9506064].}
\bib{\KontsevichLG}{M. Kontsevich, {\em unpublished}.}
\bib{\KapA}{A.~Kapustin and Y.~Li,
``D-branes in Landau-Ginzburg models and algebraic geometry,''
JHEP {\bf 0312}(2003)005 [arXiv:hep-th/0210296].}
\bib{\Orlov}{D. Orlov, 
"Triangulated categories of singularities and D-branes in Landau-Ginzburg
  models", math.AG/0302304.}
\bib{\BHLS}{I.~Brunner, M.~Herbst, W.~Lerche and B.~Scheuner,
``Landau-Ginzburg realization of open string TFT,''
arXiv:hep-th/0305133.}
\bib{\KapB}{A.~Kapustin and Y.~Li,
``Topological correlators in Landau-Ginzburg models with boundaries,''
arXiv:hep-th/0305136.}
\bib{\KapC}{A.~Kapustin and Y.~Li,
``D-branes in topological minimal models: The Landau-Ginzburg approach,''
arXiv:hep-th/0306001.}
\bib{\CILLG}{C.~I.~Lazaroiu, 
``On the boundary coupling of topological Landau-Ginzburg models,''
arXiv:hep-th/0312286.}
\bib{\DiacLG}{
S.~K.~Ashok, E.~Dell'Aquila and D.~E.~Diaconescu,
``Fractional branes in Landau-Ginzburg orbifolds,''
arXiv:hep-th/0401135.}
\bib{\HoriLG}{K.~Hori, ``Boundary RG flows of N = 2 minimal models,''arXiv:hep-th/0401139.}


\bib{\Sonoda}{H.~Sonoda, ``Sewing Conformal Field Theories. 2,''
Nucl.\ Phys.\ B {\bf 311}(1988)417.}
\bib{\Lewellen}{D.~C.~Lewellen,
``Sewing constraints for conformal field theories on surfaces with
boundaries,'' Nucl.\ Phys.\ B {\bf 372}(1992)654.}
\bib{\MooreSegal}{G. Moore and G. Segal, unpublished; 
see http://online.kitp.ucsb.edu/online/mp01/}
\bib{\Moore}{G.~W.~Moore,
``Some comments on branes, G-flux, and K-theory,''
Int.\ J.\ Mod.\ Phys.\ A {\bf 16}(2001)936 [arXiv:hep-th/0012007].}
\bib{\CILtop}{C.~I.~Lazaroiu,
 ``On the structure of open-closed topological field theory in 
 dimensions'',
Nucl.\ Phys.\ B {\bf 603}(2001)497
[arXiv:hep-th/0010269].}
\bib{\NatanzonA}{A.~Alexeevski, S.~Natanzon,
``Non-commutative extensions of two-dimensional topological field theories and
Hurwitz numbers for real algebraic curves,'' math.GT/0202164.}
\bib{\NatanzonB}{S.M.Natanzon,
``Extension cohomological fields theory and noncommutative Frobenius manifolds'',
math-ph/0206033.}

\bib{\StasheffA}{J.~D.~Stasheff, ''On the homotopy associativity of
    H-spaces, I.'', Trans. Amer. Math. Soc. {\bf 108}(1963)275.}
\bib{\StasheffB}{J.~D.~Stasheff, ''On the homotopy associativity of
    H-spaces, II.'', Trans. Amer. Math. Soc. {\bf 108}(1963)293.}
\bib{\KontsevichHMS}{M. Kontsevich, "Homological Algebra of Mirror Symmetry", 
Proceedings of the International Congress of Mathematicians
(Zurich, 1994) 120--139, Birkhauser, Basel, 1995. [arXiv:alg-geom/9411018].}
\bib{\Fukayainfty}{K.~Fukaya, ``Morse homotopy, $A_\infty$ categories and Floer homologies,'' 
in Proceedings of the 1993 Garc Workshop on Geometry and Topology, Lecture Notes 
Series, vol.~{\bf 18} (Seoul National University, 1993), pp.~1--102}
\bib{\Fukayabook}{K.~Fukaya, Y.~G.~Oh, H.~Ohta and K.~Ono, 
``Lagrangian intersection Floer theory~--- anomaly and obstruction,'' preprint
available at \\http://www.kusm.kyoto-u.ac.jp/$\sim$fukaya/fukaya.html}
\bib{\Fukayarev}{K.~Fukaya, ``Deformation theory, homological algebra and mirror symmetry,'' 
in "Geometry and Physics of Branes" (Como, 2001), pp.~121--209, 
Ser. High Energy Phys. Cosmol. Gravit. (IOP, Bristol, 2003)}
\bib{\Fukayactg}{K.~Fukaya and Y.~G.~Oh, 
"Zero-loop open strings in the cotangent bundle and Morse homotopy", 
Asian J. Math.  {\bf 1} (1997) no. 1, 96--180.}
\bib{\FukayamirrorA}{K.~Fukaya, ``Floer homology and mirror symmetry I,'' 
Proceedings of the Winter School on Mirror Symmetry, Vector Bundles and Lagrangian
Submanifolds (Cambridge, Massachusetts, 1999), pp.~15--43, 
AMS/IP Stud. Adv. Math. {\bf 23} (Amer. Math. Soc., Providence, RI, 2001)}
\bib{\FukayamirrorB}{K. Fukaya, "Floer homology and mirror symmetry II", 
in "Minimal surfaces, geometric analysis and symplectic geometry",
Adv. Stud. Pure Math {\bf 34}, Math. Soc. Japan, Tokyo 2002.}


\bib{\KSV}{
T.~Kimura, J.~Stasheff and A.~A.~Voronov,
``On operad structures of moduli spaces and string theory,''
Commun.\ Math.\ Phys.\  {\bf 171}(1995)1 [arXiv:hep-th/9307114].}
\bib{\Voronov}{A.~A.~Voronov,
``Topological field theories, string backgrounds and homotopy algebras,''
Adv. Appl. Clifford Algebras  {\bf 4}(1994),  Suppl. 1, 167--178. [arXiv:hep-th/9401023].}
\bib{\KVZ}{T.~Kimura, A.~A. Voronov; G.~J.~Zuckerman,
``Homotopy Gerstenhaber algebras and topological field theory'',
in  {\em Operads: Proceedings of Renaissance Conferences (Hartford,
CT/Luminy, 1995)},  pp. 305--333, Contemp. Math. {\bf 202}, published by 
Amer. Math. Soc., Providence, RI, 1997.}
\bib{\Markl}{M.~Markl,
``Loop homotopy algebras in closed string field theory,''
Commun.\ Math.\ Phys.\  {\bf 221}(2001)367 [arXiv:hep-th/9711045].}
\bib{\Penkava}{M.~Penkava, A.~Schwarz,
``$A\sb \infty$ algebras and the cohomology of moduli spaces.'', 
Amer. Math. Soc. Transl. Ser. {\bf 2} 169 [arXiv:hep-th/9408064].}

\bib{\Sullivan}{D.~Sullivan, 
``Open and Closed String field theory interpreted in classical Algebraic
Topology'', math.QA/0302332.}

\bib{\Voronovsc}{A.~A.~Voronov, ``The Swiss-Cheese Operad'', math.QA/9807037.}
\bib{\VoronovG}{A.~A.~Voronov, ``Homotopy Gerstenhaber algebras'',
Conference Moshe Flato 1999 (G. Dito and D. Sternheimer, eds.), vol. 2. Kluwer
Academic Publishers, the Netherlands, 2000, pp. 307-331 [math.QA/9908040]}
\bib{\VoronovHochshild}{P. Hu, I. Kriz, A. A. Voronov, 
``On Kontsevich's Hochschild cohomology conjecture'', math.AT/0309369.}
\bib{\Tamarkin}{ D.~E.~Tamarkin, 
``Deformation complex of a d-algebra is a (d+1)-algebra'',
 math.QA/0010072.}

\bib{\AS}{
S.~Axelrod and I.~M.~Singer,
``Chern-Simons Perturbation Theory. 2,''
J.\ Diff.\ Geom.\  {\bf 39}(1994)173
[arXiv:hep-th/9304087].}


\bib{\CaVa}
{F.~Cachazo and C.~Vafa,
``N = 1 and N = 2 geometry from fluxes,''
arXiv:hep-th/0206017.}

\bib{\Cardyir}{
J.~L.~Cardy,
``Boundary Conditions, Fusion Rules And The Verlinde Formula,''
Nucl.\ Phys.\ B {\bf 324}(1989)581.}

\bib{\Lazarev}{A.~Lazarev,
``Hochschild Cohomology and Moduli Spaces of Strongly Homotopy
Associative Algebras''[arXiv:hep-th/0204062].}

\bib{\rastelli}{
D.~Gaiotto and L.~Rastelli,
 ``A paradigm of open/closed duality: Liouville D-branes and the Kontsevich
model,''arXiv:hep-th/0312196.}

\bib{\ADKMV}{M.~Aganagic, R.~Dijkgraaf, A.~Klemm, M.~Marino and C.~Vafa,
``Topological strings and integrable hierarchies,'' arXiv:hep-th/0312085.}

\bib{\HoriTC}{K.~Hori, ``Linear models of supersymmetric D-branes,''
arXiv:hep-th/0012179.}

\bib{\Tomasiello}{
A.~Tomasiello,
``A-infinity structure and superpotentials,''
JHEP {\bf 0109}(2001)030 [arXiv:hep-th/0107195].}

\bib{\MayrLoc}{
P.~Mayr,
``Summing up open string instantons and N = 1 string amplitudes,''
arXiv:hep-th/0203237.}

\bib{\MayrFF}{
P.~Mayr,
``N = 1 mirror symmetry and open/closed string duality,''
Adv.\ Theor.\ Math.\ Phys.\  {\bf 5}(2002)213
[arXiv:hep-th/0108229].}

\bib{\Polishchuk}{A.~Polishchuk,
``Homological mirror symmetry with higher products,''
Proceedings of the 
Winter School on Mirror Symmetry, Vector Bundles and Lagrangian Submanifolds 
(Cambridge, MA, 1999), pp 247--259, 
AMS/IP Stud. Adv. Math. {\bf 23}, Amer. Math. Soc., Providence, RI, 2001
[arXiv:math.AG/9901025].}

\bib{\mirbookA}{Ed.~S.~T.~Yau, {\em Essays on mirror manifolds}.
\newblock International Press, 1992.}

\bib{\mirbookB}{Ed.~B.~R.~Greene~et al, {\em Mirror symmetry {II}}.
\newblock International Press, 1997.}

\bib{\mirbookC}{Ed.~K.~Hori~et al, {\em Mirror symmetry}.
\newblock American Mathematical Society, 2003.
\newblock ISBN 0821829556.}

\bib{\VerlindeWarner}{E.~Verlinde and N.~P.~Warner,
``Topological Landau-Ginzburg matter at c = 3,''
Phys.\ Lett.\ B {\bf 269}(1991)96.}

%

\catcode`\@=11
\def\mref#1{\ifx\und@fined#1{Need to supply reference \string#1.}\else #1 \fi}
\catcode`\@=12

\end{document}